\documentclass[11pt]{article}
\usepackage{graphicx} % Required for inserting images
\usepackage{amsfonts}
\usepackage{amsmath}
\usepackage{hyperref}
\usepackage{amssymb}
\usepackage[percent]{overpic}
\usepackage{tikz}
\usepackage{amsthm}
\usepackage{verbatim}
\usepackage{grffile}

\usepackage{amsmath,amsfonts,amscd,a4wide}
\usepackage{amsmath}
\usepackage{amssymb}
\usepackage{graphicx}
\usepackage{amsthm}
\usepackage{color}
\usepackage{color}
\usepackage[margin=0.75in]{geometry}

\usepackage{color}
\usepackage{transparent}
\usepackage[skip=10pt,font=footnotesize]{caption}
\usepackage{subcaption}
\usepackage{enumitem}
 {\begin{trivlist} \item[]{\bf Proof. }}%
 {\hspace*{\fill}$\rule{.4\baselineskip}{.4\baselineskip}$\end{trivlist}}

 {\begin{trivlist}\item[]\textbf{Acknowledgments.}}{\end{trivlist}}

\makeatletter\@addtoreset{figure}{section}\makeatother

\makeatletter \@addtoreset{equation}{section} \makeatother

\newcommand{\R}{\mathbb{R}}
\newcommand{\C}{\mathbb{C}}

        \newcommand{\tl}[1]{\tilde{#1}}

        \newcommand{\beq}{\begin{equation}}
        \newcommand{\eeq}{\end{equation}}
        \newcommand{\ba}{\begin{align}}
        \newcommand{\ea}{\end{align}}

        \newcommand{\ri}{\mathrm{i}}
        \newcommand{\br}{\mathrm{br}}
        
        \newcommand{\rlin}{\mathrm{lin}}
        
        \newcommand{\re}{\mathrm{e}}
        
        \newcommand{\eps}{\varepsilon}

        \newcommand{\rre}{\mathrm{Re}}

        \newcommand{\tmu}{\tilde \mu}
        \newcommand{\tomega}{\tilde \omega}

                                \newcommand{\rss}{\mathrm{ss}}
                                       
                                                \newcommand{\ru}{\mathrm{u}}

\title{Patterned fronts in the wake of a parameter ramp in the complex Ginzburg Landau equation}
\author{ Ryan Goh\thanks{Department of Mathematics and Statistics, Boston University, 665 Commonwealth Ave., Boston,  MA 02215, USA; \texttt{rgoh@bu.edu}.}, Benjamin Krewson\thanks{Department of Mathematics and Statistics, Boston University, 665 Commonwealth Ave., Boston,  MA 02215, USA;}, 
Nilay Patel\thanks{Department of Mathematics, University of California-Berkeley, 970 Evans Hall, MC 3840, Berkeley, CA 94720-3840, USA; }, 
Kiersten Ratcliff\thanks{Department of Mathematics, University of Alabama at Birmingham, University Hall, Room 4005,
1402 10th Avenue South,
Birmingham AL 35294-1241, USA; }}

\begin{document}
\maketitle
\begin{abstract}
We study of the formation of pattern-forming fronts in the presence of a rigidly-propagating parameter ramp which is slowly-varying in space.  In the context of the prototypical supercritical complex Ginzburg-Landau equation, we show that not only the leading order front interface, but also the selected spatial wave number is governed by the transition of the ramp between absolute and convective instability. The slow ramp then induces a further delay of the front interface and perturbation of the selected wave number, controlled by the slow passage near a complex fold of strong- and weak-stable eigenspaces.  To analyze the behavior near this fold, we perform a multiple scales analysis to predict the higher-order front interface delay in terms of zeros and poles of a complex Airy quotient inner solution. We confirm these predictions with numerical continuation of heteroclinics in the associated traveling wave equation. We also numerically characterize their spectral stability, finding accumulation of eigenvalues consistent with previous results on slow absolute spectrum. We then show the leading-order absolute/convective instability heuristic accurately describes selected wave numbers in an analogous slowly-ramped Swift-Hohenberg equation.

    \textbf{Keywords: pattern formation, invasion front, slow parameter ramp, geometric singular perturbation theory, dynamic bifurcation} 
    
\textbf{Mathematics Subject Classification:  34E15, 34E13 , 35B25, 35B36}% 35B36,35B32,
\end{abstract}

\section{Introduction}
%%Introduction

Parameter ramp heterogeneities have been shown to mediate and control pattern formation in a variety of settings.  This includes fluid experiments, such as Taylor-Couette flow  with a varying distance between inner and outer cylinder walls \cite{cannell83}, fluid flow past a slowly-varying obstacle \cite{hunt91,couairon99}, and Rayleigh-B\'enard convection with a spatial variation in the temperature difference between between upper and lower plates \cite{steinberg1985pattern}. It also arises in heterogeneous quenching in the light-sensing CDIMA chemical reaction \cite{miguez2006effect,konow2019turing}, and in signaling gradients in biological pattern formation \cite{HISCOCK2015408,digit,feng21}. In these contexts, parameter ramps have been found to significantly narrow the band of admissible stripe wave numbers compared with the full Busse balloon of stable wave numbers in the un-ramped homogeneous system. Additionally, heterogeneities which propagate as time evolves have been shown to suppress the formation of defects and \emph{select} the orientation and wave number of both striped and spotted patterns. These are sometimes called \emph{quenching} or \emph{directional solidification} heterogeneities, and are often used as an experimental model for growth processes in pattern forming systems; see \cite{goh2023growing} for a review. 

The mathematical analysis of such phenomena arose in the context of the aforementioned fluid experiments, using phase-diffusion \cite{kramer1982wavelength, rehberg1987rayleigh}, Ginzburg-Landau \cite{pomeau1983pattern}, and Swift-Hohenberg type equations \cite{riecke1987perfect}. To summarize these results, consider an example of the latter, posed in one spatial dimension
\beq\label{e:sh}
u_t = -(1+\partial_x^2)^2 u + \mu(x) u - u^3, \qquad u\in \R, x\in \R.
\eeq 
In the homogeneous system, with $\mu\equiv \mu_0,$ constant with $0<\mu_0\ll1$, there exists a family of stable periodic stripes $u_p(x;k)\approx \sqrt{4(\mu_0 - \kappa^2)/3} \cos(k x),$ for $\kappa^2 = 1 - k^2$  in the Eckhaus-stable regime  $|\kappa| < \sqrt{\mu_0/3} + \mathcal{O}(\mu_0)$, implying that wave numbers $k$ must roughly lie in a $\mathcal{O}(\mu_0^{1/2})$-wide band around 1. A model parameter ramp considered in many of the works is a slowly-varying heterogeneity $\mu$ which monotonically moves the system through ``criticality" at $\mu = 0$, with $\mu(x)\rightarrow \mu_\pm $ as $x\rightarrow\pm\infty$ respectively. Here $\mu_+<0$ so that $u = 0$ is stable and patterns suppressed, while $\mu_->0$ allowing pattern forming instabilities. Take for example $\mu(x) = -\mu_0 \tanh(\eps x)$ with ramp slope $0<\eps \ll1$. 

The heterogeneity acts as a ``wave number selecting" effective boundary condition or ``pinning" condition which leads to patterned front solutions connecting a periodic stripe $u_p(x+\varphi;k)$ at $x = -\infty$ and $u = 0$ at $x = +\infty,$ where the asymptotic phase $\varphi$ of the pattern at $x = -\infty$ determines its asymptotic wave number $k$ via what is known as a \emph{strain-displacement} relation \cite{morrissey2015characterizing,pomeau1981wavelength,scheel2018wavenumber}. 
For small $\eps>0$, the range of wave numbers traced out by this relation is exponentially small in $\eps$, that is $|k - 1|\leq \mathcal{O}(\re^{-C/\eps})$ for some $C>0$ ; see Fig. 4.1 of \cite{goh2023growing}. This is contrasted with the opposite limit of an infinitely steep ramp, $\eps\gg1$, where the selected wave number band is wider, with $|k-1|\leq \mu_0/16 + \mathcal{O}(\mu_0^{3/2})$; see \cite{scheel2018wavenumber}. 

We also note such ramps have been considered with subcritical nonlinearities \cite{pomeau1983pattern,krause2024pattern} where the Maxwell point determines transition between $u = 0$ and $u = u_p$, as well as stripes and spots in multiple spatial dimensions for various geometries \cite{malomed1993two,hoyle1995steady,paul2002rayleigh}. The work \cite{kuske1997pattern} gives a general Ginzburg-Landau modulational theory for pattern-forming systems with slowly-varying geometry. The recent work \cite{krause2024pattern} uses WKB-type approaches to study patterns for a variety of heterogeneities, and \cite{dalwadi2023universal} derives and studies modulational equations for spatio-temporal ramps.

Our work is most closely connected to that of \cite{malomed1993ramp} which studied pattern-forming fronts in the complex Ginzburg Landau equation with a quenching ramp heterogeneity, $\mu(x - ct)$. It showed that a unique wave number is selected by the quenching speed and gives implicit integral formulas for the wave number which can be asymptotically approximated in various simplifying limits. Our previous works \cite{goh2024pitchfork,goh2023fronts} considered non-patterned asymptotically constant fronts in the Allen-Cahn equation with slowly-varying quenching heterogeneity. This setting can be viewed as the leading-order real-part of the normal-form modulation equation for \eqref{e:sh} with a quench of the form $\mu(x - ct)$.  In these works, geometric singular perturbation theory (GSPT) was used to rigorously establish fronts and prove that the rigid movement of the quench with speed $c$ and slow ramp slope $\eps$ induce a delay in the onset of the front interface - where the front becomes large amplitude - in the wake of the heterogeneity. At leading order, we showed that the front interface is located where $\mu$ passes from convective to absolute instability. At next order, we showed that the slow-variation of the ramp induces an $\eps$-dependent correction controlled by a dynamic slow passage near a bifurcation; for $c\ll \mathcal{O}(\eps^{1/3})$  a pitchfork at $\mu =0$, while for $c = \mathcal{O}(1)$ in $\eps$, a fold in the projective linearized dynamics at the absolute-to-convective (AC) transition. 

We continue this line of study by considering pattern-forming fronts and characterizing the dependence of the selected wave number on quenching speed $c$ and ramp slope $\eps$. As it is one of the simplest pattern-forming systems, we consider the prototypical supercritical cubic Complex Ginzburg Landau (CGL) equation \eqref{e:cgl}, where the corresponding pattern-forming front can be studied as a heteroclinic traveling wave solution in a finite-dimensional spatial dynamical system. Using the AC instability transition, we give a leading-order prediction for the selected temporal frequency, spatial wave number, and front location for a large range of speeds in the limit of small ramp slope. We describe the spatial structure of the front using a GSPT framework, where the small amplitude dynamics ahead of the front interface, $A\sim0$, are governed by a complex projective Riccati equation with slowly-varying coefficients, while the large amplitude dynamics are given by slow amplitude variation along a plane-wave solution. The heteroclinic orbit arises via a transverse unfolding of invariant manifolds in the temporal frequency parameter, indicating that a given $c,\eps$ pair selects a locally unique spatial wave number.  We describe the front interface using an Airy quotient inner solution approximation of the corresponding complex projective Riccati equation.  This accurately predicts the behavior of the front interface and higher-order bifurcation delay. We then confirm our predictions and investigate higher-order corrections to front location and selected wave number using numerical continuation of front profiles. We also numerically explore the spectral stability of such fronts. We find that, when the far-field selected periodic state is diffusively stable, eigenvalues associated with slow-absolute spectrum accumulate on the origin from the left, indicating spectral stability.  We remark that, while our numerical and asymptotic results accurately characterize pattern-forming fronts in this setting, we have left rigorous existence of these fronts using GSPT and geometric desingularization to future study.

To show the applicability of our AC instability heuristic prediction for the wave number in other patterned systems, we accurately predict the front and pattern selection dynamics in a Swift-Hohenberg equation with a similar slowly-varying parameter ramp.  We thus expect this heuristic to give accurate leading order predictions for super-critical pattern formation in a dissipative system with a slowly-varying but rigidly propagating parameter ramp heterogeneity. In short, our main finding can be summarized as follows: 

\emph{For a monotonic quenching heterogeneity $\mu(x-ct)$ with speed $c$ and small ramp slope $\eps$, the $\mu$ value for which the trivial state is marginally absolutely unstable determines not only the leading-order front interface location, but also the leading-order temporal oscillation frequency of the front. This frequency, via an appropriate dispersion relation evaluated at $\mu_-:=\lim_{\xi\rightarrow-\infty}\mu(\xi)$, determines the leading-order spatial wave number of the asymptotic pattern selected in the wake. The small ramp slope $\eps$ then induces higher order corrections to interface location and wave number.}

This work is organized as follows.  Section \ref{s:cgl} introduces the ramped CGL equation, gives preliminary information on the constant-coefficient CGL needed for our study, derives the leading-order wave number prediction, and lists the generic assumptions in CGL for which our characterization is valid. Section  \ref{s:het} then describes the multi-scale structure of the patterned fronts, investigating the slow manifolds both ahead and behind the front interface. It also derives and analyzes an Airy quotient function inner solution for the front interface, and provides a prediction for the higher order, $\eps$-dependent, front interface delay in terms of complex zeros and poles of the Airy function. Section \ref{s:numcont} then provides the results of numerical continuation of fronts and compares them to the predictions made in previous sections. Section \ref{s:stab} provides the results of numerical computation of spectra of the associated linearization about these fronts. Finally, Section \ref{s:sh} derives the AC instability transition for an analogous slowly-ramped Swift-Hohenberg equation and shows it accurately predicts front interface and selected wave number. Section \ref{s:disc} discusses our results and areas of current and future study.

\section{Slowly-varying ramped CGL Equation}\label{s:cgl}

The prototypical super-critical complex CGL equation
\begin{equation}\label{e:cgl0}
    A_t=(1+\ri\alpha)A_{xx}+\mu A-(1+\ri\gamma)|A|^2A, 
\end{equation}
with $A:\mathbb{R}\times\mathbb{R}_+\rightarrow \mathbb{C}$ and $\alpha,\gamma,\mu\in\mathbb{R}$, arises as the modulation equation for systems at the onset of a (supercritical) oscillatory instability \cite{CGL_Lambda_Stability, mielke2002ginzburg}. Due to the explicit form of its periodic wave trains, and the gauge symmetry $A\mapsto \re^{\ri\theta}A$, it serves as an accessible but useful prototype for the study of a variety of coherent structures in various parameter regimes and spatial dimensions, including periodic patterns, traveling fronts, defects, domain walls, spirals, and vortices \cite{CGL_Lambda_Stability,van1992fronts}. %Hence we use it to investigate the effect of a parameter ramp on pattern-forming fronts.%We once again remark, as cited above, that such a ramp has been considered by \cite{malomed1993ramp}.    

We shall focus on a spatio-temporally \emph{ramped} version of \eqref{e:cgl0} where the stability of the trivial equilibrium state $A\equiv0$ is moderated by a traveling, or ``quenching" parameter heterogeneity
\begin{equation}\label{e:cgl}
A_t = (1+\ri\alpha)A_{xx}+\mu(x-ct) A-(1+\ri\gamma)A|A|^2, 
\end{equation}
with $\mu(\xi) = \tanh(-\eps \xi)$. Here, for $\xi=x-ct<0$,  $\mu>0$ and the trivial state is locally unstable, while for $\xi>0$,  $\mu<0$ and the trivial state is locally stable (and patterns suppressed). Hence, the heterogeneity travels across the spatial domain with quenching speed $c$, exciting a pattern-forming instability in its wake. In this work, we focus on the slowly-varying regime $0<\eps \ll1$ - the sharp regime is discussed in \cite[\S 5.3]{goh2023growing} and references therein. The heterogeneity  satisfies $\mu\rightarrow \pm 1$ as $\xi\rightarrow \mp \infty$, allowing for fronts which converge to a fixed coherent structure with constant amplitude at $\xi = -\infty$.  We choose this specific form of $\mu$ for technical reasons. Namely, it satisfies the ODE $\mu_\xi = -\eps (1 - \mu^2)$, allowing us to augment the resulting traveling wave ODE \eqref{e:cgl-tw} and work with an autonomous system in a compact region of phase space; see also \cite{hasan23} for related work on bifurcation- and rate-induced tipping of fronts in heterogeneous environments. We expect similar phenomena and results to occur for general functions $\mu$ which are monotonically decreasing, slowly-varying, and asymptotically constant to fixed positive and negative limits as $\xi$ goes to $\mp \infty$ respectively.

\begin{figure}[h!]
    \centering
    \includegraphics[width=0.5\linewidth,trim={0 0cm 0 0.5cm},clip]{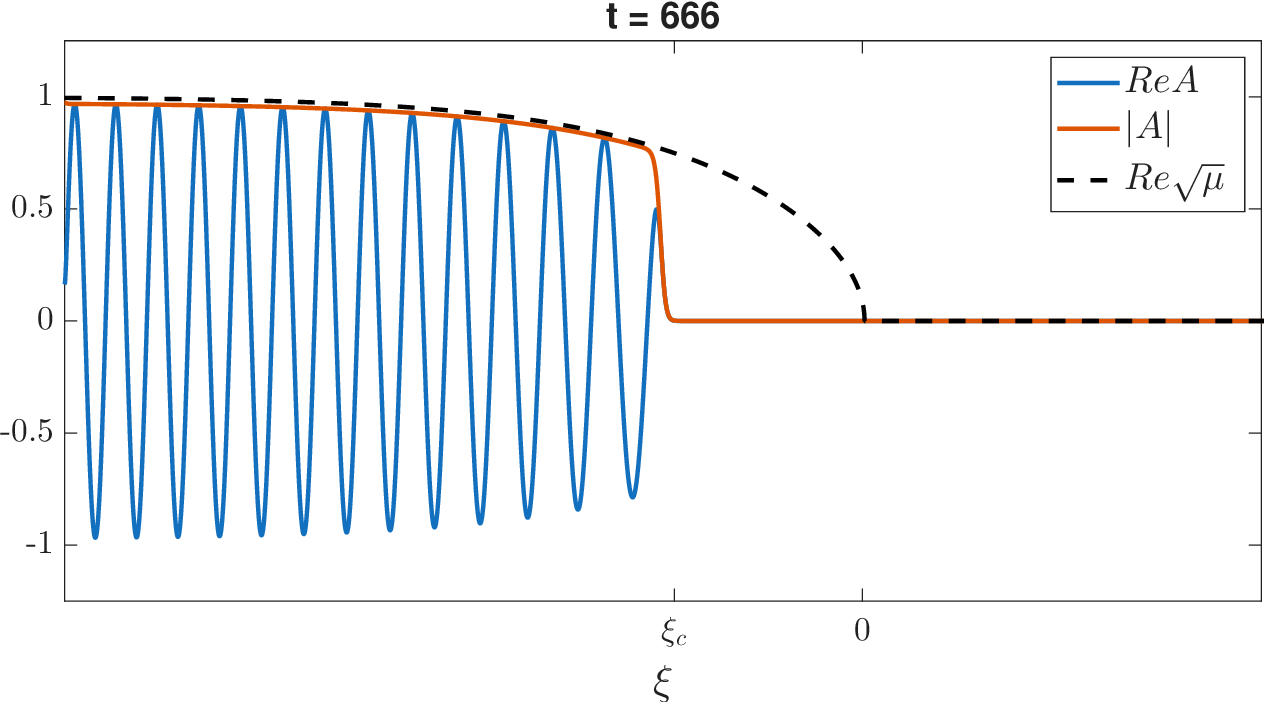}
    \caption{Direct numerical simulation of \eqref{e:cgl} using fourth-order finite differences in space with mesh size $dx = 0.1$ and balanced Strang-splitting in time with step $dt = 0.005$; Parameters $\alpha = -0.1,\gamma = -0.3, c = 1.5,\eps = 0.005$, with $\mathrm{Re}\, A(\xi)$ (blue), $|A(\xi)|$ (orange), and  $\mathrm{Re} \sqrt{\mu(\xi)}$ (dashed black).}
    \label{f:cgl_prof}
\end{figure}

We seek front solutions which connect a periodic plane wave to the trivial homogeneous equilibrium and propagate with the quenching speed $c$.  We take advantage of the gauge invariance of \eqref{e:cgl} under the transformation $A\mapsto \re^{\ri\theta} A$ to de-tune temporal oscillations and consider fronts of the form $A(\xi,t)= e^{-\ri\omega t} A_f(\xi)$ where $A_f$ solves the traveling wave ODE
\beq
    \label{e:cgl-tw}
    0 = (1+\ri\alpha)\partial_{\xi}^2 A_f + c\partial_{\xi} A_f + (\mu(\xi)+\ri \omega) A_f - (1+\ri\gamma) A_f |A_f|^2,
\eeq
with the asymptotic limits
\beq \label{e:cgl_twbc}
\lim_{\xi\rightarrow+\infty} A_f(\xi) = 0,\qquad\qquad
\lim_{\xi\rightarrow-\infty}| A_f(\xi) - r_p \re^{\ri k \xi}|. 
\eeq
Here $k$ denotes the wave number, and $r_p$ the amplitude, of the periodic plane wave solution in the asymptotic limit $\mu\rightarrow\mu_-:= \lim_{\xi\rightarrow-\infty} \mu(\xi) = 1$.   These satisfy the nonlinear dispersion relation \eqref{e:nlom} derived below. An example front solution  is depicted in Figure \ref{f:cgl_prof}.As mentioned above, we are interested in how the asymptotic wave number $k$ and  front location, $\xi_f:= \inf\{\xi\,| \, |A_f(\xi)\leq \delta\}$ for some fixed $\delta>0$ small but $\mathcal{O}(1)$ in $\eps$, is determined by the quenching speed $c$ and slope $\eps$. Also, we denote $\mu_f = \mu(\xi_f)$ to be the $\mu$-value at the front interface location. In this figure, we observe for $\xi<\xi_f$ that $|A|$ follows $\sqrt{\mu(\xi) - k^2}$ where $k$ is the asymptotic wave number. 

Before beginning our analysis of \eqref{e:cgl-tw}, we first collect some needed information about plane waves and equilibrium states in the constant coefficient CGL equation \eqref{e:cgl0}. We then conclude the section by stating our generic assumptions. %which are generic and appear to be necessary for the appearance of the desired front solutions.

\subsection{CGL Preliminaries - Plane waves and equilibrium states}
We briefly recount some properties of plane waves and equilibrium solutions of the homogeneous equation \eqref{e:cgl0}, posed in the co-moving frame $\xi = x - ct$,
\beq\label{e:cgl_c}
A_t = (1+\ri\alpha) A_{\xi\xi} + c A_\xi + \mu A - (1+\ri\gamma)A|A|^2,
\eeq
as they will be useful in the study of the limiting systems with $\mu = \pm1$ in the ramped equation \eqref{e:cgl}. First, we note for any $\mu>0$ that \eqref{e:cgl_c} possesses explicit plane wave-solutions $A_p(\xi,t) = r_p \re^{\ri( k\xi - \omega t)}$ with nonlinear dispersion relation 
\beq\label{e:nlom}
\omega = \Omega(k;\mu) :=  -(\gamma - \alpha) k^2  - c k + \gamma \mu, \qquad \qquad r_p^2 = \mu - k^2. 
\eeq
These are relative equilibria with respect to the gauge symmetry mentioned earlier. Stability of such plane waves can be studied via a modulational perturbation  $A(\xi,t) = \left[r_p(\mu) + a_+ \re^{\lambda t + \ri Q \xi} + a_- \re^{\bar\lambda t - \ri Q \xi} \right]\re^{\ri (k\xi - \Omega(k,\mu) t)}$. Setting $\mu = 1$, after linearizing in $a_\pm$ one obtains an expansion for the leading eigenvalue $\lambda(Q) = \ri d_1 Q - d_2 Q^2+\mathcal{O}(Q^3)$ for $|Q|\ll1$ and $d_1 = \left(2(\alpha - \gamma)k - c\right), d_2 = 1+\alpha\gamma - \frac{2(1+\gamma^2)k^2}{1 - k^2}$. In the Benjamin-Feir stability region $1+\gamma\alpha>0$, this gives  the  Eckhaus stability condition $k^2 < \frac{(1+\alpha\gamma)}{3+\alpha\gamma+2\gamma^2}.$%$k^2 < \frac{\mu(1+\alpha\gamma)}{3+\alpha\gamma+2\gamma^2}.$ See \cite{CGL_Lambda_Stability} for more detail and additional references.

\subsection{CGL Preliminaries - Convective and absolute instability}\label{ss:convabs}
As discussed in \cite{goh2023fronts} for the Allen-Cahn equation, when $c>0$, the transition point in $\mu$ between convective and absolute instability of the trivial state $A = 0$ in \eqref{e:cgl_c} determines the leading-order location of the front interface in the ramped equation \eqref{e:cgl}. For patterned fronts in CGL, it also governs the leading-order selected temporal frequency, and hence the asympotic wave number in the wake. This instability can be characterized using the linearized equation
\beq\label{e:cgl_cv}
v_t= (1+\ri\alpha) v_{\xi\xi} + c v_\xi + \mu v,
\eeq
with $c$ and $\mu$ constant parameters. Here, with $c>0$ large enough, or $\mu>0$ small, a small perturbation of the trivial state grows but is also convected to the left, so that $v(\xi,t)$ decays in time for each fixed $\xi$. The trivial state is \emph{absolutely unstable} if perturbations $v$ (say in $L^2(\R)$) grow both in norm and pointwise almost everywhere. Perturbations are only \emph{convectively stable} if they grow in norm but decay pointwise.   In CGL the boundary between these two regimes in $(\mu,c)$ parameter space is determined by tracking branch points $(\lambda_\mathrm{br},\nu_\br)$ of the linear dispersion relation  
\beq\label{e:dsp}
0 = d(\lambda,\nu) = (1+\ri\alpha) \nu^2 + c\nu +\mu - \lambda
\eeq
obtained by inserting $v = \re^{\lambda t + \nu x}$ into \eqref{e:cgl_cv}.%; see also \eqref{e:zl_eq}.
%In CGL, this transition is given by tracking the $(\mu,c)$ dependence of branch point(s) $(\lambda_\mathrm{br},\nu_\br)$ of \eqref{e:dsp}. 
That is, one locates $(\lambda,\nu)$ values which solve 
$
0 = d(\lambda,\nu), \,\, 0=\partial_\nu d(\lambda,\nu).
$
The transition is then given by $(\mu,c)$ values which yield \emph{marginally stable} branch points with $\mathrm{Re}\, \lambda_\mathrm{br} = 0$ \cite{vansarloos_unstable-states,holzer2014criteria}. In CGL, the relevant branch points come in a complex conjugate pair and lie at the rightmost part of the \emph{absolute spectrum} \cite{sandstede2000absolute,rademacher2007computing} in the $\lambda$-plane. Hence, absolute instability arises when the absolute spectrum intersects the open right half-plane.  In CGL, direct calculation yields
\beq
\nu_\mathrm{br} = -\frac{c}{2(1+\ri\alpha)}, \qquad \lambda_\mathrm{br} = \mu - \frac{c^2}{4(1+\ri\alpha)},
\eeq
and the marginal stability condition  $\mathrm{Re}\, \lambda_\mathrm{br} = 0$ gives absolute instability for $\mu>\mu_c$ and convectively unstable for $\mu<\mu_c$ with
\beq\label{e:muc}
\mu_c = \frac{c^2}{4(1+\alpha^2)}.
\eeq
We also remark that this transition is often given with $c$ as a function of $\mu$, as $c_\mu = 2\sqrt{\mu(1+\alpha^2)}$, the quantity known as the linear spreading speed for (pulled) free invasion fronts when $\mu$ is fixed \cite{vansarloos_unstable-states,holzer2014criteria}. We specifically define $c_\rlin =2\sqrt{1+\alpha^2}$ to be the spreading speed for the asymptotic value $\mu = 1$. We also let $\xi_c$ denote the spatial location where $\mu(\xi_c) = \mu_c.$
At the boundary \eqref{e:muc}, due to the linear dispersion parameter $\alpha$, the linearized system predicts an oscillation with frequency
\beq
\omega_c:= \mathrm{Im}\, \lambda_\br = \frac{\alpha c^2}{4(1+\alpha^2)} = \alpha \mu_c.
\eeq
%where we evaluated $\lambda_\br$ at $\mu_c$.  
We shall show below that $\mu_c$ and $\omega_c$ give the leading order front position and temporal oscillation frequency, respectively, of the ramped front in \eqref{e:cgl}. 
 
 The frequency $\omega_c$ also determines the leading-order asymptotic wave number selected in the wake of the front interface. In particular, we evaluate the nonlinear dispersion relation \eqref{e:nlom} at $\omega = \omega_c,$ and the asymptotic value $\mu = \mu_- = 1$ and solve for $k$.
% $
% \omega_c = \Omega(k;\mu_-) = (\alpha - \gamma) k^2 - c k + \gamma\mu_-
% $
Roots are given by
\begin{equation}
k_c = \begin{cases}
\frac{-c\pm \sqrt{c^2 - 4(\alpha - \gamma)(\gamma \mu_- - \omega_c)}}{2(\gamma - \alpha)}, \qquad \qquad &\alpha\neq \gamma\\
\frac{\gamma \mu_- - \omega_c}{c},  &\alpha= \gamma,
\end{cases}\label{e:kc}
\end{equation}
with the branch of the square root chosen to select a pattern with non-zero amplitude $r_c = \sqrt{1-k_c^2}>0$. A brief computation    gives that $k_c^2\leq1$ for $c>c_\dagger:= 
\mathrm{sign}(\alpha)\frac{2}{\alpha}\left( \sqrt{1+3\alpha^2 + 2\alpha^4} - (1+\alpha^2) \right),
%\begin{cases}
%\frac{2}{\alpha}\left(1+\alpha^2 - \sqrt{1+3\alpha^2 + 2\alpha^4} \right),& \qquad \alpha<0\\
%0,&\qquad \alpha = 0,\\
%%-\frac{2}{\alpha}\left(1+\alpha^2 - \sqrt{1+3\alpha^2 + 2\alpha^4} \right),& \qquad \alpha>0\\
%\end{cases}
$
which satisfies $0<c_\dagger<2\sqrt{1+\alpha^2}.$
We also chose $k_c$ so that the corresponding equilibria in \eqref{e:cgl-tw} has one-dimensional unstable direction. We note the group velocity of this wave in the co-moving frame is given as 
\beq
c_g = \frac{d\Omega}{dk}(k_c) = -2(\gamma - \alpha) k_c - c =  \mp\sqrt{c^2 +4(\gamma - \alpha)(\gamma - \omega_c)} . 
\eeq
indicating that we take the positive branch of the square root in \eqref{e:kc} so that the group velocity points to the left, away from $\xi = 0$, so that the heterogeneity acts as a source.
%For $\gamma - \alpha$ not large, the positive branch of $k_c$ has $|k_c|\leq 1$ and also has negative group velocity, traveling to the left away from $\xi = 0$, indicating the heterogeneity acts as a source. 

\subsection{CGL Preliminaries - Polar and projective spatial dynamics}\label{ss:proj}
Spatially coherent structures can be studied by seeking solutions of \eqref{e:cgl_c} of the form
\begin{equation}\label{e:cgl_tw}
    A(\xi,t) = e^{-\ri\omega t}r(\xi)e^{\ri\varphi(\xi)}.
\end{equation}
This detunes temporal oscillations and factors the gauge symmetry, yielding the coupled ODE system
\begin{equation}\label{e:cgl_ode}
    r_\xi=\kappa r,\qquad  \kappa_\xi=H(r,q,\kappa) \qquad q_\xi = G(r,q,\kappa), \qquad
\end{equation}
where $q=\varphi_\xi$ is  local wave number,  $\kappa=r_\xi/r$ a projective slope of the amplitude, and 
\begin{align}
        H(r,q,\kappa) &= q^2 - \kappa^2 - \frac{1}{1+\alpha^2}\left( \alpha(\omega + c q) + \mu + c \kappa - (1+\alpha\gamma) r^2 \right)\label{e:cgl_GH1}\\
        G(r,q,\kappa) &=  - 2 q \kappa - \frac{1}{1+\alpha^2}\left( \omega + c q - \alpha(\mu+c\kappa) - (\gamma - \alpha) r^2\right).\label{e:cgl_GH}
\end{align}
It will also be useful to consider the equivalent system obtained by performing a complex projectivized blow-up with the coordinates $z = A_\xi/A \in \C$ and $r = |A|$. This yields a complex Riccati equation for $z = \kappa + \ri q$, coupled with an equation for the amplitude
\begin{align}%\label{e:cgl_zr}
    z_\xi &= -z^2 - \frac{1}{1+\ri\alpha}\left[cz+\mu+\ri\omega -(1+\ri\gamma)r^2\right]\label{e:cgl_zr1}\\
    r_\xi &= \mathrm{Re}(z) r.\label{e:cgl_zr2}
    \end{align}
We remark when $z\rightarrow\infty$, an alternative chart via the Poincar\'e inversion $1/z$ is required.    
Also, note $G$ and $H$ give the real and imaginary parts of the $z$-equation in \eqref{e:cgl_zr1}.  Following \cite{van1992fronts}, the fixed points of \eqref{e:cgl_ode} are generally separated into two classes, those with $r = 0$ and those with $r>0, \kappa = 0$. The former are often denoted as ``linear" fixed points as they correspond to invariant subspaces of the linearized equation $0 = v_{\xi\xi} + cv_{\xi} + (\mu + \ri \omega) v$, about the trivial state. The latter correspond to plane wave solutions $r\re^{\ri (q \xi-\omega t)}$ of the full nonlinear PDE \eqref{e:cgl_c}.  

Linear equilibria take the form $(z,r)  = (z_L^\pm,0)$ with $z_L^\pm = \kappa_L^\pm + \ri q_L^\pm$ roots of
\beq\label{e:zl_eq}
0 = (1+\ri\alpha) z_L^2 + c z_L + (\mu+\ri\omega).
\eeq
Stability of these equilibria can be obtained by linearizing \eqref{e:cgl_ode} about $(\kappa_L^\pm,q_L^\pm,0)$ to find the (spatial) eigenvalues $\kappa_L^\pm,\nu_{L,\pm}, \overline{\nu_{L,\pm}}$ with
\beq
\nu_{L,\pm} = -2 z_{L}^\pm - \frac{c}{1+\ri\alpha} = \mp \sqrt{\frac{c^2}{(1+\ri\alpha)^2} - 4\frac{\mu+\ri\omega}{(1+\ri\alpha)}}.
%&= \mp\sqrt{ }
\eeq
{Under our assumptions on $c,\alpha$ and for $\mu<\mu_c$ and $\omega\sim\omega_c$, the negative branch $z_{L}^-$ is a saddle with stable subspace in the $r$-direction and repelling directions in the $r = 0$ subspace. These equilibria collide in a complex fold at $z = z_c:=-c/(2+2\ri\alpha), (\mu,\omega) = (\mu_c,\omega_c)$; see Sec. \ref{ss:sminus} below for more detail. }

Nonlinear equilibria with $r\neq0, \kappa = 0$ correspond to plane wave solutions (which are relative equilibria of \eqref{e:cgl-tw} with respect to the gauge invariance). Inserting $z_p = \ri q_p$, one finds $(r_p,q_p)$ solves the nonlinear dispersion relation \eqref{e:nlom} above with $ k=q_p$.   For $\mu,c>0$, the work \cite[\S 2.2.3]{van1992fronts} gives that one of the two roots $q_p$ of the nonlinear dispersion relation give a saddle equilibrium with one-dimensional unstable subspace and two-dimensional stable subspace. They show an equilibrium has one dimensional unstable direction as long as $\Gamma_N = 2 r_p^2\left(-\left(\frac{\alpha c}{1+\alpha^2} - 2 q_p \right)(\gamma - \alpha) - c\frac{1+\alpha\gamma}{1+\alpha^2} \right)/(1+\alpha^2)<0$, which holds for one of the wave numbers for all $c>0$.

\subsection{Assumptions}\label{ss:assump}

Returning to the ramped equation \eqref{e:cgl-tw} and the desired pattern-forming front solutions, we shall make the following assumptions. As mentioned before we consider slowly-varying ramps, $0<\eps \ll1.$ Throughout, we assume $\alpha,\gamma$ are such that plane waves in \eqref{e:cgl_c} with $\mu \equiv1$ are Benjamin-Feir stable $1+\alpha\gamma>0$. Following our previous work on Allen-Cahn, we shall fix the quenching speed $c$ to be $\mathcal{O}_\eps(1)$ and lie below the homogeneous free invasion speed for $\mu\equiv1$ so that the heterogeneity does not outrun the front. Further, we assume for $\mu = 1$ there exists a non-trivial wave train, with $r_p>0$, for the predicted wave number $k_c$ as computed in Section \ref{ss:convabs}. Putting all of these together, we assume $c_\dagger<c<c_\mathrm{lin} = 2\sqrt{1+\alpha^2}$. Finally, to allow for a simplified analysis of the inner solution near $\mu = \mu_c$ we also assume that $\alpha$ is small, but expect much of our work to hold for a broader range of $\alpha$ values.

\section{Multi-scale heteroclinic formulation}\label{s:het}
With this preliminary information on the homogeneous system and the prediction for the front interface $\mu_{f}\approx \mu_c$ and selected wave number $k_f\approx k_c$ via the linear absolute-to-convective (AC) instability transition, we now consider front solutions in the ramped traveling wave equation \eqref{e:cgl-tw} with asymptotic boundary conditions \eqref{e:cgl_twbc} and ramp $\mu(\xi) = -\tanh(\eps \xi)$ with $0<\eps \ll1$. As discussed in Section \ref{ss:proj}, we take advantage of the gauge invariance in this equation by employing projectivized coordinates $r\in \R_+, z = A_\xi/A\in\R,\mu\in \R,$ appending the ramp $\mu$ as an additional (slowly-varying) dependent variable, 
\begin{align}\label{e:ric_sys}
r_\xi &= \mathrm{Re}(z) r\notag\\
z_\xi &= -z^2 - \frac{1}{1+\ri\alpha}\left[cz + \mu + \ri\omega - (1+\ri\gamma)r^2 \right]\notag\\
\mu_\xi&=-\eps(1 - \mu^2).
\end{align}
  We shall also need the corresponding 4-dimensional real-valued system for the variables $(\kappa,q,r,\mu)$ obtained by appending $\mu_\xi = -\eps(1 - \mu^2)$ to system \eqref{e:cgl_ode}.
 \begin{equation}\label{e:rkqmu}
 r_\xi = \kappa r,\quad 
 \kappa_\xi = H(\kappa,q,r,\mu),\quad
 q_\xi = G(\kappa,q,r,\mu),\quad 
 \mu_\xi = -\eps (1 - \mu^2). %-\eps g(\mu):= -\eps (1 - \mu^2).
 \end{equation}
%{\color{blue} (The order of variables is inconsistent here.)}

Again, pattern-forming front solutions correspond, in this formulation, to heteroclinic orbits between the equilibria $(r,z,\mu) = (r_p,z_p ,1)$ at $\xi = -\infty$ and $(r,z,\mu) = (0,z_-,-1)$ at $\xi\rightarrow + \infty$; see Fig. \ref{f:front_prof} for a depiction.  The former equilibrium represents the asymptotic plane-wave, with $z_p = \ri q_p, r_p = \sqrt{1 - q_p^2}$ and $q_p$ satisfying the nonlinear dispersion relation  $\omega_f =  \Omega(q_p;1)$ in \eqref{e:nlom}, with $\omega_f$ the selected frequency.  The latter equilibrium corresponds to the stable eigenspace of the origin corresponding to exponentially decaying solutions $(A(\xi),A_\xi(\xi))$ as $\xi\rightarrow+\infty$, and thus satisfies $z_- = z_L^-$, the root of the linear dispersion relation \eqref{e:zl_eq} with negative real part.  %As seen below, the parameter $\omega$ will be varied in order to transversely unfold the relevant invariant manifolds and locate an intersection. 
We remark that the $r = 0$ plane and the $\mu = \pm1$ planes are invariant under the flow. Further, for $\eps = 0$ all $\mu$-constant planes are also invariant under the flow.

For $0<\eps\ll1$, we seek intersections of the unstable manifold, $W_\eps^\mathrm{u}(r_p,z_p,1)$, of $(r_p,z_p,1)$  with the stable manifold $W_\eps^\mathrm{s}(0,z_-,-1)$ of $(0,z_-,-1)$. %Under our generic assumption of Benjamin-Feir stability $1+\alpha\gamma>0$, and non-zero quench speed $c>0$, 
Under our assumptions, each of these manifolds is 2-dimensional with one fast direction coming from the $(r,\kappa,q)$ dynamics and one slow direction coming from the $\mu$ flow. Hence, in a 4-dimensional ambient phase space, one parameter is generically required to \emph{transversely unfold} these manifolds and locate an intersection. The frequency $\omega$ will play this role in our work. %Following our previous work on Allen-Cahn, we shall fix the ramp speed to lie below the homogeneous free invasion speed $0<c<c_\mathrm{lin} = 2\sqrt{1+\alpha^2}$ and also be $\mathcal{O}_\eps(1)$. %Further we shall choose $\alpha,\gamma$ in the Benjamin-Feir stability regime $1+\alpha\gamma>0$. 
%To allow for a simplified analysis of the inner solution near $\mu = \mu_c$ we also assume $\alpha$ is small, but expect much of our work to hold for a broader range of $\alpha$ values.  
To characterize and track these manifolds we leverage the multiple time scales in the system by using GSPT. Hence, we first describe the critical manifolds in the singular limit $\eps = 0.$

\subsection{Critical manifolds and fast fibers for $\eps = 0.$}
%Again, for $\eps = 0$ the $\mu$-constant planes are invariant. 
 For $\mu>0$ with $\eps = 0$, the plane-wave equilibria form a critical set $S_{p,0} = \{ (r_p(\mu),z_p(\mu) ,\mu)\, :\, \mu_b<\mu\leq 1\}$, where $\mu_b$ is a $c$ and $\alpha$ dependent function, given in Section \ref{ss:sp}, with $\mu_b<\mu_c$ for $\omega\sim \omega_c$. This range of $\mu$-values and $q_p(\mu)$ are chosen so that $r_p>0$ and each equilibrium in $S_{p,0}$ is a saddle with one-dimensional unstable eigenspace and thus 1-dimensional fast unstable manifold $W^\mathrm{uu}_0(r_p(\mu),z_p(\mu) ,\mu)$. Thus $S_{p,0}$ forms a normally hyperbolic invariant manifold;  see Section \ref{ss:sp}.  %It is normally hyperbolic for $\mu\in(\mu_1,1]$ for some $\mu_1\geq0$  dependent on the specific value of the parameters $\omega,c,\alpha,\gamma$; see Section \ref{ss:sp}.
  Taking the union over $\mu$, this creates a strong-unstable foliation 
$$
W_0^\mathrm{u}(r_p(1),z_p(1),1) = \bigcup_{\mu\in[\mu_b,1]} W^\mathrm{uu}_0(r_p(\mu),z_p(\mu),\mu).
$$
%The critical set $S_p$ forms a normally-hyperbolic invariant manifold with one-dimensional fast unstable fibers, and two-dimensional fast stable fibers. 
 Fenichel theory \cite{fenichel1979geometric} gives for $0<\eps\ll1$, that $W_0^\mathrm{u}(z_p(1),r_p(1),1)$ perturbs smoothly to $W_\eps^\mathrm{u}(z_p,r_p,1)$ and that the latter is smoothly foliated by a slow manifold $S_{p,\eps}:=\{(z_{p,\eps},r_{p,\eps})(\mu)\,:\, \mu\in(\mu_b+\delta,1]\}$ and its strong unstable fibers for some $\delta>0$ small. Note, the strong unstable fiber of the equilibrium $(z_p,r_p,1)$ is given by its strong unstable manifold.  While it is not necessary for our numerical continuation results, in Section \ref{ss:sp} we briefly characterize the strong unstable fibers $W_0^\mathrm{uu}(r_p(\mu),z_p(\mu),\mu)$, focusing in particular on $\mu$ values near the critical transition regime $\mu\sim \mu_c$, where we expect an intersection between $W_\eps^\mathrm{u}(r_p,z_p,1)$ and $W_\eps^\mathrm{s}(0,z_-,-1)$ to be located.

 The $r = 0$ plane itself is normally hyperbolic for all $z$ with $\rre\,z <0$ or $\rre\,z >0$ with strong stable or unstable fibers respectively pointing in the $r$-direction at leading order. In the $r = 0$ plane, again with $\eps = 0$, there is a critical set of equilibria $S_{-,0} = \{ (0,z_-(\mu),\mu)\,:\, \mu\geq-1 \}$.%, which, along with its fast stable fibers, describes the strong stable foliation $W_0^\mathrm{s}(0,z_-,-1)$. %$W_0^\mathrm{s}(0,z_-,-1)$ of $(0,z_-,-1)$. 
  Restricting to the $r = 0$ invariant plane, $S_{-,0}$ is normally hyperbolic for all $\mu\in[-1,\mu_*)$ for a $c,\alpha,\omega$-dependent value $\mu_*$, given below.  In particular, setting $r \equiv0$ and considering the $z$-linearization $v_\xi = (-2z_-(\mu) - \frac{c}{1+\ri\alpha}) v$,  the equilibrium $z_{-}(\mu)$ changes from unstable source for $\mu<\mu_*$ to stable sink for $\mu>\mu_*$ where $\mu_*$ is such that $\rre (-2z_-(\mu) - \frac{c}{1+\ri\alpha})=0$. A straightforward computation gives that $\mu_* = \omega/\alpha$. $S_{-,0}$ remains normally hyperbolic for full system including the $r$-direction, with fast stable fibers $W_0^\rss(0,z_-(\mu),\mu)$ for all $\mu$ with $\mathrm{Re}\, z_-(\mu) <0$. 
 
 For $0<\eps\ll1$, $S_{-,0}$ perturbs to a one-dimensional invariant slow manifold $S_{-,\eps}$ (i.e. a trajectory) in the $r = 0$ plane. The union, $W_0^\mathrm{s}(0,z_-,-1)$, of $S_{-,0}$ with its strong stable fibers,   perturbs to the stable manifold $W_\eps^\mathrm{s}(0,z_-,-1)$. In Section \ref{ss:sminus}, we describe the behavior of $S_{-,\eps}$ for $0<\eps\ll1$ and $(\mu,\omega) \sim (\mu_c,\omega_c)$. %Such strong stable fibers have tangent space pointing vertically in the $r$-direction.

Since both normally-hyperbolic invariant manifolds perturb smoothly for $0<\eps\ll1,$ we track $W_\eps^\mathrm{s}(0,z_-,-1)$ backwards and $W_\eps^\mathrm{u}(r_p,z_p,1)$ forwards into a neighborhood of the point $(z,r,\mu) = (z_-(\mu_c),0,\mu_c)$ to locate an intersection. %We use the parameter $\omega$ as the unfolding parameter and observe a locally unique value $\omega_{f}\sim \omega_c$ with an intersection and hence a locally unique selected wavenumber $k_{f}\sim k_c$ for the front.

\subsection{Characterization of slow manifold $S_{-,\eps}$}\label{ss:sminus}
%\begin{itemize}
%\item Thoughts about rigorous slow passage through (transcritical)? Maybe can actually do this without too much effort?
%    \end{itemize}

We next characterize the slow-manifold $S_{-,\eps}$, contained in the $r = 0$ invariant subspace of \eqref{e:ric_sys}. To ease computations, restricting to the $r = 0$ invariant plane, we make the following change of coordinates:
\beq\label{e:ztmt}
z = \tilde z +z_c, \quad \mu = \tilde \mu + \mu_c,\qquad \omega = \tilde\omega+\omega_c , \qquad \xi = -\zeta,
\eeq
obtaining the system 
\begin{align}\label{eq:zsc}
\tilde z_\zeta &=  \tilde z^2 - Q(\tilde \mu),\qquad Q(\tilde \mu) = -\frac{\tilde\mu+\ri\tilde\omega}{1+\ri\alpha}\notag\\
\tilde \mu_\zeta &= \eps(1 - (\tilde\mu + \mu_c)^2).
\end{align}
Note, that $(\tmu,\tilde\omega) = (0,0)$ corresponds to the AC instability transition at $(\mu,\omega) = (\mu_c,\omega_c)$ and $z = z_c:= - \frac{c}{2(1+\ri\alpha)}$. Also, we reversed the direction of the evolutionary time-like variable so that the critical set, $S_{-,0}$, under hypotheses given below, is attracting for $\tilde\mu<\tilde\mu_*=\tl\omega/\alpha$. %Under the assumption that $\tmu>-\tomega\alpha$, $S_{-,0}$ loses normal hyperbolicity at $\alpha\tmu =\tomega.$

\begin{figure}[htbp]
    \centering
    \hspace{-0.3in}
    \includegraphics[width=0.23\linewidth,trim={0 2.5cm 0 2.5cm},clip]{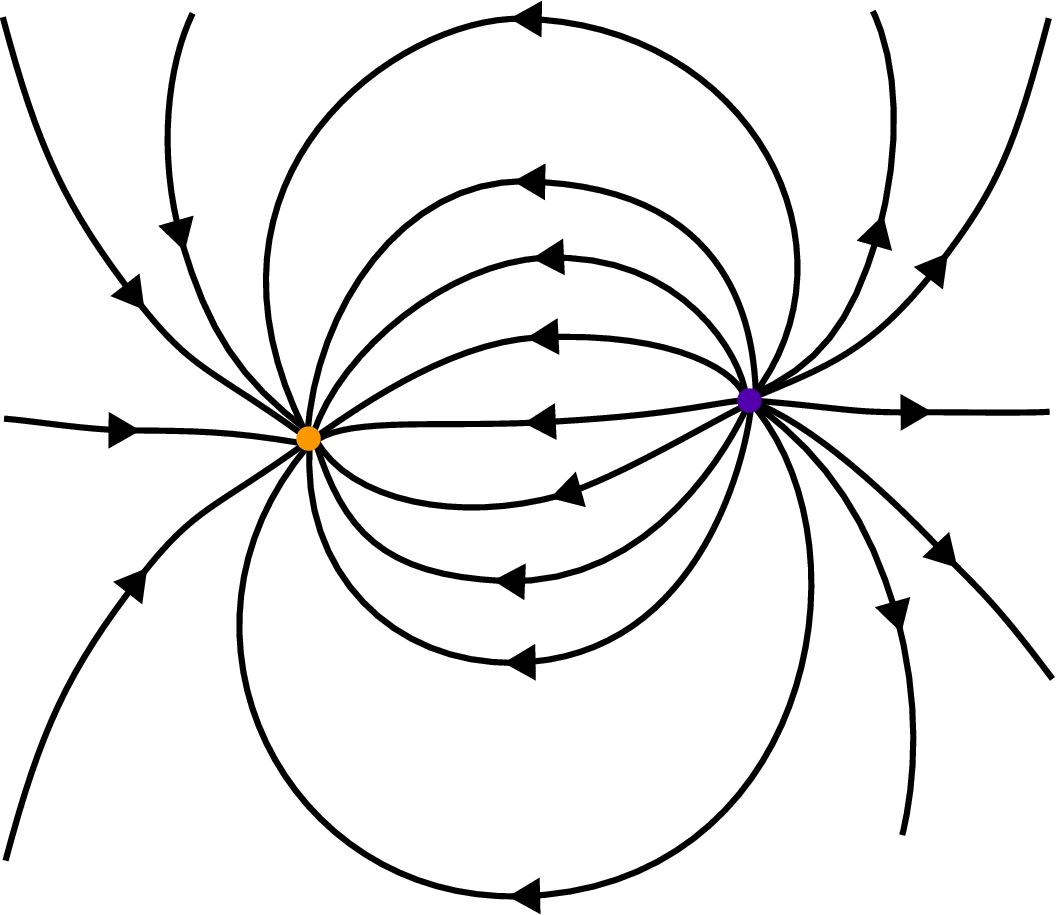}
    \includegraphics[width=0.23\linewidth,trim={0 2.5cm 0 2.5cm},clip]{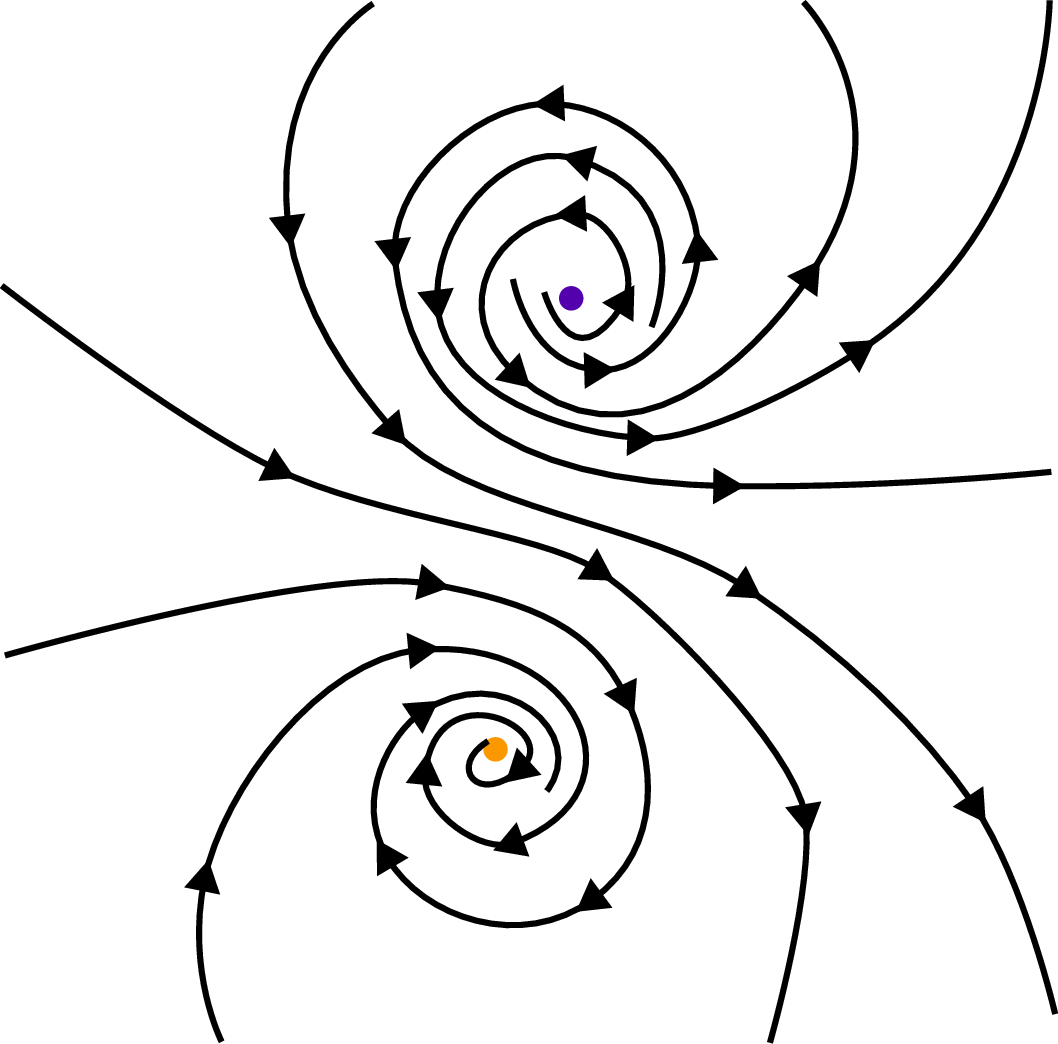}  
    \includegraphics[width=0.23\linewidth]{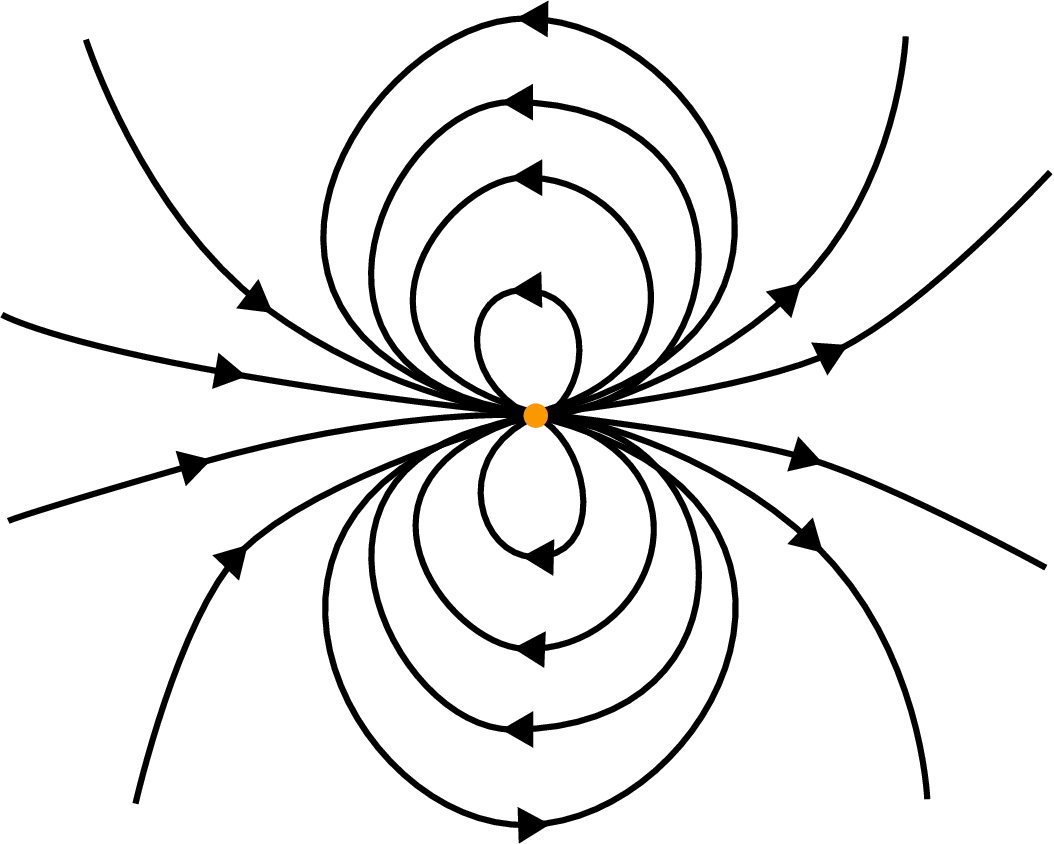}
        \includegraphics[width=0.3\linewidth]{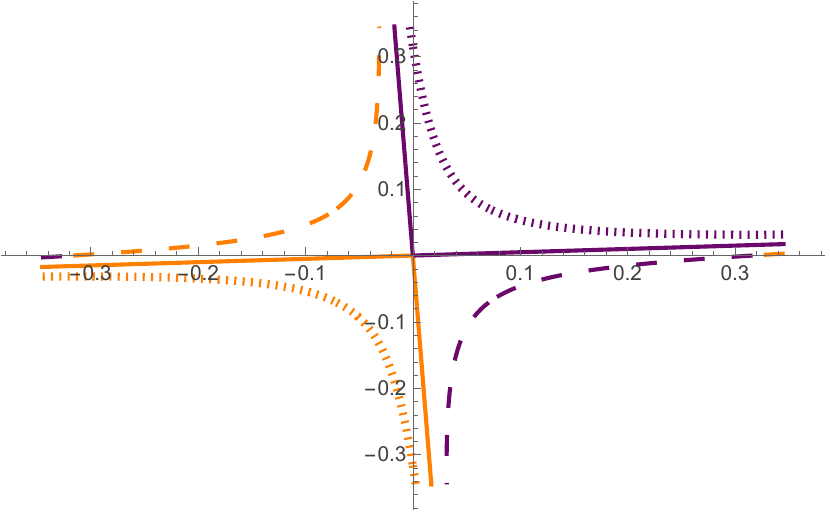}\hspace{-0.3in}
    \caption{Phase portraits of \eqref{eq:zsc} with $\eps = 0,$ for a range of $\tmu$ values $\tmu <0$ (left), $\tmu>0$ (center left) both for $\tomega<0$, and $\tmu = 0, \tomega = 0$ (center right); Right:  Critical equilibrium curves $\tilde z_{\pm,0}(\tilde \mu)$ for $\tilde\mu\sim 0$ with $\alpha = -0.1$, and $\tomega<0$ (small dashed) $\tomega = 0$ (solid), and $\tomega>0$ (large dashed). Purple curves denote $\tilde z_{+,0}(\tmu)$, orange denotes $\tilde z_{-,0}(\tmu).$}
    \label{f:zm0}
\end{figure}

 The critical equilibria for $\eps = 0$ now take the form $\tilde z_{\pm,0}(\tmu) = \pm \sqrt{Q(\tmu)}$.  The branch of the square root is chosen so that $\mathrm{Re}\, \tilde z_{-,0}(\mu) < 0$ for $\tmu$ sufficiently negative. The parameter $\alpha$ rotates the two branches which collide in a singularity at $\tmu = 0$ when $\tomega = 0$.  Variations in $\tomega$ break the equilibrium curves into hyperbola; see Fig. \ref{f:zm0}.  We note for $\alpha\tomega>0$, %\gtrless 0$, 
 the curves cross the imaginary axis for $\tmu = \tmu_* = \tomega/\alpha$ and exchange stability types. The linearization about each $\tilde z_{-,0}$ gives eigenvalues $2\tilde z_{-,0}, \overline{2\tilde z_{-,0}}$. One finds $\mathrm{Re}\,\tilde z_{-,0}(\tmu) <0 $ for all $\tmu<\tmu_* $ so that 
$
S_{-,0} = \{ (\tilde z_{-,0}(\tilde \mu), \tilde\mu) \,|\, -1-\mu_c\leq \tilde\mu < \tilde \mu_* - \delta\}
$
is normally attracting for any small $\delta>0$ fixed.

\begin{figure}[htbp]
    \centering
    \includegraphics[width=0.4\linewidth]{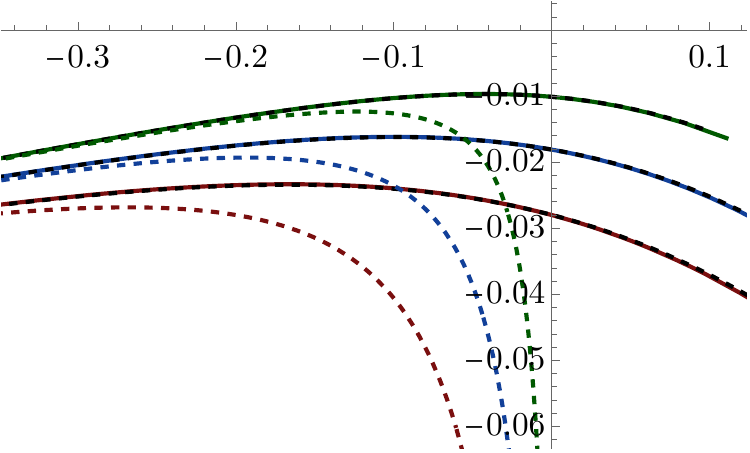}
    \caption{ Comparison in the complex $\tilde z$-plane of numerical slow manifold curves  $\tilde z_{-,\eps}(\tmu)$ (solid colors), critical equilibrium curve $\tilde z_{-,0}(\tmu)$ (dashed color), and approximate inner solution $\tilde z_{-,\eps}^\mathrm{in}(\tmu)$ (dashed black) for $\tmu\sim0$ with values $(\eps,\tomega) \approx (9.8,7.3)\times 10^{-3}$  (red), $( 4.5, -3.8)\times 10^{-3}$ (blue), and $(1.5, -1.5)\times 10^{-3}$ (green); $\alpha = -0.1,\, c = 1$. The $\tilde\omega$-values are $\mathcal{O}(\eps^{2/3})$ and were determined by numerical continuation of the full problem, see Sec. \ref{s:numcont} below. }
    \label{f:zme}
\end{figure}

%Fenichel theory gives that $S_{-,0}$ smoothly perturbs 
As mentioned above, for $0<\eps\ll1$, this smoothly perturbs to a 1-D invariant slow manifold $S_{-,\eps}$, which can be described as a graph,  $\tilde z_{-,\eps}(\tilde \mu),$ over the $\tilde \mu$ variable. This manifold is described by a single, slowly-evolving trajectory which converges to the equilibrium $(\tilde z,\tilde \mu) = (\tilde z_{-,0}(-1-\mu_c),-1-\mu_c)$ as $\zeta\rightarrow-\infty$. Figure \ref{f:zme} depicts a numerical representation of this slow manifold found by direct simulation (solid color), along with the $\eps = 0$ critical set (dashed color). Here we observe that the slow manifold $\tilde z_{-,\eps}$ deviates from the critical manifold $\tilde z_{-,0}(\tmu)$ for small $\tmu<0$, before the non-normally hyperbolic point.    

To understand this divergence, we expand the graph $\tilde z_{-,\eps}(\tilde \mu)  = \sum_{j =0}^n h_j(\tilde \mu)\eps^j$ to order $n>0$ and use the standard invariance condition in \eqref{eq:zsc}, 
$$
(1-(\tilde\mu + \mu_c)^2) \sum_{j = 0}^n \partial_{\tilde \mu} h_j(\tilde \mu) \eps^{j+1}  = \partial_\zeta\left( \tilde z_{-,\eps}(\tilde \mu)  \right) = \left( \tilde z_{-,\eps}(\tilde \mu)  \right) ^2 - Q(\tmu) =  \left(\sum_{j = 0}^n h_j(\tilde \mu) \eps^j\right)^2 -Q(\tmu), 
$$
to determine the coefficients in the expansion. The first three coefficients are found to be
\begin{equation}
    h_0(\tilde \mu) = \tilde z_{-,0}(\tilde \mu), \quad 
    h_1(\tilde \mu) = \frac{-(1-(\tilde\mu+\mu_c)^2)}{4 h_0(\mu)^2 (1+\ri\alpha)},\quad
    h_2(\tilde \mu) = \frac{-(1-(\tilde\mu+\mu_c)^2)\partial_{\tmu} h_1(\tmu) - h_1(\tmu)^2}{2h_0(\tmu)}.
\end{equation}
Due to the branch points in derivatives and negative powers of $h_0(\tmu)$, we observe that the general coefficient $h_j(\tmu)$ has a complex pole at $\tmu = -\ri\tomega$ of order $(3j - 1)/2$ for $j = 1,2,3...$. This leads to a divergent Taylor series for $\tmu$ lying in the complex ball of radius $\mathcal{O}(\eps^{2/3})$ centered at $-\ri\tomega$. When $\tomega$ is also $\mathcal{O}(\eps^{2/3})$ small, this ball can intersect the real $\tmu$-axis and induce poor approximation of the slow manifold for real $\tilde\mu\sim0$.%; see Fig. \ref{f:zme}. %{\color{blue} Figure depicting this? Might not be worth it. Mabge give some citations?} 

\subsubsection{Inner Expansion}\label{ss:inn}
To more accurately describe the slow manifold in a neighborhood of $\tilde \mu = 0$  we perform a formal asymptotic analysis to describe the slow manifold for $0<\eps\ll1$. %(recall this corresponds to $\mu \sim \mu_c, z\sim -c/(2+2\ri\alpha)$)
 We define $\Lambda = \tmu + \ri \tomega,$ and make the critical scaling 
$$
\tilde z = \eps^{1/3}\hat z,\qquad  \tomega = \eps^{2/3} \hat\omega, \qquad \tmu = \eps^{2/3} \hat \mu,\qquad \Lambda = \eps^{2/3} s,
$$
obtaining
\begin{equation}
\hat z_\zeta = \eps^{1/3}\left( \hat z^2 + a s \right),\qquad 
\hat s_\zeta = \eps^{1/3}\left(1-\mu_c^2 - 2\eps^{2/3} (s - \ri\hat\omega\mu_c) + \eps^{4/3} (s - \ri\hat\omega)\right).
%\hat\mu_\zeta = \eps^{1/3}\left(1-\mu_c^2 - 2\eps^{2/3} (s - \ri\hat\omega\mu_c) + \eps^{4/3} (s - \ri\hat\omega)\right).
\end{equation}
We remark this is akin to computing the re-scaling chart in geometric blow-up. For $\eps$ small, and $\tmu\sim0$, we approximate $s_\zeta \approx C \eps^{1/3}$ with $C = 1 - \mu_c^2,$ and make the coordinate change $\partial_\zeta =  C\eps^{1/3}\partial_s$, to obtain the scaled Riccati equation
\begin{equation}\label{e:zhatmu}
C \hat z_s = \hat z^2 + bs,
\end{equation}
with $s\in \R + \ri\hat\omega$ and $b = 1/(1+\ri\alpha)$.  We then set $\hat z(s) = -C u'(s)/u(s)$, and scale the (complex) dependent variable $\hat s = \theta s$ with $ \theta %= (-a^2 b)^{1/3}
 = \left( -C^2(1+\ri\alpha) \right)^{-1/3}$, to obtain the classic Airy equation $u_{\hat s\hat s} - \hat s u = 0$, with general solution $u(\hat s) = c_1 \mathrm{Ai}(\hat s) + c_2 \mathrm{Bi}(\hat s)$ for arbitrary constants $c_1,c_2\in\C$ and the standard Airy functions $\mathrm{Ai}(\hat s), \mathrm{Bi}(\hat s)$. In order to match with the correct outer solution, we chose the cube root branch of the constant $\theta$,  with $\mathrm{Re}\,\theta <0$. For $\alpha$ small, this corresponds to $\mathrm{Arg}\, \theta \approx -\pi.$  The general solution for \eqref{e:zhatmu} is then
\begin{equation}
\hat z(s) = -\theta C \frac{c_1 \mathrm{Ai}'(\theta s ) + c_2 \mathrm{Bi}'(\theta s) }{c_1 \mathrm{Ai}(\theta s) + c_2 \mathrm{Bi}(\theta s)}. 
\end{equation}
To match with the attracting outer solution for $\mathrm{Re}\, s\ll-1$ we impose the asymptotic boundary condition $\hat z(s) \approx -\sqrt{-s b}$ and use the asymptotic expansions of the Airy functions and their derivatives \cite[\S 9.7]{NIST:DLMF} to determine $c_2 = 0$, giving $\hat{z}(s) = -\theta C \mathrm{Ai}'(\theta s)/\mathrm{Ai}(\theta s),$ an Airy quotient function. 

In the unscaled coordinates for \eqref{eq:zsc}, the inner approximation to the slow manifold is then given by 
\begin{equation}\label{e:tzin}
  \tilde z_{-,\eps}^{\mathrm{in}}(\tmu)  =   - \theta C \eps^{1/3}  \frac{\mathrm{Ai}'\big(\theta  (\tmu + \ri\tomega)\eps^{-2/3}\big)}{\mathrm{Ai}\big(\theta  (\tmu + \ri\tomega)\eps^{-2/3}\big)},\quad \quad \tmu\sim0
\end{equation}
and hence for the original equation \eqref{e:ric_sys}
\begin{equation}\label{e:zin_unc}
     z_{-,\eps}^{\mathrm{in}}(\mu)  =   - \theta C \eps^{1/3}  \frac{\mathrm{Ai}'\big(\theta \eps^{-2/3}  (\mu - \mu_c + \ri(\omega - \omega_c))\big)}{\mathrm{Ai}\big(\theta \eps^{-2/3} (\mu - \mu_c + \ri(\omega - \omega_c))\big)} - \frac{c}{2(1+\ri\alpha)},\qquad \mu\sim\mu_c.
\end{equation}
As seen in Figure \ref{f:zme}, we find this asymptotic inner solution (dashed black) agrees quite well with the numerical slow manifold for  $\tilde\mu \approx 0$ for $\eps$ small and $\tomega = \mathcal{O}(\eps^{2/3}).$ As will be shown below in Figures \ref{f:front_prof} and \ref{f:front_eps}, it also agrees well with the $z,\mu$ components of the full heteroclinic orbit for $(z,r,\mu)\sim (z_c,0,\mu_c).$

\subsubsection{Delayed onset asymptotics}\label{ss:bdin}
We now use this inner solution to track the leading order delay $\tmu_f = \mu_f - \mu_c$ of the front interface. We proceed in a similar manner to \cite{goh2023fronts,krupa01}, where the analogous delay is predicted by the vertical asymptote (i.e. blow-up time) of a real-valued Riccati equation. In our case,  $\tilde z$ varies in the complex domain. Thus generic solutions wind around the Riemann sphere and do not blow up to infinity. In other words, there is not a unique value of $ \tmu$ for which all solutions blow up as seen in the real case \cite{krupa01,MishchenkoRozov1980}. Further, the non-normally hyperbolic point $\tmu_*$ is $\tomega$-dependent leading to a parameter dependent point from which to measure the bifurcation delay of the slow manifold.   

To estimate $\tmu$ for $\tilde\omega\sim0$, we track the value,  $\tilde\mu_0$, such that  $\mathrm{Re}\, \tilde z_{-,\eps}^{\mathrm{in}}(\tilde \mu_0)  = 0$. In un-scaled coordinates, this corresponds to the value $\mu_0$ where $\mathrm{Re}\, z_{-,\eps}^{\mathrm{in}}(\mu_0)  = \mathrm{Re}\, z_c:= \mathrm{Re}\,\left\{ -c/(2+2\ri\alpha)\right\}$, the real part of the fold point $z_c$ for $\omega = \omega_c$, that is $\tilde \omega = 0$.  Note we recall for $\eps = 0$, that the parameters $(\tmu,\tomega) = (0,0)$ give a degenerate equilibrium in the $\tilde z$-plane in which all trajectories are homoclinics; see Fig. \ref{f:zm0} center right. Hence we expect the $\tilde z$ solution to leave a neighborhood of the origin after it passes the ghost of this point at $\mathrm{Re}\, \tilde z = 0$, for $\tilde \omega$ small.

Without loss of generality, take $\alpha<0$ ($\alpha>0$ is analogous). To determine $\mu_0$, it thus suffices to track the location $\mathrm{Re}\,\hat z(s_0) = 0$ of the scaled solution. By known values of the Airy function and its derivative \cite[\S 9.18]{NIST:DLMF} (see also \cite[App. A]{melrose1987boundary}), we have $\hat z(0) = \theta C 3^{1/3} \Gamma(2/3)\,/\,\Gamma(1/3)$, with the standard Gamma function $\Gamma(s)$. For $\alpha<0$ small, this value lies near the negative axis in the 3rd quadrant. As $\mathrm{Re}\, \hat z(s)$ increases as $\mathrm{Re} \,s$ increases from 0, we expect $s_0$ to have positive real part, that is $\hat\mu_0>0$.

\begin{figure}[htbp]
    \centering
    \includegraphics[width=0.55\linewidth,trim={1.5cm 0cm 0 0cm},clip]{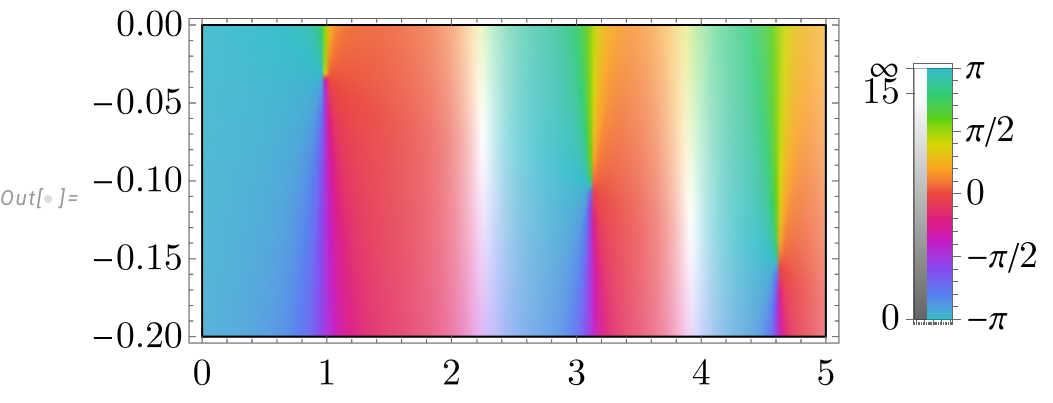}
     \includegraphics[width=0.4\linewidth,trim={0cm 0cm 0 0cm},clip]{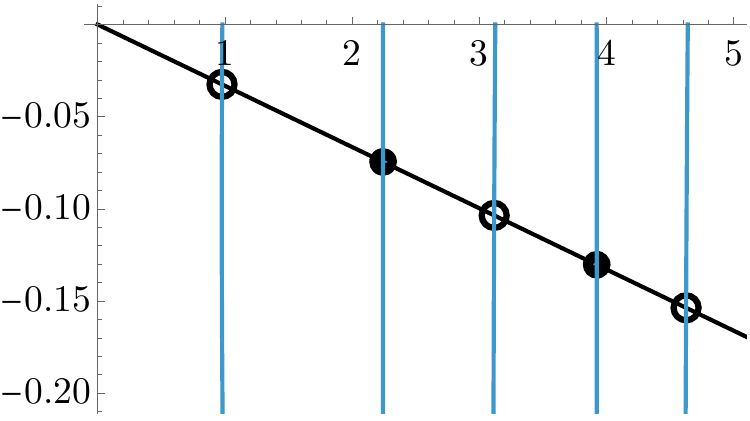}
    \caption{Left: Plot in the complex $s$-plane of the function $\hat z(s)$ for $\alpha = -0.1$, lightness corresponds to $|\hat z|$ while color denotes $\mathrm{Arg}\, \hat z$, see scale on side; Right: Plot of the line $\mathfrak{L}_\theta$ (black) along with zeros  $ s_{z,j} = \tilde \sigma_j/\theta$ (open dots) and poles $s_{p,j} = \sigma_j/\theta$ (closed dots) of $\hat z$ and contours $\mathfrak{R}_j$ for $j = 1,2,3$ (blue).    }
    \label{f:ztzp}
\end{figure}

We track intersections with the imaginary axis by locating where $\mathrm{Arg}\, \hat{z} = \pm \pi/2.$ These can be found by studying the zeros,  $  s_{z,j} = \tilde \sigma_j/\theta$, and poles, $s_{p,j} = \sigma_j/\theta $, of $\hat{z}(s)$, where $\tilde \sigma_j,\sigma_j$ are the  zeros of $\mathrm{Ai}'$ and $\mathrm{Ai}$ respectively.  Note zeros of $\mathrm{Ai}$ and $\mathrm{Ai}'$ are all negative, real, and algebraically simple \cite[\S 9.9]{NIST:DLMF}. Zeros and poles of $\hat z$ thus line on the line $\mathfrak{L}_\theta = \{\ell\theta^{-1}\,,\, \ell>0\}$ both of which are are depicted in the right plot of Figure \ref{f:ztzp} along with the connected contours $\mathfrak{R}_{j}$ defined by $\mathrm{Re}\,\hat z = 0$, which connect each $s_{z,j}, s_{p,j}$ pair.

As seen in the left plot of Figure \ref{f:ztzp}, if we fix $\hat \omega\sim \mathrm{Im}\, s_{z,j}$ the horizontal contour $s = \hat\mu+\ri\hat\omega$ passes near the zero $s_{z,j}$ for some positive $\hat \mu$.  Through this passing, $\hat z(s)$ crosses  $\mathfrak{R}_j$ and thus also the imaginary axis, with  $\mathrm{Arg}\,\hat z(s)$ increasing by roughly $\pi$. It then becomes large amplitude near the pole $s_{p,j}$, while changing another $\pi$ in argument.

For $\alpha = 0$ (not pictured), we observe $\mathfrak{L}_\theta = \mathbb{R}_+$ and each $\mathfrak{R}_j$ is symmetric about the real axis (by symmetry of $\mathrm{Ai}$), with vertical tangent line at $s_{z,j}$ and $s_{p,j}$ (by an argument using the Cauchy-Riemann equations and the Riccati equation \eqref{e:zhatmu}). For $|\alpha|\ll1$, the slope of $\mathfrak{L}_\theta$ is small (with angle roughly given by $-\frac{\mathrm{Arg}\,{1+\ri\alpha}  }{3}$). We observe from the right plot that, in this case, the contours $\mathrm{Re}\,\hat z = 0$ (blue) are nearly vertical in a neighborhood of the real axis  and hence also the first several zero/pole pairs.  For a horizontal contour $s = \hat\mu+\ri\hat\omega$ with fixed $\hat \omega\approx \mathrm{Im}\, s_{z,0}$,  taking a vertical approximation of the $\mathrm{Re}\,\hat z = 0$ contour near $s_{z,0}$, we thus find the delay point $\hat \mu_0 := s_0 - \ri\hat\omega$ as the first crossing of $\mathfrak{R}_0$, satisfying
\begin{align}
\hat\mu_0 \approx  \mathrm{Re}\,  s_{z,0}=  \mathrm{Re}\,\{ \tilde\sigma_0/\theta\},
\end{align}
where we recall the known value of the first zero of $\mathrm{Ai}'(s)$ is  $\tilde \sigma_0 \approx -1.01879$. We observe that a $\tomega$ value less than $\mathrm{Im}\,  s_{z,0}$ is required in order to have $\mathrm{Arg}\, \hat z$ increase through $-\pi/2$ (as opposed to $\pi/2$) as  $\hat\mu+\ri\hat\omega$ passes near $s_{z,0}$. %We also observe that $\hat z$ becomes large amplitude when passing through the next contour near the pole $s_{p,0} = \mathrm{Re}\,\sigma_0/\theta$. %Using the known value $\tilde \sigma_0 = -1.01879...$, and approximating that the contour $\mathrm{Re}\hat z(s) = 0$ is vertical, we obtain

Translating back to the $\tilde z$ coordinates we thus predict the second-order correction to the front interface delay to be
\begin{align}\label{e:mu0}
\tilde \mu_0 = \hat\mu_0\eps^{2/3}.
\end{align}
For our example value $\alpha = -0.1$, using this vertical contour approximation, we obtain $\hat\mu_0\approx 0.977824$. %Recall that $\mu_0 = \mu_c + \tilde\mu_0.$
This prediction agrees well with the value measured from the results of numerical continuation; see Sec. \ref{ss:num-cgl} and Figure \ref{f:front_eps} below. An exact formula for $\hat\mu_0$ is $\hat\omega$-dependent and requires a more precise characterization of the contour $\mathfrak{R}_0$, or numerical evaluation of its intersection with the horizontal line $\hat\mu+\ri\hat\omega$. An intermediate step, which we we do not pursue here, is that one could estimate $\hat\mu_0$ by finding the intersection of $\hat\mu+\ri\hat\omega$ with the linear or quadratic approximation of the contour $\mathfrak{R}_0$ at $s_{z,0}$. The tangent vector and curvature here can be computed using the Riccati equation \eqref{e:zhatmu} and the Cauchy-Riemann equations.  The $\tomega$-dependence here also means the delay $\tmu_0$ is in fact coupled to the nonlinear transverse unfolding of the heteroclinic intersection in $\tilde \omega$ and a rigorous analysis there is required to understand both precisely. %affects the higher order corrections of the front interface delay. 

\subsection{Characterization of fast unstable fibers $W_0^\mathrm{uu}(r_p(\mu),z_p(\mu),\mu)$ }\label{ss:sp}

We next briefly characterize the fast unstable fibers $W_0^\mathrm{uu}(r_p(\mu),z_p(\mu),\mu)$, with a particular focus on $\mu$ values near $\mu_c$ where we expect and numerically observe, the leading-order intersection $W_\eps^\mathrm{u}(r_p,z_p,1)\cap W_\eps^\mathrm{s}(0,z_-,-1)$ to occur. 

To do this we first give more detail on the $\mu$-range where the equilibria is non-trivial and normally hyperbolic with one dimensional unstable manifold. We fix $\omega = \omega_c$, and $c>0$. %that $\alpha,\gamma,$ are such that the the family of equilibrium $r_p(\mu),q_p(\mu)$ is non-trivial (i.e. $r_p^2 = \mu - q_p^2 >0$) and has one-dimensional unstable subspace for a range of $\mu$ values containing the interval $[\mu_c,1]$. 
Non-triviality, $r_p>0$, can be found in a straight-forward but cumbersome computation. We solve the dispersion relation \eqref{e:nlom} for $q_p(\mu)$ with square root branch chosen so that $q_p(1) = k_c$. We then find $r_p^2 = \mu - q_p(\mu)^2\geq0$ for $\mu\geq \mu_b$ with $\mu_b$ implicitly defined by $\mu_b - q_p(\mu_b)^2 = 0$. Solving this last equation, we obtain $\mu_b = \mu_c \left( 2 + 3\alpha^2 - 2\sqrt{1+3\alpha^2 + 2\alpha^4} \right)/\alpha^2$.  It can readily be found that $\left( 2 + 3\alpha^2 - 2\sqrt{1+3\alpha^2 + 2\alpha^4} \right)/\alpha^2 < 1$ for all $\alpha$ and thus we can conclude $0<\mu_b<\mu_c$ for all $\alpha$ and any $c>0$. By our choice of equilibrium branch for $q_p$, we also obtain the desired normal hyperbolicity. Smooth dependence on parameters gives these properties for $\omega$ in a neighborhood of $\omega_c$ as well.

\begin{figure}[htbp]
    \centering
    \hspace{-0.2in}
               \includegraphics[width=0.33\linewidth,trim={2cm 0cm 0.0cm 1cm},clip]{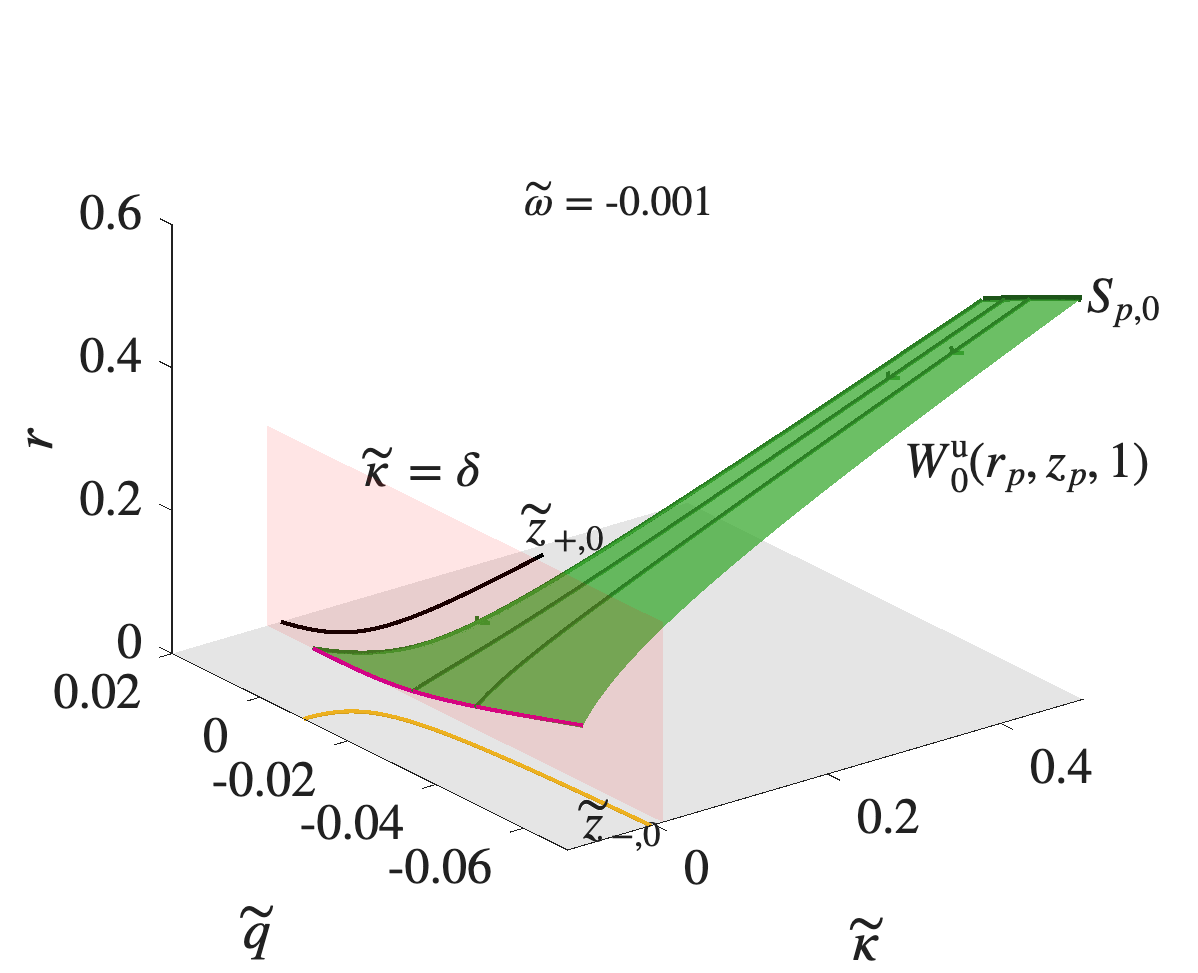}\hspace{-0.0in}
               %\vspace{-0.1in}
    \includegraphics[width=0.33\linewidth,trim={2cm 0cm 0.25cm 2cm},clip]{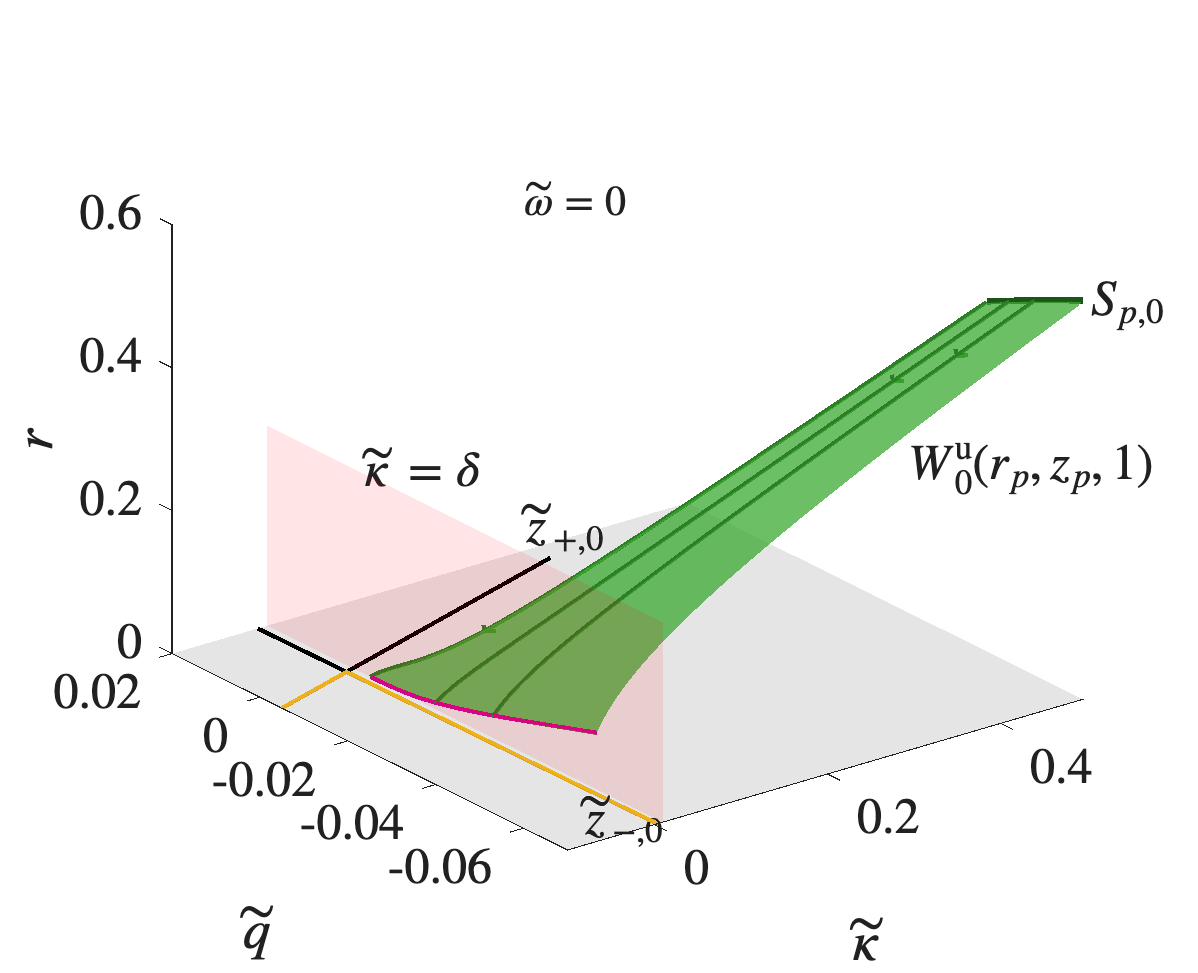}\hspace{-0.0in}
    \includegraphics[width=0.33\linewidth,trim={2cm 0cm 0.25cm 2cm},clip]{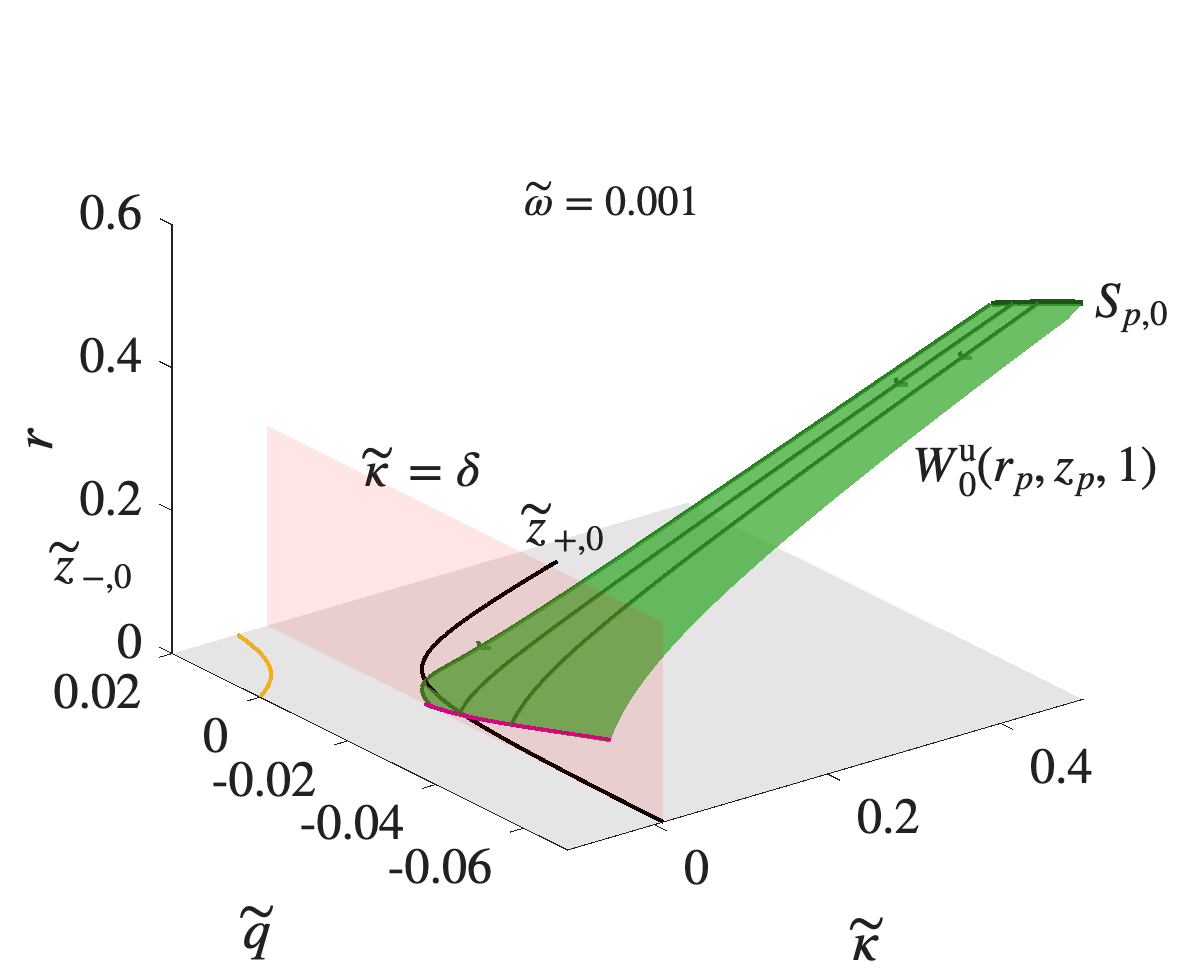}\hspace{-0.1in}
    \caption{Singular unstable manifold $W^\mathrm{u}_0(r_p,\tilde z_p,1-\mu_c)$ (green surface) in the $\tilde z,r$ space with the critical set $S_{p,0}$ (dark green line), for $\tmu\in(0,0.08)$ solved up to the section $\tilde\Sigma_\delta = \{\tilde \kappa = \delta\}$ (red). The fast fibers $W^\mathrm{uu}_0(r_p,\tilde z_p,1-\mu_c)$ (green curves) with larger $\tilde q$ values correspond to larger $\tmu$ values;  Also plotted are the critical sets $S_{-,0}$ (yellow) and $S_{+,0}$ (black) for a range of $\tmu$ values near 0;  Plots give several values of $\tomega =\{-1,0,1\}\times 10^{-3}$ (left to right, top to bottom); all with $c = 1, \alpha = -0.1,\gamma = -0.3$. }
    \label{f:zpme}
\end{figure}

\begin{figure}[htbp]
    \centering
    \hspace{-0.2in}
          \includegraphics[width=0.65\linewidth,trim={2cm 0cm 1.25cm 2cm},clip]{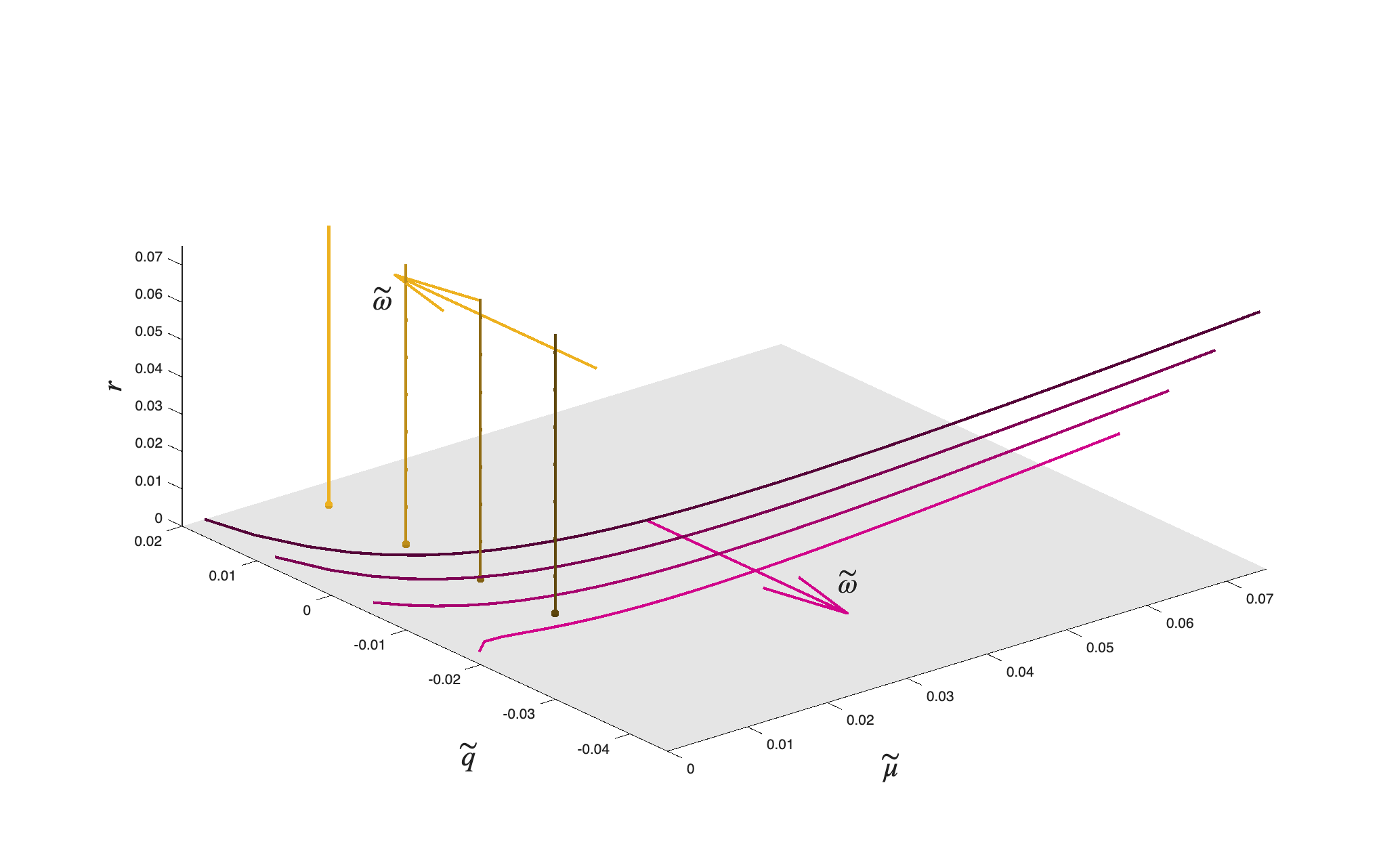}\hspace{-0.2in}
    \caption{Depiction of the leading order unfolding of  $W_\eps^\mathrm{u}(r_p,z_p,1)\cap\tilde\Sigma_\delta$ (magenta curves) and $W_\eps^\mathrm{s}(0,z_-,-1)\cap\tilde\Sigma_\delta$ (yellow curves) for $\tilde\omega = [-2,-1,0,1]\times 10^{-3}$; with $c = 1, \alpha = -0.1, \gamma = -0.3$, inner solution $\tilde z_{-,\eps}^\mathrm{in}(\tmu_\delta)$ intersection with $\tl\Sigma_\delta$ (yellow dots) and vertical fibers in $r$ (vertical yellow lines); the lightness of each curve in a set increases with $\tilde\omega$. }
    \label{f:lead_match}
\end{figure}

Returning to the structure of the unstable fibers, we once again use the change of coordinates in \eqref{e:ztmt}, so that system \eqref{e:ric_sys} takes the form
\begin{align}
r_\xi = \mathrm{Re}\,\{ \tilde z - \frac{c}{2(1+\ri\alpha)} \} r,\qquad 
\tilde z_\xi = -\tilde z^2 + Q(\tmu) + \frac{1+\ri\gamma}{1+\ri\alpha} r^2,\qquad
\tmu_\xi = -\eps(1 - (\tmu + \mu_c)^2).
\end{align}
In these coordinates the $(r_p,z_p)(\mu)$ equilibria take the form $\tilde z_p(\tmu) = \ri q_p+c(2+2\ri\alpha)^{-1},$ where $q_p$ satisfies the nonlinear dispersion relation $(\gamma - \alpha)q_p^2 + c q_p  - (\gamma - \alpha)\mu_c+ (\tomega - \gamma\tmu)=0$ in shifted coordinates and $r_p^2 = \mu_c+\tmu - q_p^2.$ Figure \ref{f:zpme} depicts a collection of these fibers, projected into the $r,\tilde z$ space, for $\tmu\gtrsim0$ for several fixed values of $\tomega \sim 0$. Each fiber was computed using numerical shooting from a small perturbation of each equilibria $(\tilde z_p(\tmu),r_p(\mu))$ in the direction of the one-dimensional strong unstable eigenspace. 
We observe that the manifold has decaying $r$ component as the trajectory converges into a neighborhood of the origin $\tl z= 0$. Indeed, for $\tomega = 0$, the $\tmu = 0$ fiber converges to the nilpotent equilibrium $\tl z_{\pm,0}(0) = 0$. This can be confirmed rigorously using a straight-forward trapping region argument and the strong exponential decay of the $r$ component for $\tilde\kappa$ near zero. Smooth dependence on parameters then implies, for $\tomega\sim0$, the union $\cup_{\tilde\mu\gtrsim0}W_0^\mathrm{uu}(r_p,\tilde z_p,\tmu)$ of these fibers intersect the 3D section $\tilde \Sigma_\delta = \{\tilde \kappa = \delta\}$ with $\delta>0$ small, in a smooth curve $(r_{0}^\ru,\tilde q_0^\ru)(\tmu;\omega)$ with $r$ component converging to zero as $\tmu\rightarrow0^+$; 
see magenta curves in Figure \ref{f:lead_match}. 

Since $S_{p,0}$ is normally hyperbolic for $\tmu$ in this range, Fenichel theory gives that the manifold smoothly perturbs for $0<\eps\ll1$ and hence %by smooth dependence of the manifold, 
we expect its intersection with $\tilde\Sigma_\delta$ to perturb smoothly to a curve $(r_{\eps}^\ru,\tilde q_\eps^\ru)(\tmu;\omega)$. To visualize the overall heteroclinic intersection, we also approximate $W_\eps^\mathrm{s}(0,\tilde z_-,-1-\mu_c)$ in a neighborhood of $(r,\tilde z,\tmu)\sim (0,0,0)$ and track its intersection with $\tilde \Sigma_\delta$.
We recall this manifold is foliated by the slow manifold $S_{-,\eps}$ and strong-stable fast fibers which are, at leading-order, vertical in the $r$ direction. 
In Figure \ref{f:lead_match}, we approximate $S_{-,\eps}\cap\tilde \Sigma_\delta$  using the inner solution $\tilde z_{-,\eps}^\mathrm{in}$ from \eqref{e:tzin}, numerically evaluated at $\tmu_\delta := \mathrm{arg}_{\tmu}\{ \mathrm{Re}\,\tilde  z_{-,\eps}^\mathrm{in}(\tmu) = \delta \}$ (yellow dots) for the same range of $\tilde \omega$ values near $0$.  We then approximate the strong-stable fibers with vertical lines in $r$ (yellow).

 In the same plot, we also track the variation of the curves $(r_{0}^\ru,\tilde q_0^\ru)(\tmu;\omega)$ in $\tilde\omega$ (magenta). We observe that, as $\tomega$ increases, the two sets of curves (yellow and magenta) cross each other, indicating a transverse unfolding of the two manifolds. That is, there exists a locally unique value $\tomega_f$ for which $W_\eps^\mathrm{u}(r_p,z_p,1)\cap  W_0^\mathrm{s}(0,z_-,-1)$ is non-empty. Indeed, in this figure we chose $\eps = 0.0015...,c = 1,\alpha = -0.1,\gamma = -0.3$ and observe that the two sets of curves cross near $\tilde\omega \approx -1\times 10^{-3}$. Numerical continuation results of the full heteroclinic, reported in the next section, found $\tilde\omega_f = -1.54\times10^{-3}$, an $\mathcal{O}(\eps)$ distance away. 
 
 We anticipate a rigorous proof of the leading-order intersection observed above will require computation of a Melnikov type splitting distance to unfold in $\tilde \omega\sim0$ and take into account the variation of each manifold's intersection with $\tilde \Sigma_\delta$.

\section{Numerical continuation results}\label{s:numcont}
Using numerical continuation, we now investigate the desired heteroclinic orbits, the selection of $\omega$ and $k$, and their dependence on $c$ and $\eps$. In sum, we find that the front location $\mu_f$ and selected frequency $\omega_f$ are accurately predicted by the AC instability transition values $(\mu_c,\omega_c)$ in the limit $\eps \rightarrow0^+$. The heteroclinic orbit exhibits a multi-scale structure, following the slow manifolds $S_{-,\eps}$ for $\mu<\mu_c$ and $S_{p,\eps}$ for $\mu>\mu_c$, outside of the transition regime for $\mu\sim \mu_c.$ We also investigate the $\mathcal{O}(\eps^{2/3})$ correction to the AC prediction is controlled by the slow passage of $S_{-,\eps}$ near the complex Airy point $z_c = -c/(2+2\ri\alpha)$ in the $r = 0$ dynamics for $(\omega,\mu) = (\omega_c,\mu_c).$ Throughout, we choose the example parameters $\alpha = -0.1, \gamma = -0.3$, which satisfy our assumptions above, but we expect similar results for a broad range of parameter values. 

\subsection{Numerical approach}
 We compute and continue solutions of \eqref{e:rkqmu} in AUTO07p \cite{doedel2007auto} as a boundary value problem in a large bounded domain $\tl\xi\in[-L/2,L/2]$ of length $L = 1000$ using the scaled variable $\tilde \xi = \eps\xi$ (so that $\mu_{\tilde\xi}$ is $\mathcal{O}(1)$) with parameter-dependent boundary conditions $z(-L/2) = z_L^-,\,\, z(-L/2) = \ri q_p$ and an integral phase condition which fixes the location of $\mu  = 0$ in the computational domain. Note $q_p$ is determined via the system parameters and the nonlinear dispersion relation. The frequency $\omega$ is solved for at each continuation step and either $\eps$ or $c$ serves as the continuation variable. An initial guess of the front is obtained by direct numerical simulation of \eqref{e:cgl} with $\alpha = \gamma = 0$ (i.e. real Ginzburg-Landau), $\eps = 0.25$   on a moderate computational domain where the selected front is asymptotically constant and purely real. This profile gives initial guesses  for $r$ and $\kappa$ with $q\equiv0$. After converging to a solution, we then homotope to the desired computational domain length and dispersion parameters $\alpha$ and $\gamma$. We are then able to continue either in the ramp speed $c$ or slope $\eps$. In the following section, we use the parameters $\alpha=-0.1, \gamma = -0.3$ as they satisfy our assumptions given in Section \ref{ss:assump}.  Source codes used to produce the computational results of this work can be found at the GitHub repository \url{https://github.com/ryan-goh/cgl-ramped-patterns}.

\subsection{Numerical Continuation Results in CGL}\label{ss:num-cgl}

\begin{figure}[htbp]
    \centering
    %\vspace{-0.5in}
   \hspace{-0.2in} \includegraphics[trim={0.0cm 0 0 5.5cm},clip,width=0.70\linewidth]{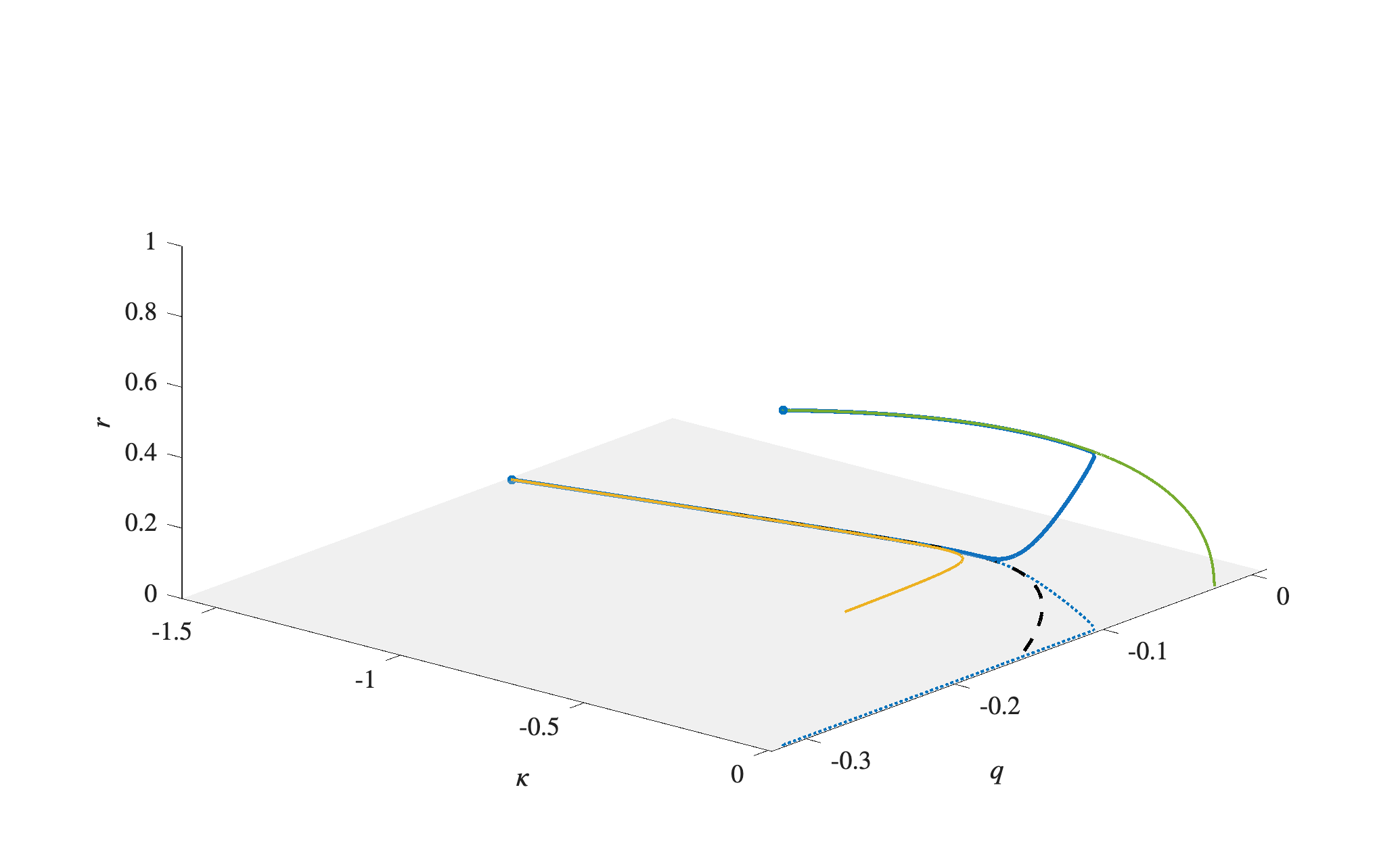}\hspace{-0.2in}
    \includegraphics[width=0.3\linewidth]{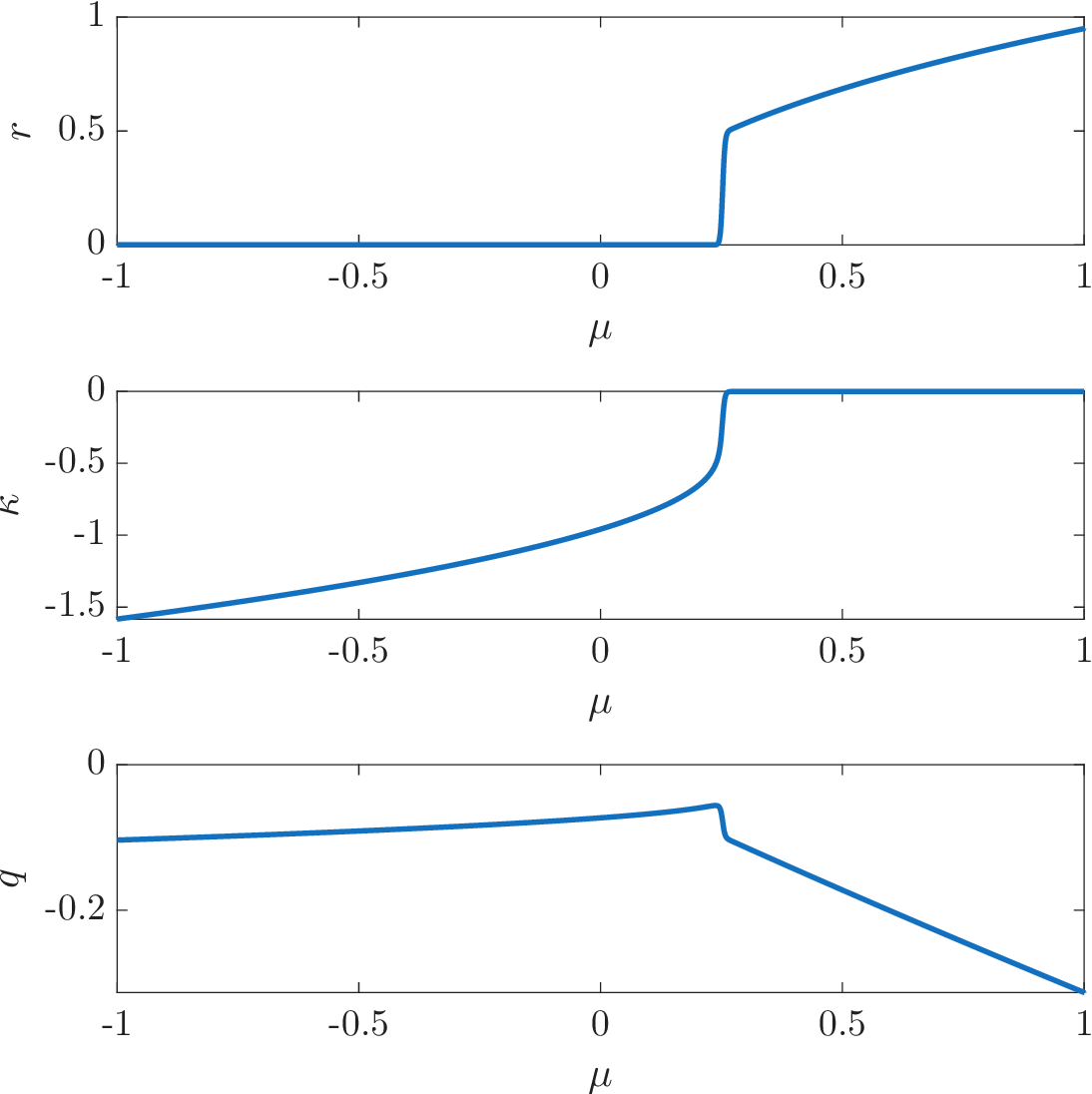}\hspace{-0.2in}\\
     \includegraphics[width=0.45\linewidth]{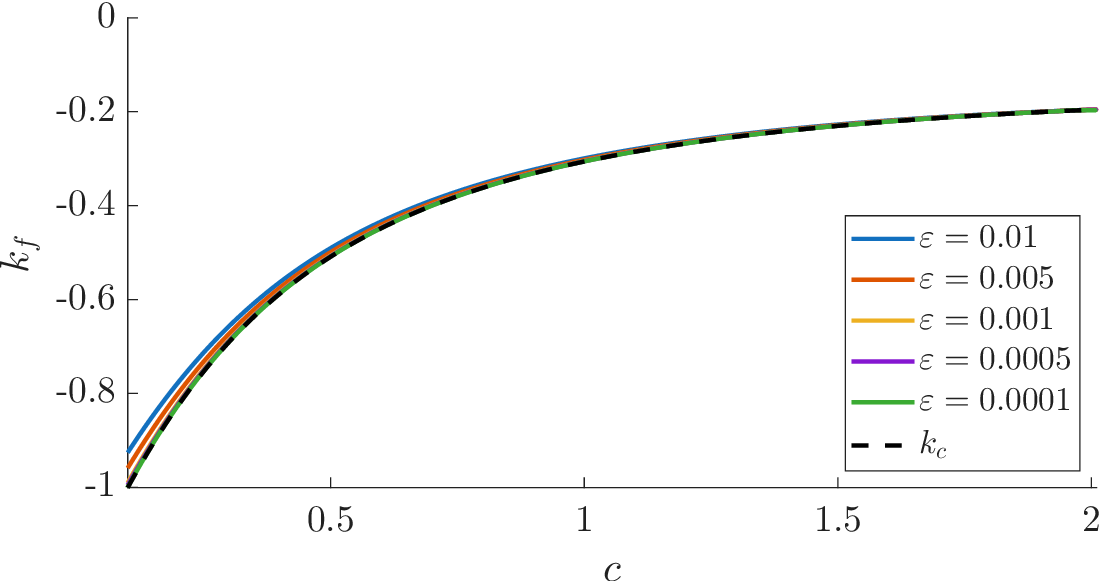}
      \includegraphics[width=0.45\linewidth]{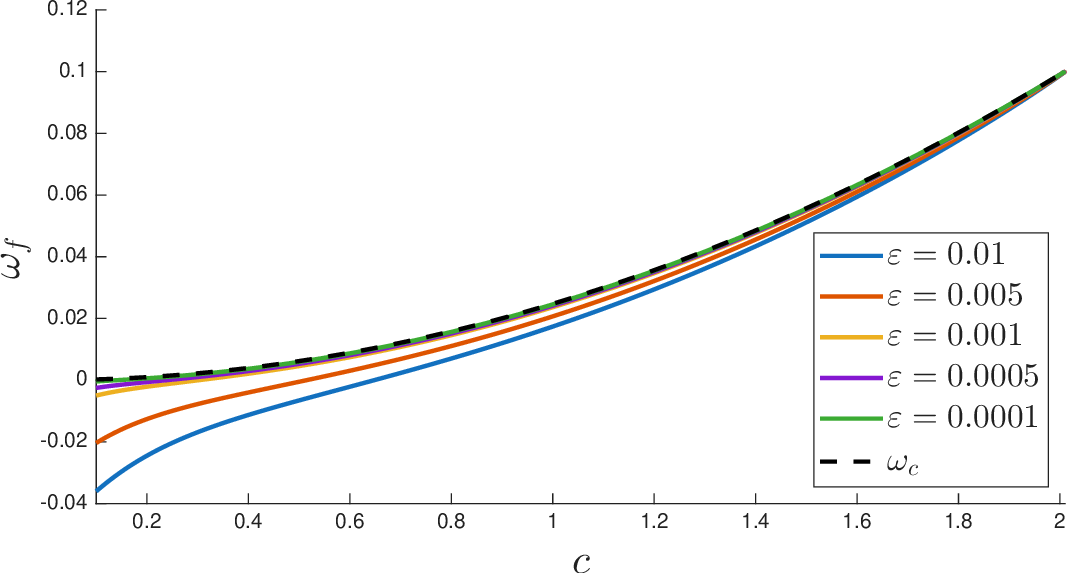}
    \caption{Top left: Projection of the heteroclinic trajectory of \eqref{e:rkqmu} in $(r,\kappa,q)$ space (solid blue) between $\mu=\pm1$ equilibria (blue dots), along with projection (dotted blue) onto the $r = 0$ plane (grey), plotted with $\eps = 0$ critical manifolds, $S_{-,0}$ (yellow) and $S_{p,0}$ (green), and the inner solution expansion \eqref{e:zin_unc} for the slow manifold $S_{-,\eps}$ (dashed black), parameter values $\alpha = -0.1,\gamma = -0.3,c = 0.967,\eps  = 0.001$, so $\mu_c\sim 0.231$; Top right: Trajectory components $r,\kappa,q$ plotted against $\mu(\xi)$; Bottom left: asymptotic selected wave number curves $k_f$ against $c$ for a selection of ramp slopes $\eps$ (colors) along with the AC prediction $k_c$ (dashed black); Bottom right: plot of asymptotic selected frequency $\omega_f$ against $c$ along with AC prediction $\omega_c$ (dashed black) for same values of $\eps$. Note: $c$ regime restricted away from $c=0$ as selected wave number has $|k|>1$ and hence does not correspond to a patterned front with non-zero amplitude $\sqrt{1-k^2}$ at $\xi = -\infty.$}
    \label{f:front_prof}
\end{figure}

Figure \ref{f:front_prof} depicts the projection of the heteroclinic orbit into $r,\kappa,q$-space, along with the dependence of each component on $\mu$ for a fixed $\eps$ small. We observe for $\mu\sim1$, the trajectory follows closely along $S_{p,0}$ (green) before jumping via a fast transition near $\mu\sim \mu_c$ with $\kappa$ leaving a neighborhood of 0, and touching down on top of the $r = 0$ plane. For $\mu$ below $\mu_c$, it follows $S_{-,0}$ (yellow) and further follows the slow-manifold $S_{-,\eps}$ (approximated in dashed black) for a larger $\mu$ interval including some values above $\mu_c$. 
We continue these fronts in $c$ for a range of fixed $\eps$ values, solving for the selected frequency $\omega_f$, from which wave number $k_f$ can be computed.  The bottom plots of Figure \ref{f:front_prof} show that the absolute spectrum prediction $\omega_c$ (dashed black) accurately predicts the selected frequency for $c$ values $\mathcal{O}_\eps(1)$, bounded away from $c = 0$, and less than the linear spreading speed $c_\rlin =2\sqrt{1+\alpha^2}$. Further, we find the frequency curves $\omega_f(c)$ converge pointwise onto $\omega_c$ as $\eps\rightarrow0^+$ for all $0<c<c_\rlin$. This is consistent with the behavior observed in the Allen-Cahn equation \cite{goh2023fronts,goh2024pitchfork} where the absolute spectrum governs the front transition for all $c\gtrsim \eps^{1/3}$ (while Painleve-II type dynamics govern the $c\lesssim \eps^{1/3}$ regime).   Figure \ref{f:c_front} gives the amplitude profiles $r$ against $\mu$ for a range of $c$ values, and plots the measured front location $\mu_f$ (i.e. where $r(\xi) = \delta$ for some fixed $\delta>0$ small). We once again find that $\mu_c$ (dashed black) accurately predicts the front location for a range of speeds when $\eps$ is small.
\begin{figure}[h!]
    \centering
    \includegraphics[width=0.45\linewidth]{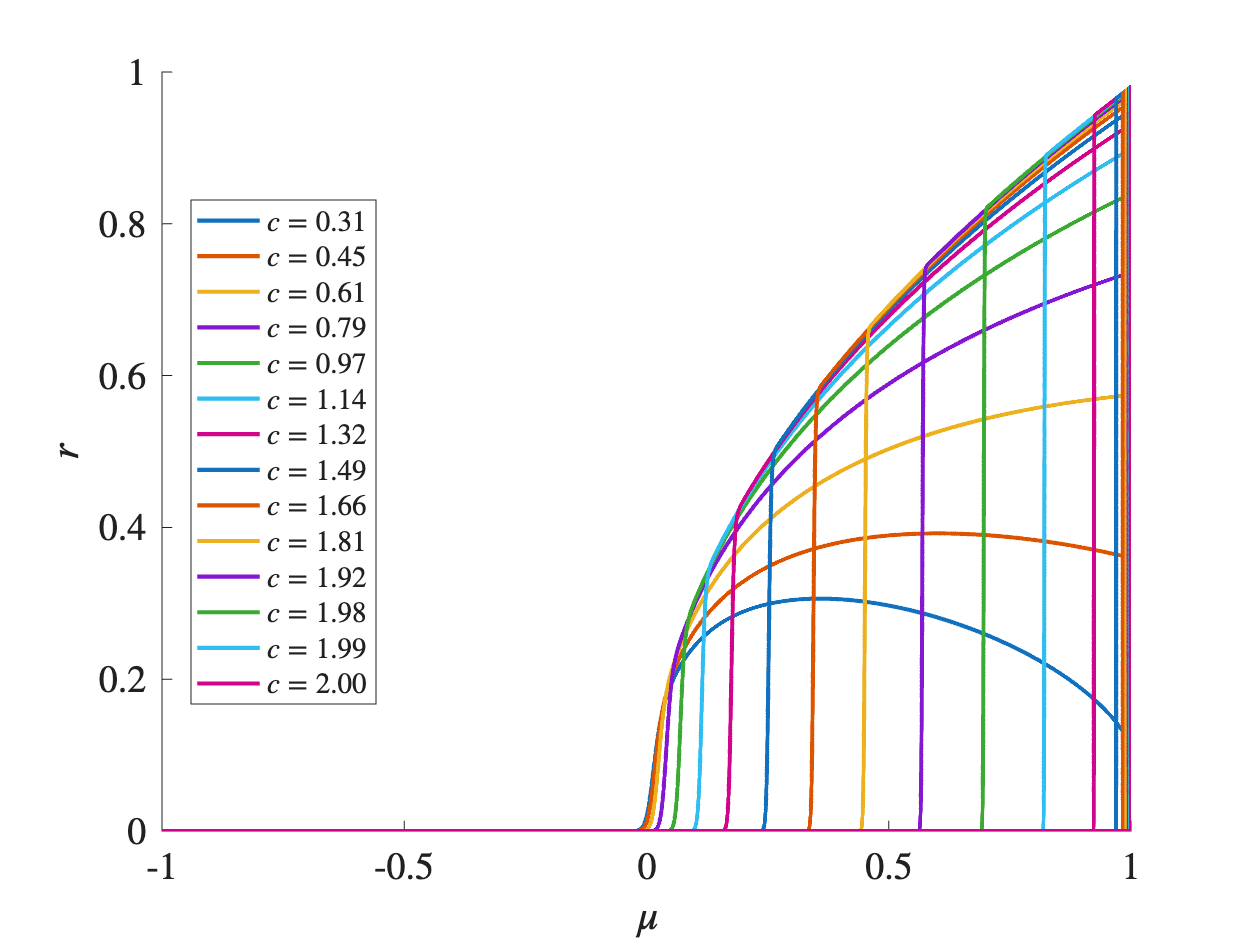}
    \includegraphics[trim={0.4cm 0 0 0},clip,width=0.37\linewidth]{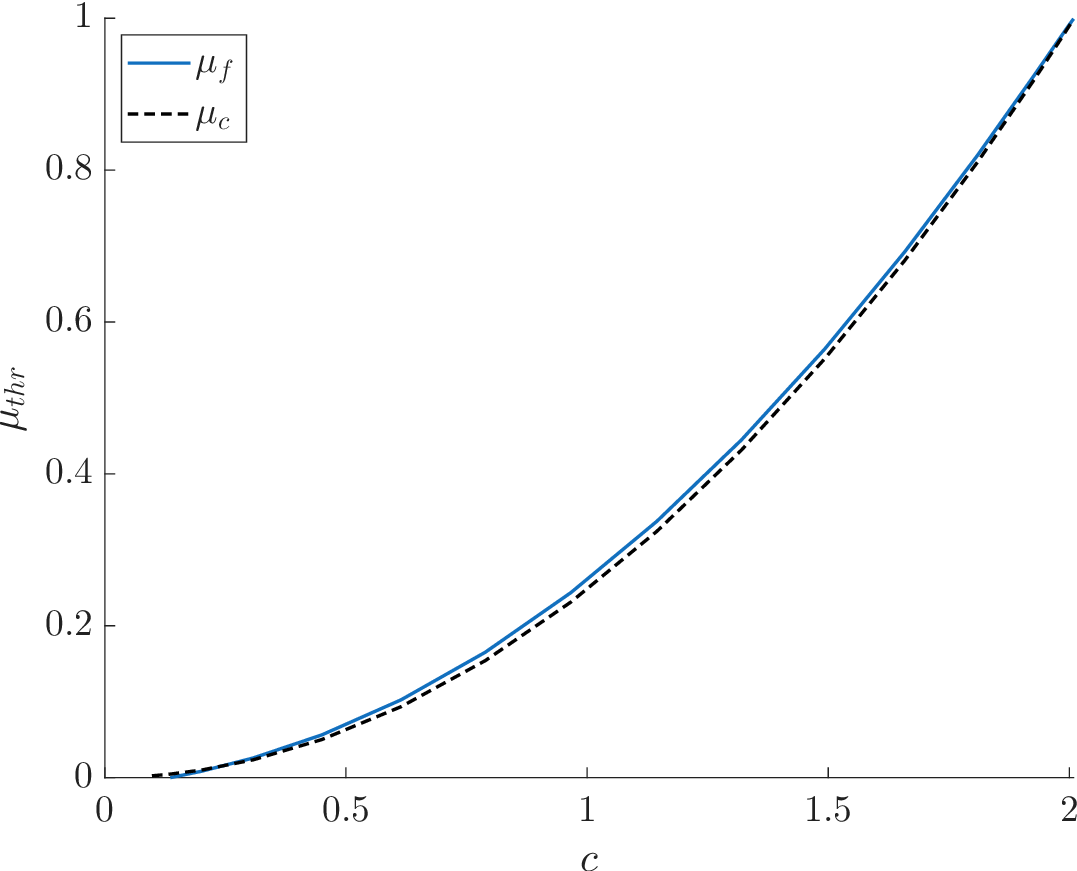}\\
    \caption{Left: Plot of front amplitudes $r(\xi)$ against $\mu(\xi)$ for a range of $c$ values (recall $\mu$ is decreasing in $\xi$ from $1$ to $-1$); Right: measurement of front position $\mu_{f} = \mu(\xi_f)$ for $\delta = 0.01$ (blue), plotted against AC prediction $\mu_c$ (dashed black); pparameters same as Fig. \ref{f:front_prof} top right.  }
    \label{f:c_front}
\end{figure}

\begin{figure}[h!]
    \centering
    \hspace{-0.3in}
    \includegraphics[width=0.35\linewidth]{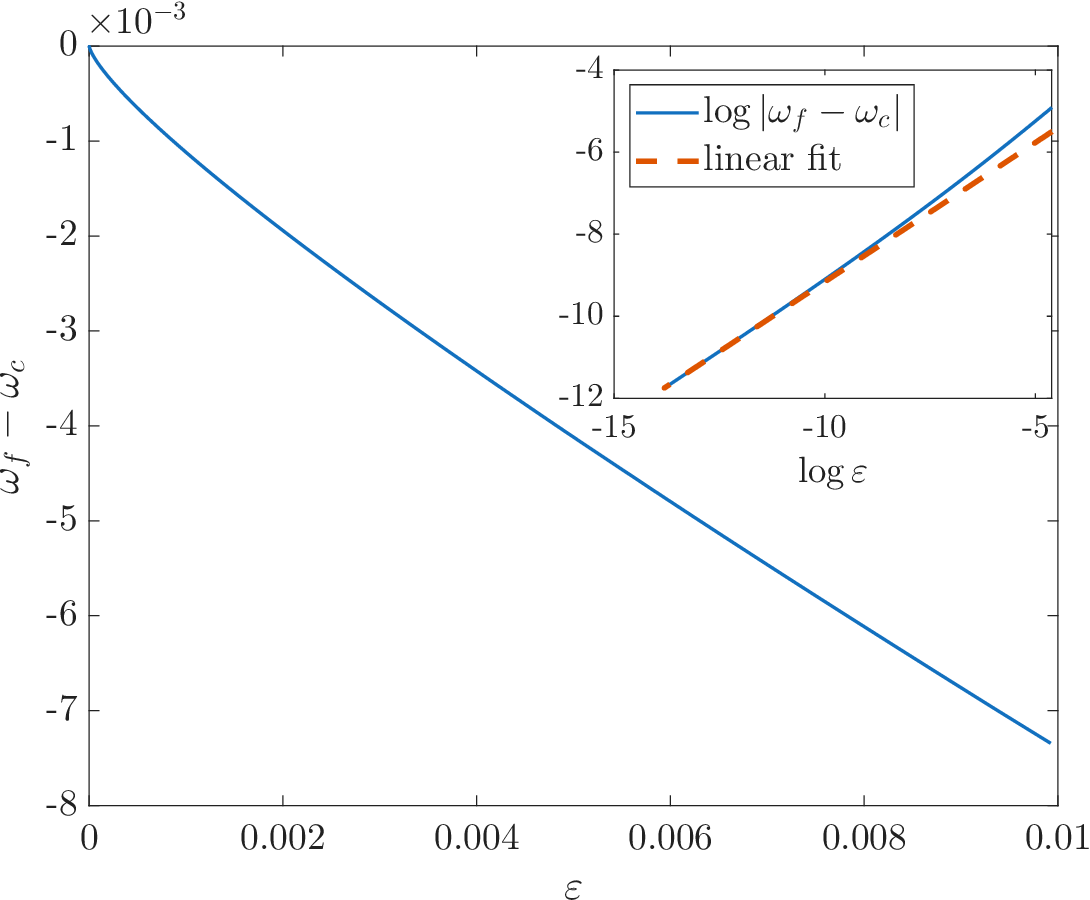}\hspace{-0.15in}
     \hspace{-0.0in} \includegraphics[width=0.35\linewidth]{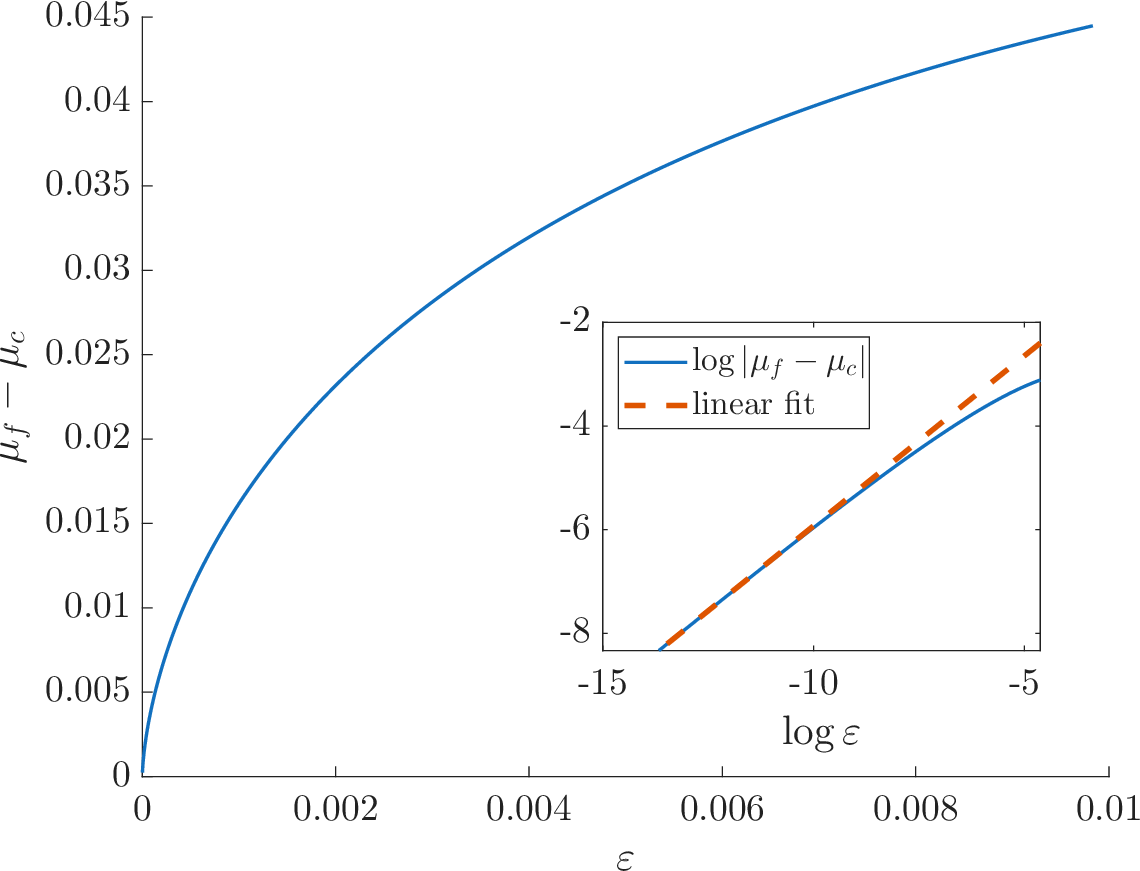}\hspace{-0.15in}
          \hspace{-0.0in} \includegraphics[width=0.35\linewidth]{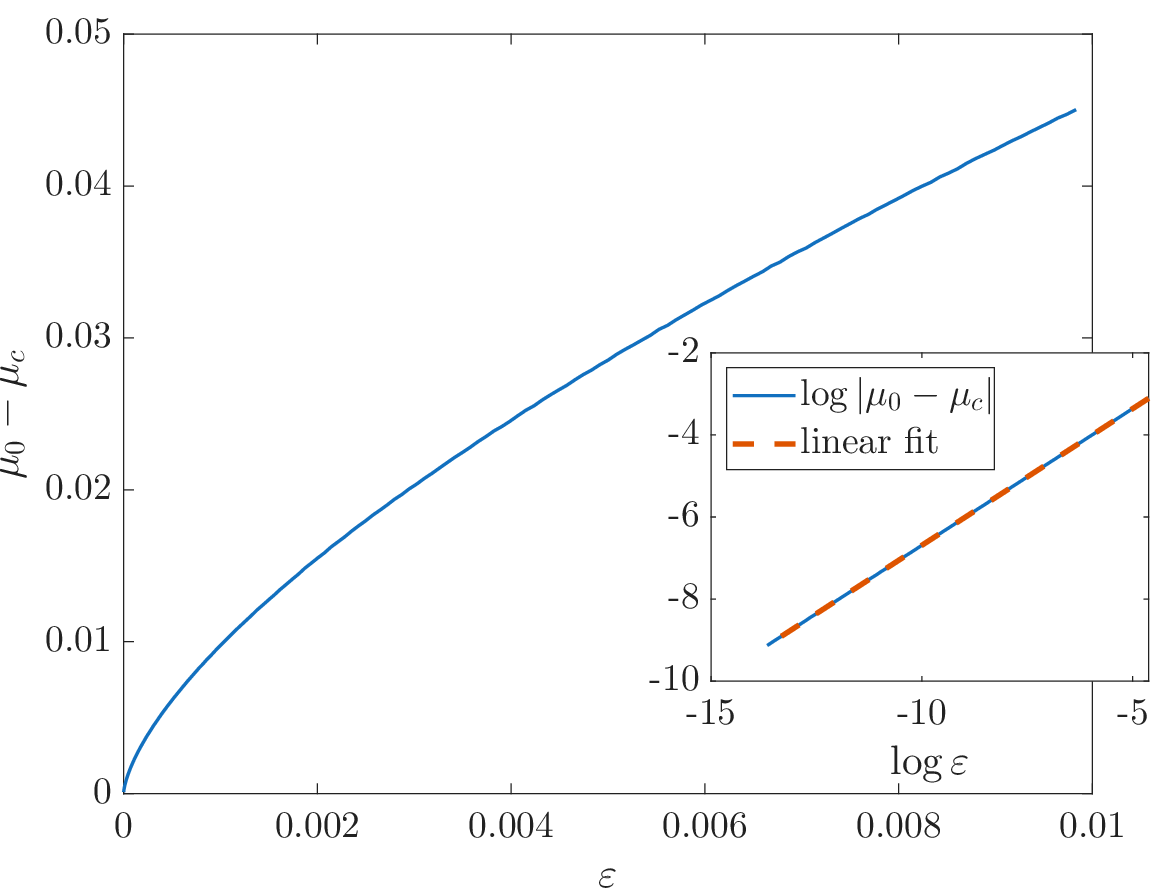}\hspace{-0.35in}\\
        \hspace{-0.2in}   \includegraphics[width=0.4\linewidth]{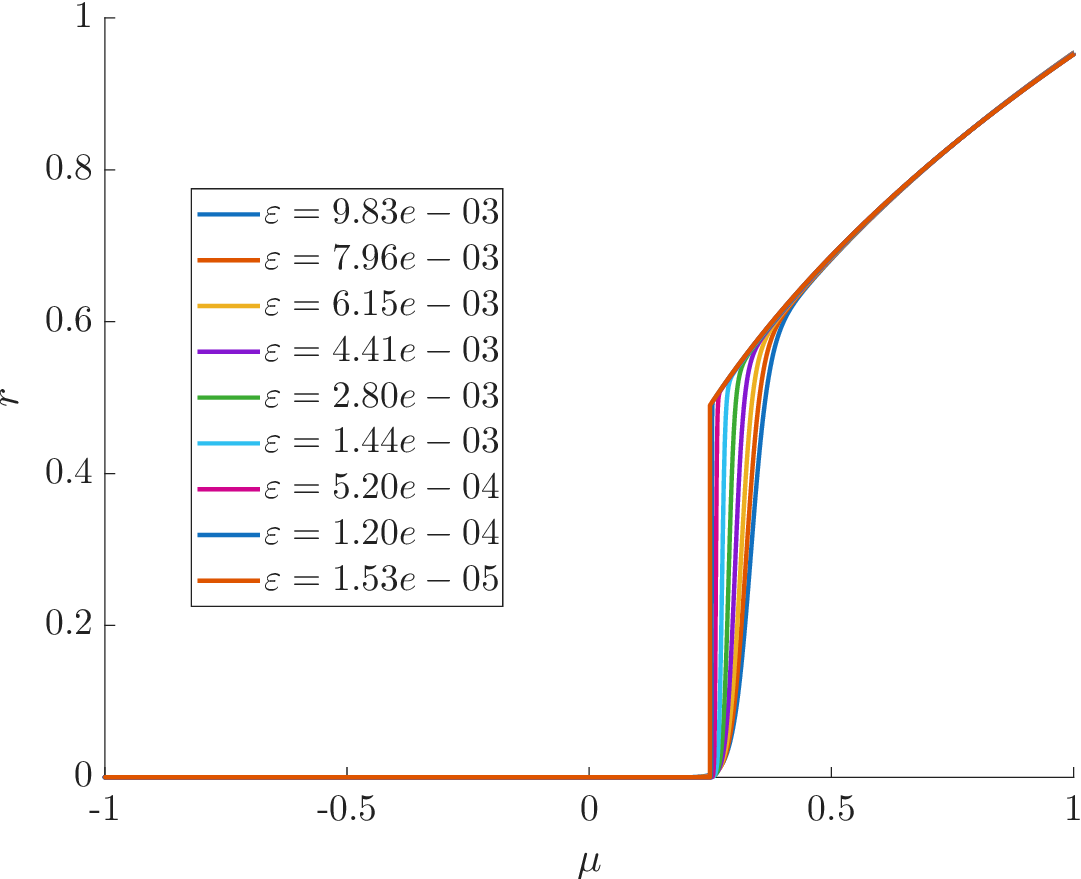}\hspace{-0.05in}
          \includegraphics[width=0.4\linewidth]{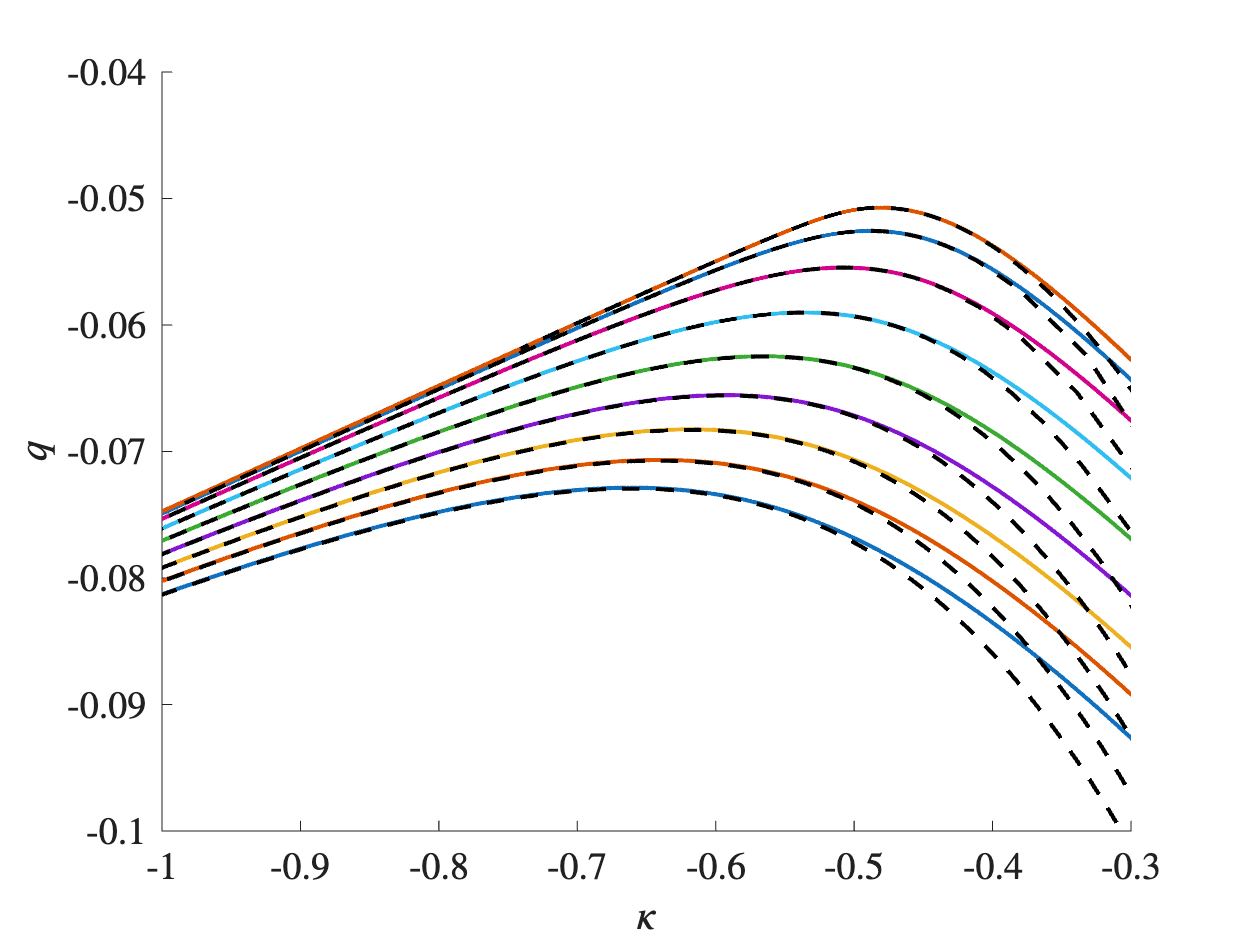}\hspace{-0.05in}
    \caption{Continuation of front in $\eps$ with $c = 1,\alpha = -0.1,\gamma = -0.3$ fixed; Top left: Descrepancy $\omega_f - \omega_c$ between selected and predicted frequency  for $\eps$ ranging between $0.01$ down to $10^{-6}$, inset gives  a log-log plot for data (blue) with linear fit line $0.679\log\eps -2.371$ (dashed orange) for the smallest values of $\eps$; Top center: Plot of front position discrepancy $\mu_f - \mu_c $, inset gives a log-log plot of data (blue) with linear fit line $0.656\log\eps+0.632$ (orange) of smallest $\eps$ data;
   Top right: Plot of discrepancy $\mu_0 - \mu_c$, inset gives log-log plot of data (blue) with linear fit $-0.02815...+ 0.666\log \eps$ (orange);
   Bottom left: Front amplitude profiles $r$ against $\mu$ for a range of $\eps$ values; Bottom right: Comparison of $\kappa,q$ solution components (colors correspond to same $\eps$ values as in bottom left plot) with the unscaled inner solution $z_{-,\eps}^\mathrm{in}(\mu)$ (dashed black) near $z_c$ for $\mu$ close to $\mu_c$ and $\omega$ values chosen from numerical continuation.
  }
    \label{f:front_eps}
\end{figure}

We further investigate the $\eps$ dependence of fronts by continuing down in $\eps$ for $c = 1$ fixed. %We continue from $\eps = 10^{-2}$ down to $\eps = 10^{-6}$. 
Figure \ref{f:front_eps} shows the convergence of $\omega_f$ and $\mu_f$ to the predicted values $\omega_c$ and $\mu_c$ with $\mathcal{O}(\eps^{2/3})$ rate. In the bottom left plot, depicting front amplitude $r$ against $\mu$, we observe an $\eps$-dependent delay in the interface, with interface location $\mu_f \geq \mu_c$ (so $\xi_f< \xi_c$), and steepening interface as $\eps \rightarrow0^+.$ Plotting on a log scale and performing a linear fit for the smallest $\eps$ values, we observe $|\mu_f - \mu_c|\sim \eps^{\beta}$ with measured $\beta = 0.656...$, consistent with the $\eps^{2/3}$ delay predicted by our inner solution asymptotics above and rigorously proved in \cite{goh2023fronts} for asymptotically constant fronts in the Allen-Cahn equation; see Fig \ref{f:front_eps} top middle. Furthermore, we also measure the location $\mu_0$ where the $z$ coordinate has $\mathrm{Re}\, z  = \kappa_c:= -c/(2+2\alpha^2)$; see Fig. \ref{f:front_eps} top row right. Performing a linear fit on the log data, we observe that $\mu_0 - \mu_c\approx 0.9722 \eps^{2/3}$, which agrees quite well with the predicted correction $0.9778\eps^{2/3}$ from the asymptotic inner solution in \eqref{e:mu0}. %Remark, 

Further, in the upper left plot of Figure \ref{f:front_eps},  we observe the expected asymptotic behavior in the selected frequency $\omega_c - \omega_f\sim \eps^{\tilde\beta }$ with a measured exponent of  $\tilde\beta = 0.679$. Finally, in the lower right plot of Figure \ref{f:front_eps} we find the inner solution $z_{-,\eps}^\mathrm{in}(\mu)$ (dashed black), given in \eqref{e:zin_unc} and evaluated using the numerically determined $\omega_f$, accurately predicts the behavior of the heteroclinic front solution (colors) when $r\sim0$, that is before the front becomes large amplitude.

\section{Spectral Stability of slowly-ramped fronts}\label{s:stab}

We briefly report on spectral stability of the patterned fronts found above. Following from the intuition outlined in Remark 8.1 of \cite{goh2023fronts}, for $c$ values where the selected asymptotic wave train is diffusively stable - that is has (essential) spectrum in the closed left half-plane touching the imaginary axis only via a quadratic tangency at $\lambda = 0$ -  we expect patterned-fronts to also be  stable.  To summarize this remark for convenience, the slowly-varying spatial ramp induces a patterned front with a plateau region where $A$ is small-amplitude and close to the temporally unstable state $A = 0$ for a large $\xi$-region, where $\mu(\xi)\in(0,\mu_f)$. For the sub-interval $\mu\in(0,\mu_c)$, the homogeneous state $A = 0$ is only convectively unstable in \eqref{e:cgl_cv}, while for $\mu\in(\mu_c,\mu_f)$ the state is absolutely unstable. Scaling arguments, as well as the numerical results depicted above in Sec. \ref{f:front_eps}, indicate that the latter region has $\mathcal{O}(\eps^{2/3})$ width in $\mu$. Since $\mu'(\xi) = \mathcal{O}(\eps)$ in this region, the corresponding $\xi$-interval of absolute instability has length $L = \mathcal{O}(\eps^{-1/3})$. By general theory on absolute spectrum \cite{sandstede2000absolute} we expect all but finitely many eigenvalues to accumulate onto absolute spectrum of the homogeneous state with rate $L^{-2} = \mathcal{O}(\eps^{2/3})$ as $\eps\rightarrow0^+$.  Next, since the projective winding frequency (roughly how fast trajectories $z(\xi)$ wind around the Riemann sphere) goes to zero as  $\mu\rightarrow\mu_c$, we expect no intersections for $\mathrm{Re}\,\lambda\geq0$ and therefore no unstable point spectrum; see also \cite{carter2021pulse} for related phenomena and \cite{goh2020spectral} for a rigorous analysis of these eigenvalues in the sharp ramp case $\eps = \infty$ for $c\lesssim c_\mathrm{lin}$.

\begin{figure}[htbp]
    \centering
    \hspace{-0.2in}
    \includegraphics[width=0.4\linewidth]{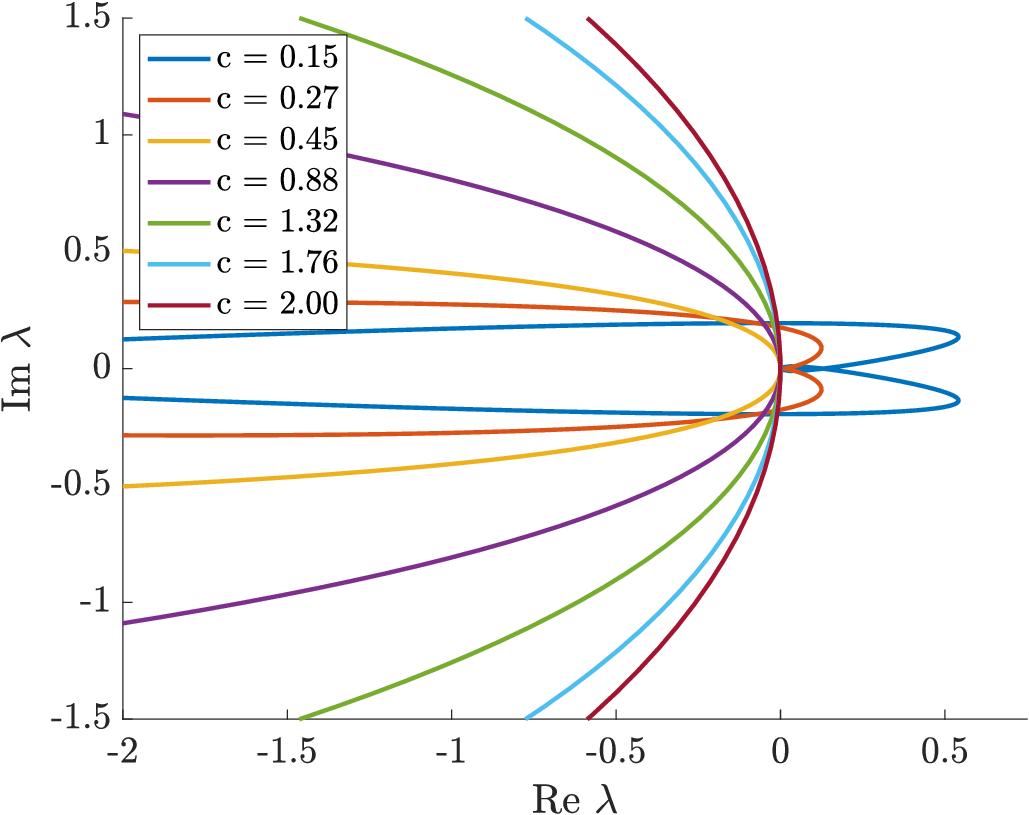}\hspace{-0.05in}
     \includegraphics[width=0.4\linewidth]{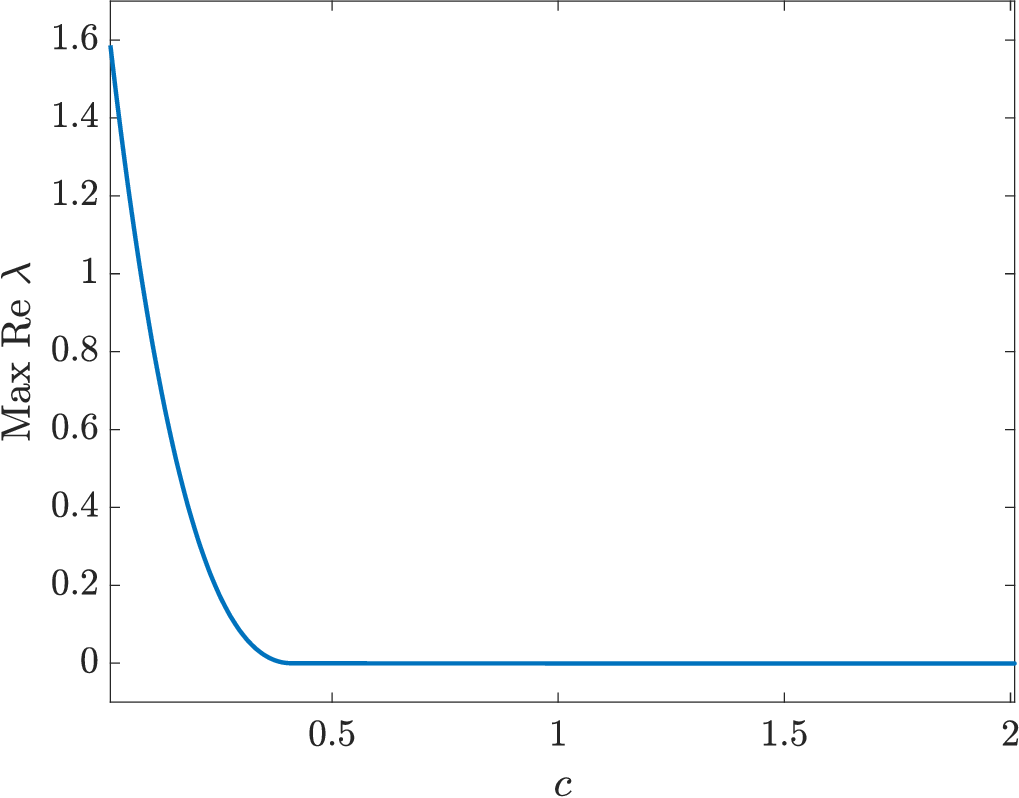}\hspace{-0.05in}
    \caption{Left: Plot of right most Fredholm boundary $\lambda_{-,+}(\ell)$ of the essential spectrum for $\mathcal{L}$ about the front for $\eps = 0.005$ and a range of $c$ values; Right: Plot of $\max_\ell \mathrm{Re} \lambda_{-,+}(\ell)$ for a range of $c$. For $(\alpha,\gamma) = (-0.1,-0.3)$, Eckhaus unstable wave numbers are selected by the ramp for $c\leq c_\mathrm{eck}\approx 0.41$, the value where $k_c^2 = k_{eck}^2$.}
    \label{f:essspec}
\end{figure}

To study spectral stability, we perturb the patterned-front solution $\re^{-\ri\omega_f t}A_f(\xi) = r_f(\xi)\re^{\ri\left(\varphi_f(\xi) - \omega_f t \right)} $ in \eqref{e:cgl} with the ansatz $A(\xi,t) =\left(r_f(\xi)+R(\xi,t)\right)\re^{\ri\left(\varphi_f(\xi)-\omega_f t + \phi(\xi,t) \right)} $  as given in \cite{beck2014nonlinear}; see also \cite{goh2020spectral} for an equivalent perturbation scheme. One obtains a nonlinear equation for the variables $U = (R, r_f \phi)$ of the form $U_t = \mathcal{L}_f U + \mathcal{N}(R,\phi)$ with $\mathcal{N}(0,0) = 0, D_{R,\phi}\mathcal{N}(0,0) = 0$ and with linear operator $\mathcal{L}_f = D_2 \partial_\xi^2 + (c I_2 - 2q_f D_1)\partial_\xi + D_0(\xi)$ for 
$$
D_2 = \begin{pmatrix}
 1 && -\alpha \\ \alpha && -1   
\end{pmatrix},\,
D_1 = \begin{pmatrix}
 \alpha && 1 \\ -1 && \alpha   
\end{pmatrix},\,
D_0(\xi) = 
\begin{pmatrix}
 -q_f^2 - \alpha q_{f,\xi} + \mu - 3 r_f^2 && \alpha (\kappa_{f,\xi} + \kappa_f^2) + 2 q_f \kappa_f  \\
 \omega_f - \alpha q^2_f + q_{f,\xi} - 3 \gamma r_f^2 + c q_f &&  -(\kappa_{f,\xi} + \kappa_f^2) + 2 \alpha q_f\kappa_f   - c \kappa_f
\end{pmatrix},
$$
where we recall that $\kappa_f = r_{f,\xi}/r_f,\, q_f = \varphi_{f,\xi}.$  The essential spectrum of $\mathcal{L}_f$ posed in $L^2(\R)$ is found by studying the Fredholm boundaries $\sigma_F(\mathcal{L}_\pm)$ for $\mathcal{L}_\pm:= \lim_{\xi\rightarrow\pm\infty} \mathcal{L}_f$ \cite{kapitula2013spectral}. In standard fashion, these curves are found by replacing $\partial_\xi$ with $\ri \ell, \ell\in\R $ and computing the pairs of eigenvalues $\lambda_{\pm,j}(\ell), j = \pm$ for the resulting matrices
$$
M_\pm(\ell) = -\ell^2 D_2 + (c I_2 - 2 q_{f,\pm}) D_1\ri\ell +D_{0,\pm}
$$
with $D_{0,\pm} =\lim_{\xi\rightarrow\pm\infty} D_0(\xi)$, $q_{f,\pm} = \lim_{\xi\rightarrow\pm\infty} q_f(\xi)$; recall $\lim_{\xi\rightarrow\pm\infty} \mu(\xi) = \mp 1.$
A straight-forward computation gives that the pair of eigenvalues $\lambda_{+,j}(\ell)$ for $M_+(\ell)$ with $\mu = -1$ both satisfy $\mathrm{Re}\, \lambda_{+,j}(\ell) <0$ for all $\ell\in\R$. Thus, essential (in)stability of the front is determined by the eigenvalues $\lambda_{-,j}(\ell)$ which,  after a straight-forward but somewhat involved computation, are found to be
\beq\label{e:l_ess}
\lambda_{-,j}(\ell) = -\ell^2 + (c+2\alpha k_f)\ri\ell + (1-k_f^2)\left(-1  \pm \sqrt{1-2\gamma\rho_-(\ell) - \rho_-(\ell)^2}\right),\qquad \rho_-(\ell) = \frac{\alpha\ell^2 + 2 k_f \ri\ell}{1 - k_f^2}.
\eeq
with $j = \pm$ denoting the two branches of the square root.
Examples of these curves are depicted in Figure \ref{f:essspec} for a range of $c$ values using the wave number values $k_f$ from the previous numerical continuation for a fixed small $\eps$ value. 

We find, for this specific set of parameters, that the selected wave number is essentially unstable for small speeds, with $k_f^2> k_{eck}^2:= \frac{1+\alpha\gamma}{3+\alpha\gamma+2\gamma^2}$ the Eckhaus stability boundary, transitioning to diffusive stability for speeds $c$ larger than some value $c_\mathrm{eck}$. This value can be approximated in the limit $\eps\rightarrow0^+$ using the leading order AC wave number prediction by solving for $c$ in $k_c^2 = k_{eck}^2$ where we recall $k_c$ is defined in formula \eqref{e:kc}. For the example parameters $(\alpha,\gamma) = (-0.1,-0.3)$ we find $c_\mathrm{eck} \approx 0.4090$, consistent with what is observed in Figure \ref{f:essspec}.

We study point spectrum of $\mathcal{L}_f$ by discretizing the operator using fourth-order finite differences on a large uniform spatial mesh with Neumann boundary conditions. The $\xi$-dependent coefficients of this operator are evaluated using solution profiles from AUTO continuation which are interpolated onto the uniform mesh in $\xi$. MATLAB's \texttt{eigs} command was then used to compute eigenvalues of the discretized operator matrix.

 The work \cite{sandstede2000absolute} gives that eigenvalues of $\mathcal{L}_f$, posed on a large bounded domain of length $L$ with separated boundary conditions, converge onto the point spectrum of the unbounded domain operator and  the absolute spectrum of the asymptotic states $\mathcal{L}_\pm$ as $L\rightarrow+\infty$. Despite this, when computing eigenvalues of the discretized operator, the essential spectrum of the asymptotic states can induce numerical pseudo-spectrum and thus spurious eigenvalues \cite{dodson2024efficient}. For diffusively stable essential spectrum, this effect may cause difficulties locating point spectrum near the origin. 

To deal with such pseudo-spectrum and locate point spectrum, we instead consider the exponentially conjugated operator $\mathcal{L}_{f,\eta} = \re^{\eta \xi}\mathcal{L}_{f}\re^{-\eta \xi}$ which, for a range of positive $\eta$ values, shifts essential spectrum to the left. When the asymptotic states are either diffusively stable or only convectively unstable, a weight $\eta$ can be chosen to move the essential spectrum, and thus numerical pseudo spectrum, into the left-half plane, leaving the numerical eigenvalue problem well-conditioned near $\lambda = 0$; see \cite{dodson2024efficient,sandstede2000absolute} for more detail. 
\begin{figure}[htbp]
    \centering
    \hspace{-0.2in}
    \includegraphics[width=0.55\linewidth]{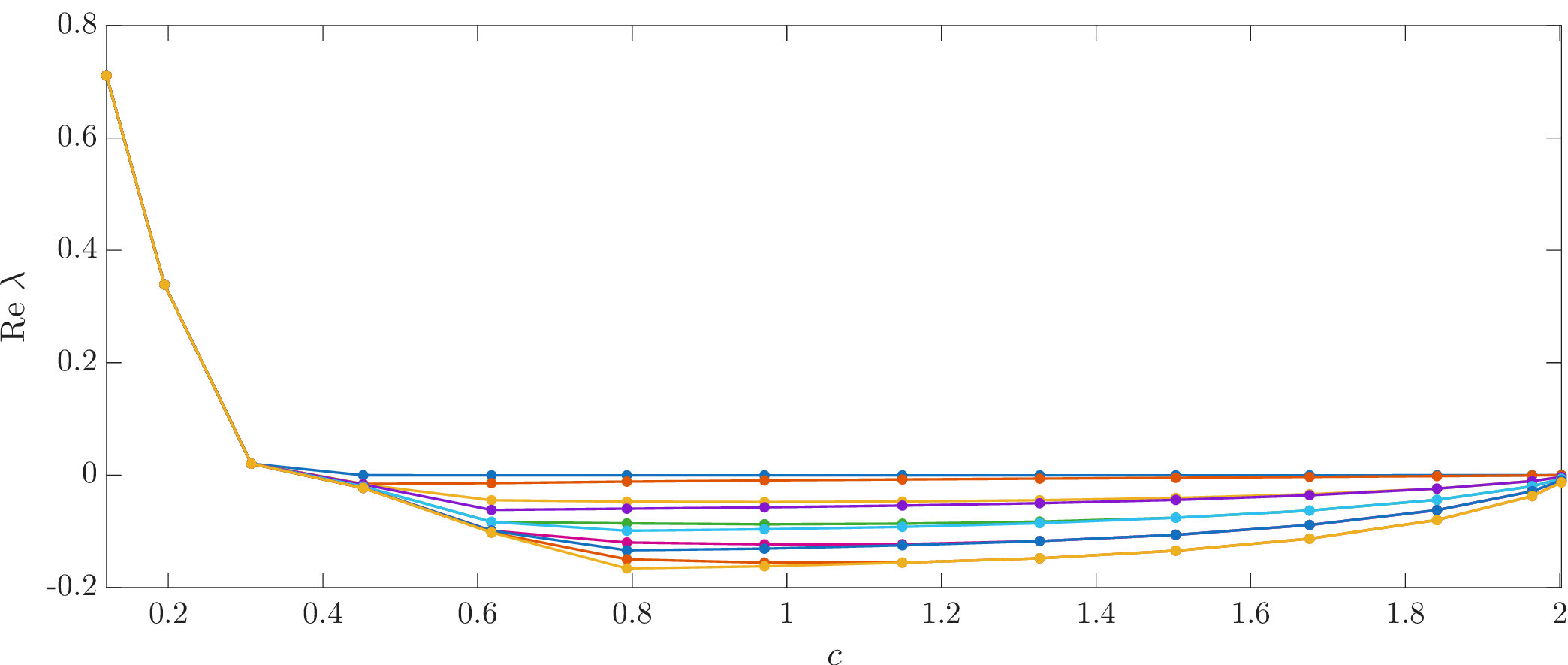}\hspace{-0.05in}\\
     \hspace{-0.2in}\includegraphics[width=0.33\linewidth]{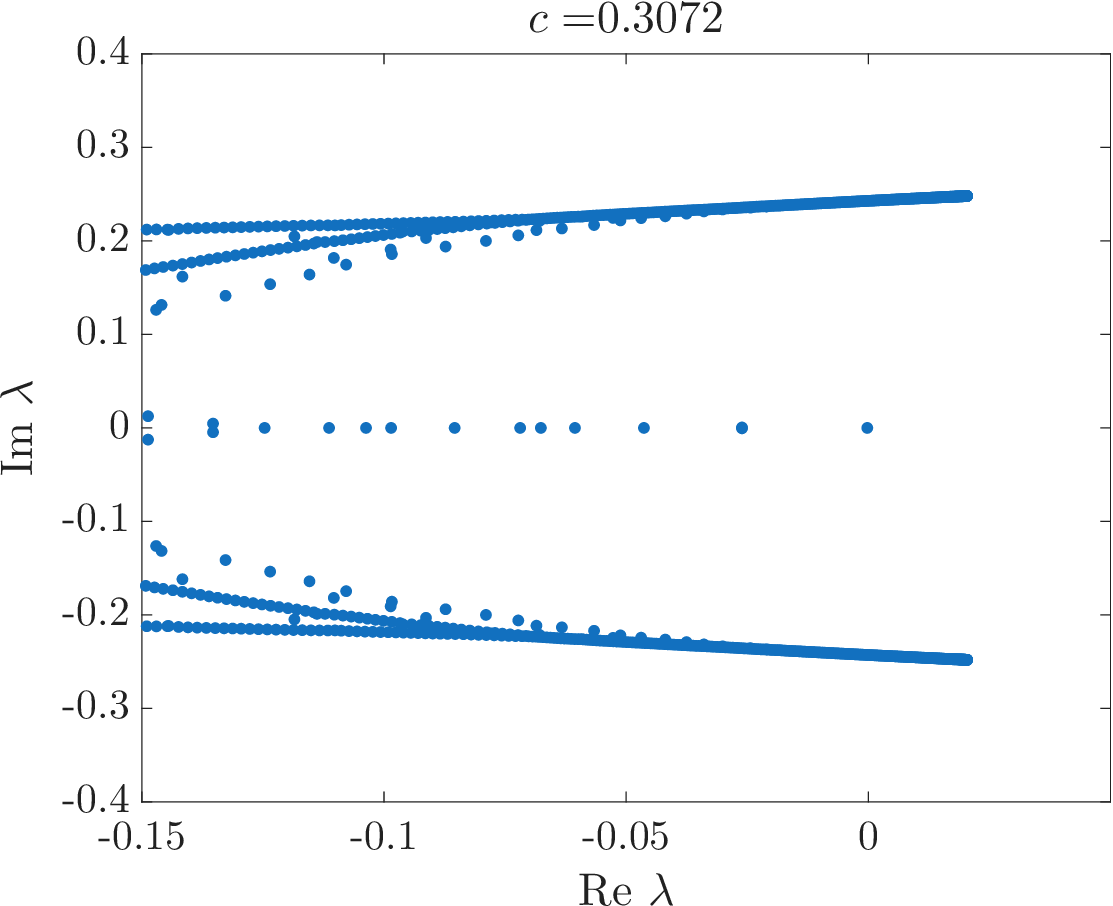}\hspace{-0.05in}
     \includegraphics[width=0.33\linewidth]{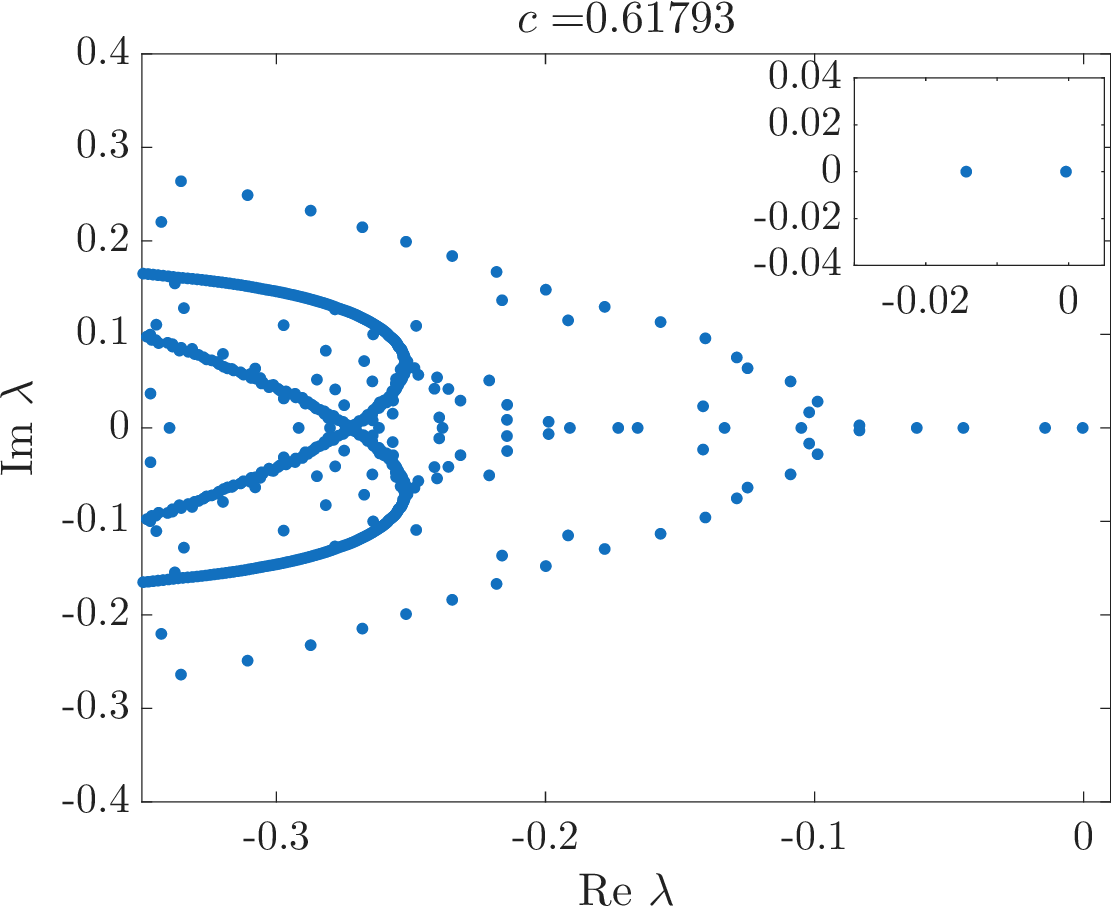}\hspace{-0.05in}
      \includegraphics[width=0.33\linewidth]{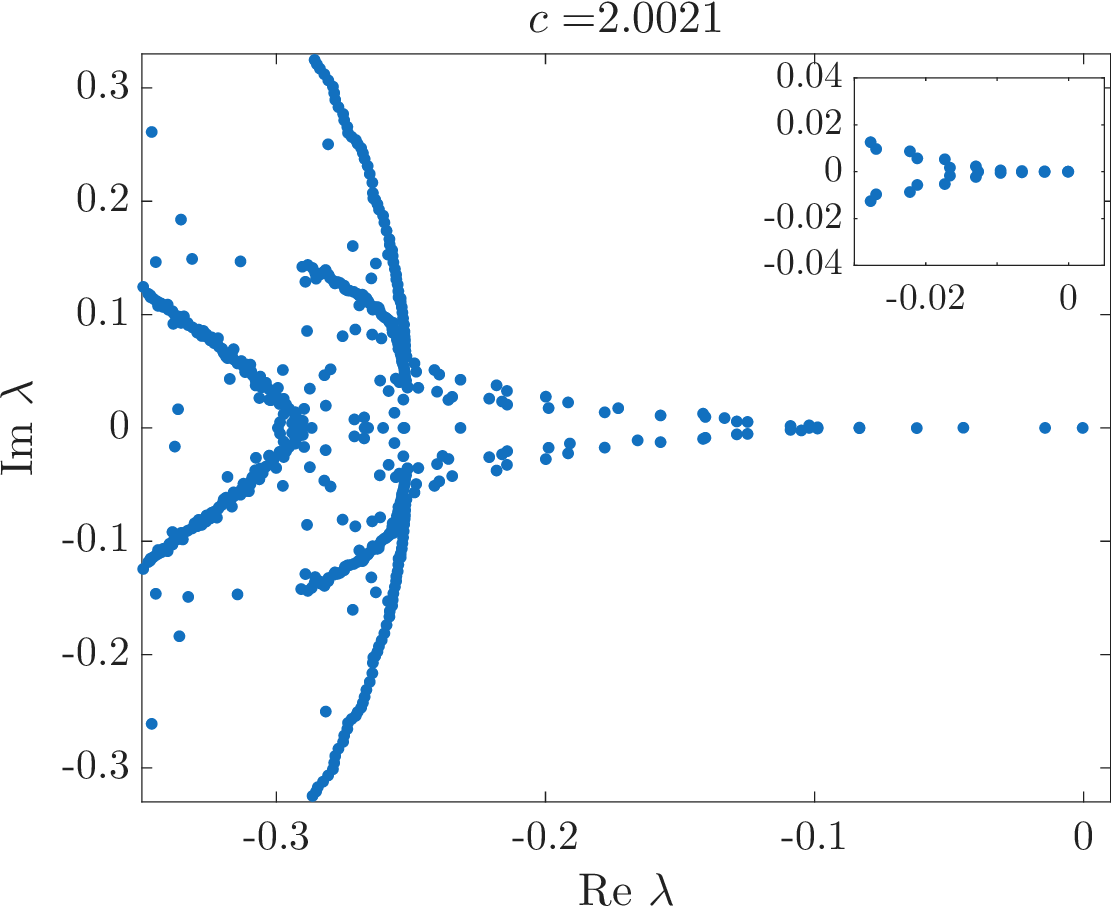}\hspace{-0.2in}
    \caption{Point spectrum for numerical discretization of $\mathcal{L}$; Top: plot of ten most positive values of $\mathrm{Re}\,\lambda$ plotted against $c$; Bottom: plot of numerical spectrum for $c = 0.307,0.617, 2.002$ (left to right); insets give zoom in near the origin. In all, $\eps = 0.005$, $\alpha = -0.1,\gamma = -0.3$, the linearization was evaluated on solution profiles from AUTO continuation which were interpolated onto a uniform mesh $\xi\in[-3000,1000]$ with $dx = 0.1$ and spatial derivatives evaluated with fourth-order finite differences and Neumann boundary conditions, the exponential weight was chosen to stabilize pseudo spectrum as much as possible, $\eta = 0.05$ to $\eta = 0.25$ for $c$ values less than 0.4 and $\eta = 0.8$ otherwise.  }
    \label{f:spec}
\end{figure}

\begin{figure}[htbp]
    \centering
    \hspace{-0.2in}
    \includegraphics[width=0.55\linewidth]{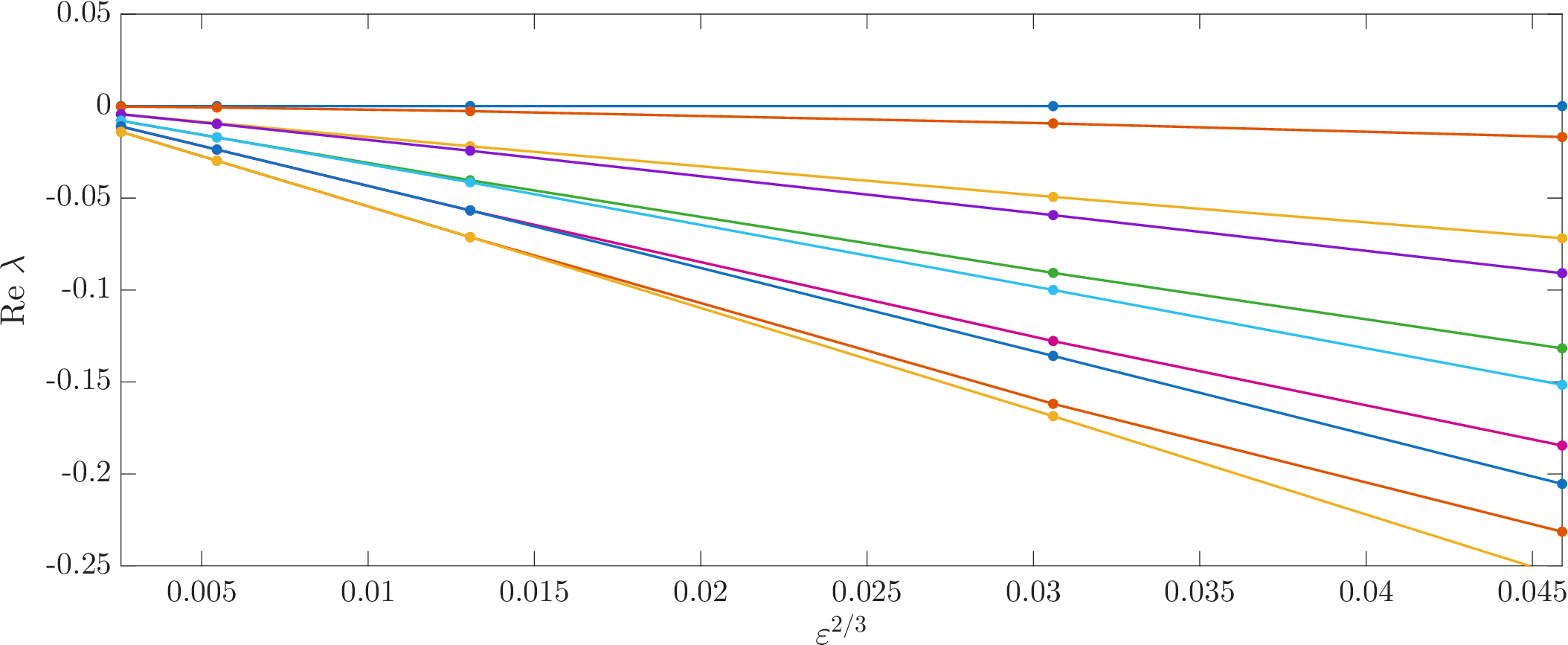}\hspace{-0.05in}\\
    \hspace{-0.2in}
    \vspace{0.0in}
    
           \includegraphics[width=0.33\linewidth]{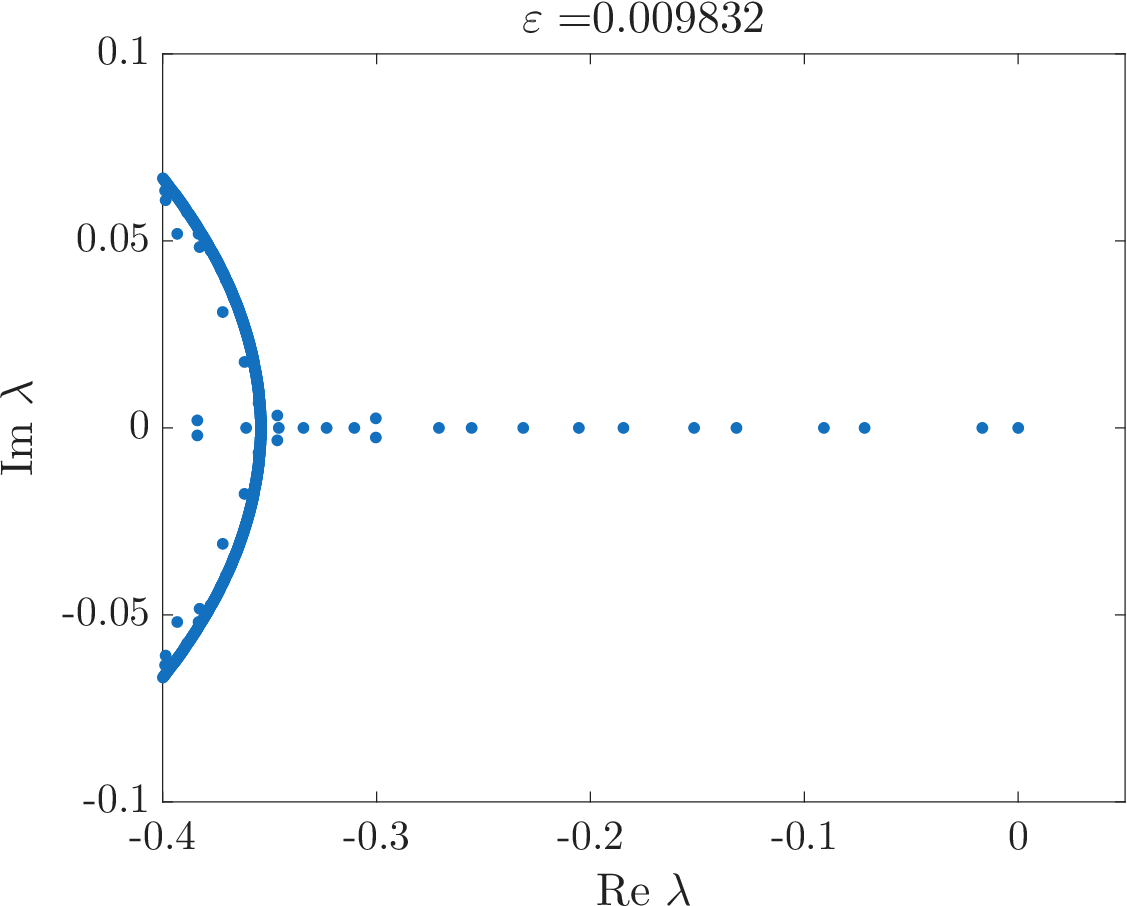}\hspace{-0.05in}
     \includegraphics[width=0.33\linewidth]{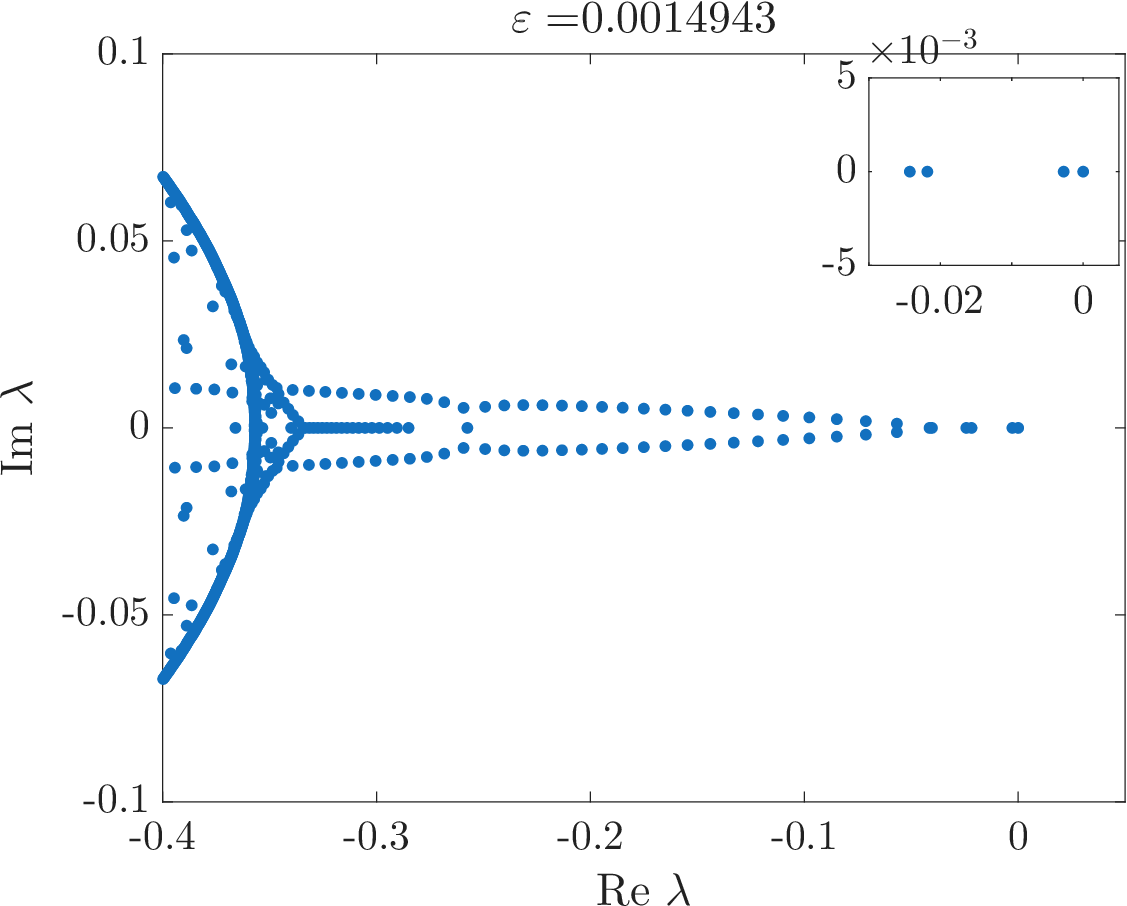}\hspace{-0.05in}
     \includegraphics[width=0.33\linewidth]{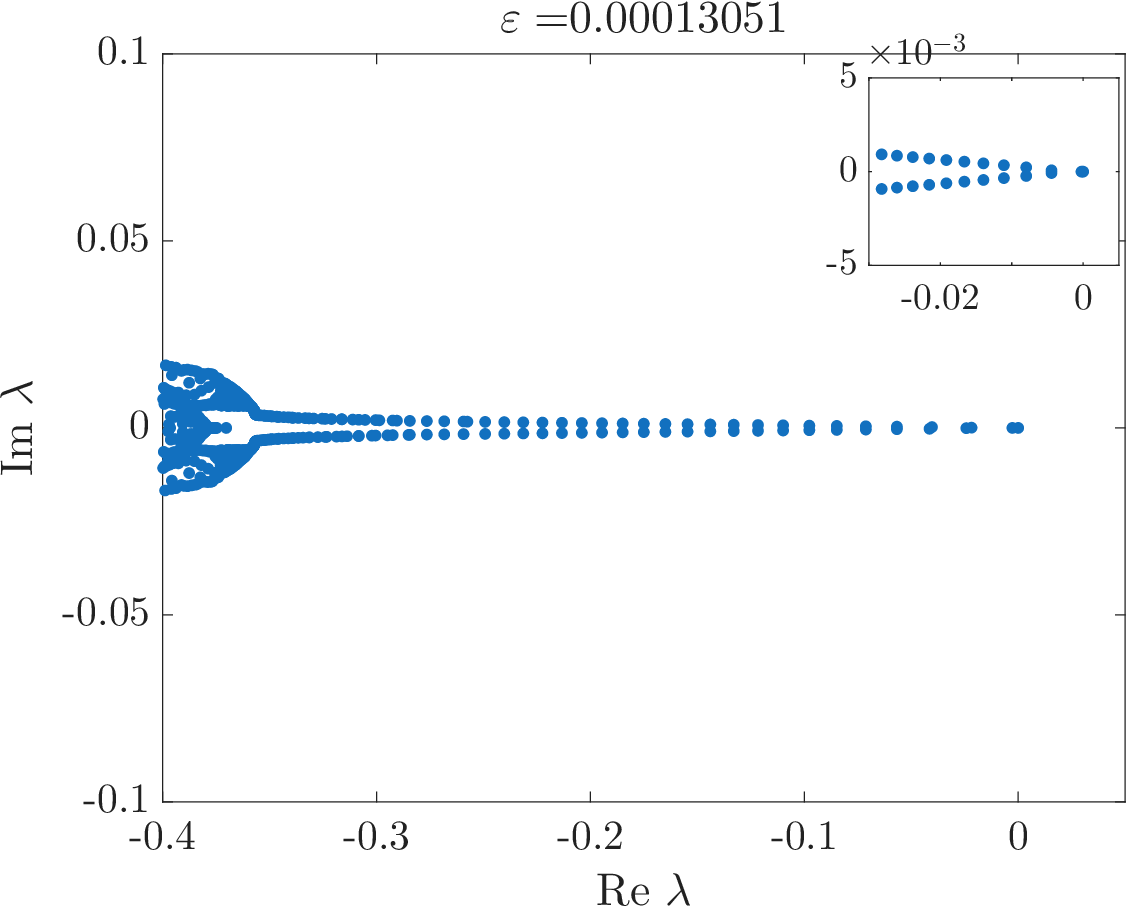}\hspace{-0.05in}
    \caption{Point spectrum for numerical discretization of $\mathcal{L}$; Top: ten most positive values of $\mathrm{Re}\,\lambda$ plotted against $\eps^{2/3}$; Bottom: plot of numerical spectrum for $\eps =0.0098,0.0015, 0.00013,$ from left to right; insets give zoom in near the origin. In all, $c =1$, $\alpha = -0.1,\gamma =-0.3$, %coefficients of $\mathcal{L}$ were evaluated on solution profiles from AUTO continuation which were interpolated onto a uniform mesh of large enough size to fit slow ramp, with 
   exponential weight  $\eta = 0.5$,  grid size $dx = 0.2$, and domain length chosen to fit the ramp for each $\eps$. }
    \label{f:spec_eps}
\end{figure}

Figure \ref{f:spec} depicts the results of our computations. The top figure gives the ten most positive values of $\mathrm{Re}\,\lambda$ for a range of $c$ values, with suitably chosen $\eta.$  For this particular choice of parameters, we find that fronts are (diffusively) stable for  $c\gtrsim0.41$. With exponential weights shifting pseudo spectrum to the left, we find a simple eigenvalue at $\lambda = 0$ corresponding to the bounded, but non-localized gauge invariance mode $U_0 = (0,r_f)$, and no unstable point spectrum.  For smaller values of $c$, we find that no weight $\eta>0$ can stabilize pseudo spectrum, and eigenvalues accumulate onto unstable absolute spectrum of the $\xi = -\infty$ far-field state; see for example Fig. \ref{f:spec} bottom left. This indicates the slow ramp for such $c$ values and this particular choice of parameters selects an absolutely unstable wave train in its wake, and hence we expect the resulting front to be unstable in any exponentially weighted space.

Figure \ref{f:spec_eps} depicts numerical spectrum for $c = 1$ fixed and $\eps$ varied. The top plot depicts the 10 most positive values of  $\mathrm{Re}\,\lambda$, plotted against $\eps^{2/3}$. We once again observe the simple gauge symmetry eigenvalue at $\lambda = 0$ and all other eigenvalues lying strictly in the left half plane and accumulating to $\lambda = 0$, like $\mathrm{Re}\lambda \sim -\eps^{2/3}$, consistent with our heuristic prediction made above.

\section{Slowly ramped patterns in Swift-Hohenberg}\label{s:sh}
To evidence the utility of our AC instability heuristic, we derive it for pattern-forming fronts in a slowly-ramped super-critical Swift-Hohenberg (SH) equation and compare with numerical continuation results for a range of ramp speeds $c>0$ and small $\eps$.  In particular, we consider the SH equation in the co-moving frame $\xi = x - ct$,
\beq\label{e:shc}
u_t = -(1+\partial_\xi^2)^2 u + c u_\xi+  \mu(\xi) u - u^3, 
\eeq
 with $\mu(\xi) = -\rho_0 \tanh(\eps\xi)$ for some fixed $\rho_0>0$ small.   At $\xi \rightarrow-\infty$, where $\mu\equiv\rho_0$, recall the equation supports a family of periodic wave solutions $u_p(k\xi+\omega t;k) $ with spatial wave number $k$ satisfying $0<1-k^2<\rho_0$. Since these rolls are temporal equilibria in the stationary frame, they are time-periodic with frequency $\omega = ck$ in the co-moving frame. %For small $\rho_0>0$, they satisfy $u_p(\theta;k) =  \frac{2}{\sqrt{3}}\sqrt{\rho_0 - \kappa^2} \cos(k\theta) + \mathcal{O}(\rho_0).$ At asymptotic $\xi\rightarrow+\infty$, where $\mu\equiv - \rho_0$, the trivial state $u = 0$ is nonlinearly stable. 

\begin{figure}[htbp]
    \centering
    \hspace{-0.2in}
         \includegraphics[width=0.4\linewidth]{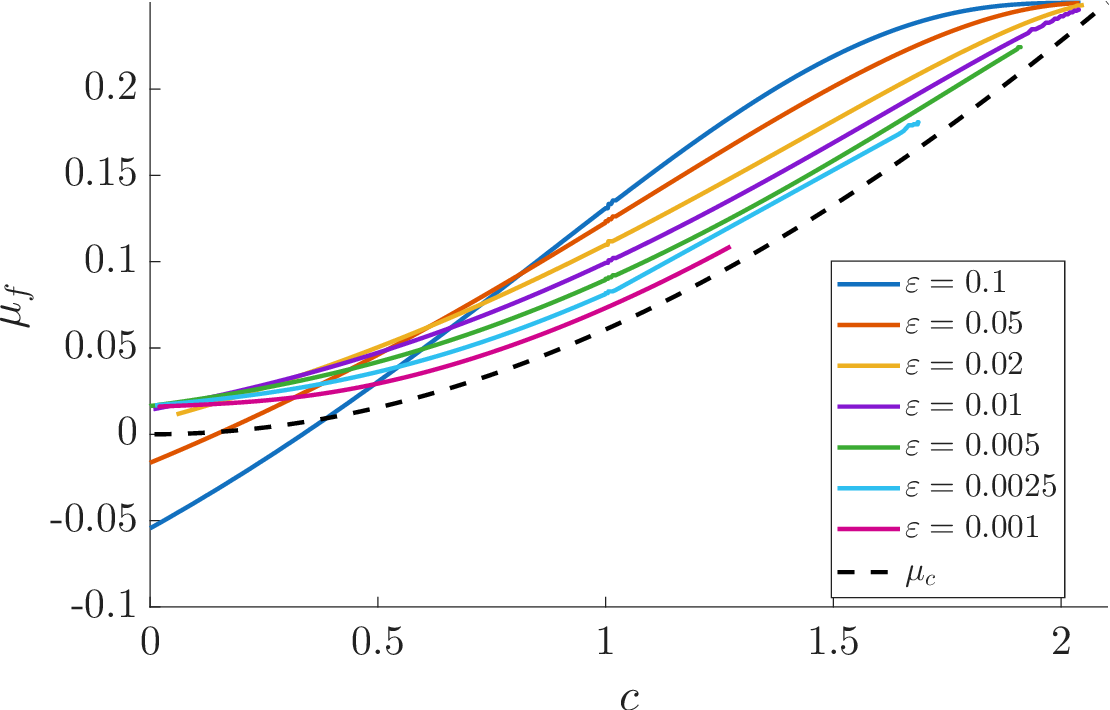}\hspace{-0.05in}
     \includegraphics[width=0.4\linewidth]{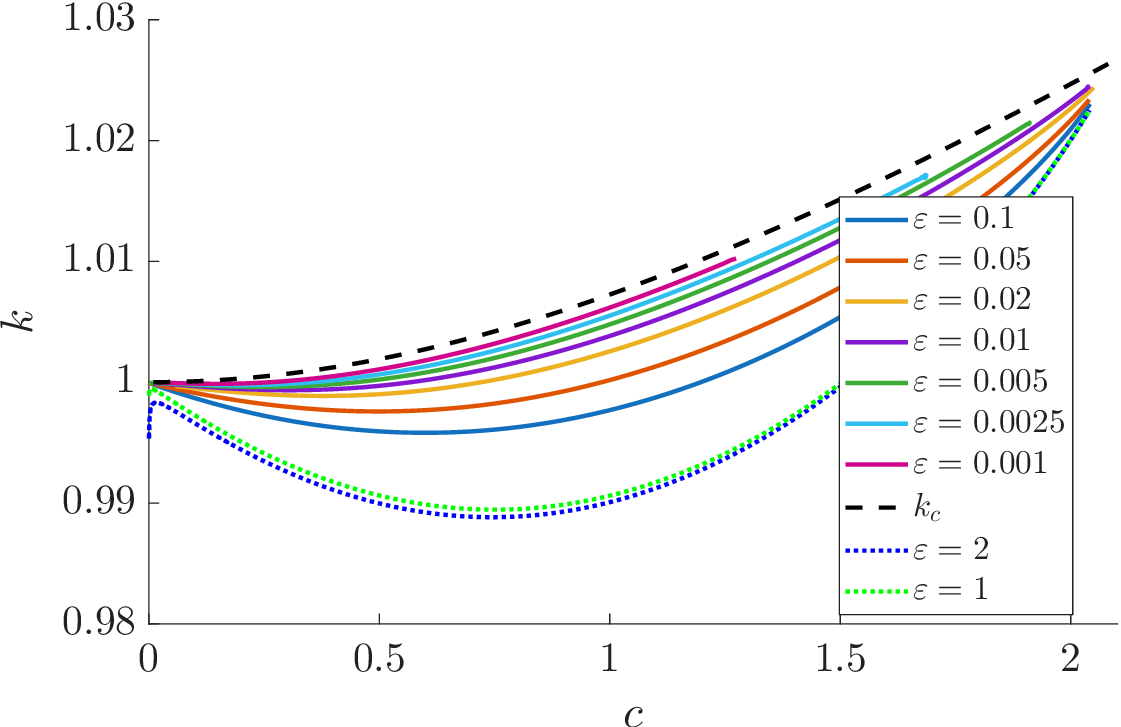}\hspace{-0.05in}
    \includegraphics[width=0.75\linewidth]{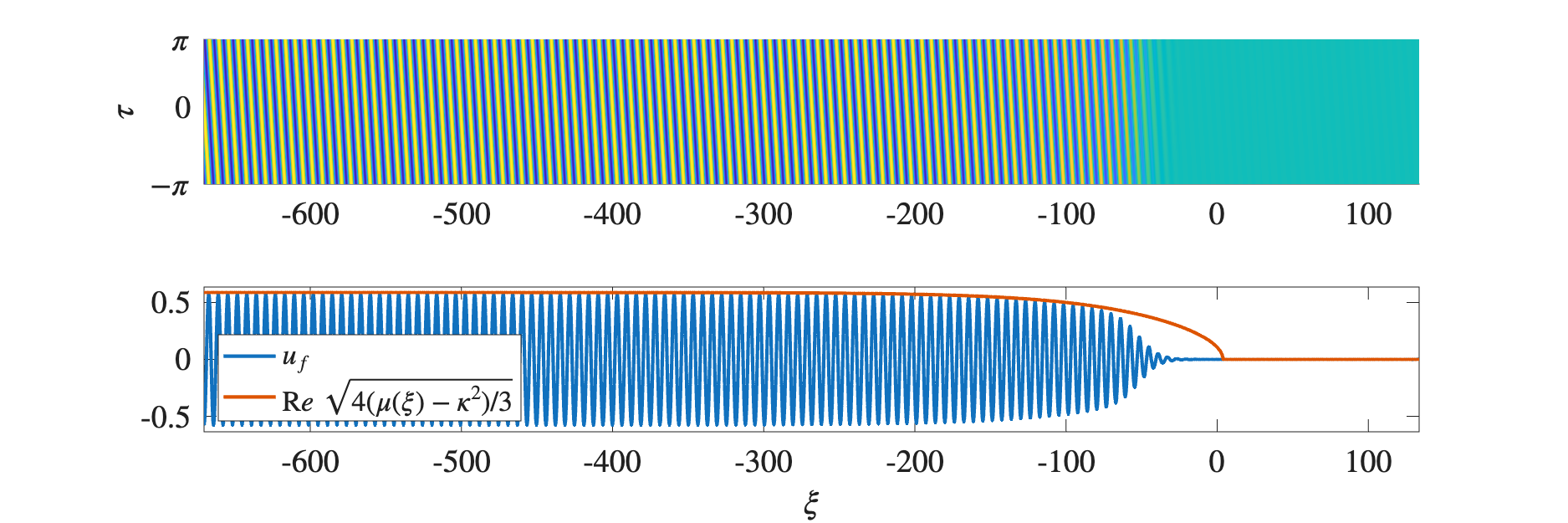}\hspace{-0.2in}
    \caption{ Top left: Front interface location $\mu_f$ (solid color) of fronts in \eqref{e:shmtw} and AC prediction $\mu_c$ (dashed black) plotted against $c$ for range of $\eps$ values. Top Right: wave number selection curves $k_f(c)$ (solid color) for a range of $\eps$ values along with the AC prediction $k_c$ (dashed black), sharp ramp curves $\eps = 1,2$ are given in dashed green and blue, curves stopped when continuation broke down; Bottom: solution profile in $(\xi,\tau)$-domain along with solution cross-section for $\tau = 0$ (blue) with slow amplitude $\mathrm{Re}\sqrt{4(\mu(\xi) - ( 1 - k_f^2)^2)/3}$ (orange), (orange), here $c = 1.024,\rho_0 = 0.25,$ and $\eps = 0.009$. }
    \label{f:sh}
\end{figure}

Following \cite[\S 1.3]{goh2023growing} (but restricting to only one spatial dimension), we seek time-periodic, or ``modulated", traveling wave solutions $u_f$ which are 1:1 resonant with the frequency $\omega$ and satisfy the following asymptotic boundary conditions. Restricting to $2\pi$-periodic functions in the scaled variable $\tau = c k t$, we obtain
\begin{align}
0 &= -(1+\partial_\xi^2)^2 u_f + (c \partial_\xi + c k \partial_\tau) u_f + \mu(\xi) u_f - u_f^3,\notag\\ 
0&=\lim_{\xi\rightarrow-=\infty} u_f(\xi,\tau) - u_p(k \xi + \tau;k),\quad 0= \lim_{\xi\rightarrow+\infty} u_f(\xi,\tau) ,\quad u_f(\cdot,\tau)  = u_f(\cdot,\tau+2\pi).\label{e:shmtw}
\end{align}
A depiction of such a front profile is given in the bottom plot of Figure \ref{f:sh}. Generically, such fronts are Fredholm index -1 in a suitable exponentially-weighted space and hence we expect there to be a locally unique wave number $k_f$ for each ramp speed $c>0$ as in CGL above; see \cite{goh2023growing,goh2018pattern} for more detail. 

We once again can use the AC instability transition to compute the corresponding $\mu_c$ and $k_c$ values and predict the front interface location $\mu_f$ and selected wave number $k_f.$ Following the same general approach as for CGL in Section \ref{ss:convabs}, this transition can be determined by inserting $v = e^{\lambda t + \nu x}$ into the constant-coefficient linear equation $v_t = -(1+\partial_\xi^2)^2 v + cv_\xi + \mu v,$ obtaining the dispersion relation $0 =d(\lambda,\nu) := -(1+\nu^2)^2 + c\nu + \mu - \lambda.$  For a given positive $\mu$ value, a standard pinched double root computation \cite{vansarloos_unstable-states} gives the linear spreading speed and  selected wave number as 
\beq
c_\mathrm{lin}(\mu) = \frac{4\left(2 + \sqrt{1+6\mu} \right)\sqrt{-1 + \sqrt{1+6\mu}}}{3\sqrt{3}},\qquad 
k_\mathrm{lin}(\mu) =  \frac{3}{8} \frac{\left( 3 + \sqrt{1+6\mu} \right)^{3/2}}{2+\sqrt{1+6\mu} },
\eeq
and selected frequency $\omega_\mathrm{lin}(\mu) = c_\mathrm{lin}(\mu) k_\mathrm{lin}(\mu)$. Fixing $c\in(0,c_\mathrm{lin}(\rho_0))$, to find the AC transition in $\mu$, we could again compute pinched double roots $(\lambda_\mathrm{br},\nu_\mathrm{br})(\mu)$ and seek
$$
\mu_c:= \mathrm{argmin}_{\mu\in(0,\rho_0)} \{ \mathrm{Re}\, \lambda_\mathrm{br}(\mu) > 0 \}.$$
 In our case, this is equivalent to inverting the formula $c = c_\mathrm{lin}(\mu)$. Using a computational algebra software (Mathematica), one finds
\begin{equation}
\mu_c = 
\frac{\left(8+g(c)\right)^2-216 c^2}{48 g(c)}
%\frac{1}{48} \left(\frac{8 \left(8-27 c^2\right)}{g(c)}+g(c)+16\right), \quad 
%+\frac{1}{3}-\frac{3538944 c^2-1048576}{786432 \sqrt[3]{729 c^4-4320 c^2+3 \sqrt{3} \sqrt{19683 c^8+139968 c^6+331776 c^4+262144 c^2}-512}}+\frac{1}{48} \sqrt[3]{729 c^4-4320 c^2+3 \sqrt{3} \sqrt{19683 c^8+139968 c^6+331776 c^4+262144 c^2}-512}
\end{equation}
for $g(c) = \sqrt[3]{-512-4320 c^2+729 c^4+3 \sqrt{3} c  \left(27 c^2+64\right)^{3/2}}$, taking the real-valued branch of the cube root. Then, as in CGL, we compute the corresponding temporal frequency $\omega_c = \omega_\mathrm{lin}(\mu_c)$ using the above formula which then yields the selected wave number $k_c = \omega_c/c.$ These predictions are compared to numerical continuation results in Figure \ref{e:sh}.

To numerically approximate these fronts, we employ the farfield-core numerical continuation approach used in \cite[\S 4.2]{goh2023growing};  note that work performed preliminary computation of fronts in the slowly-ramped \eqref{e:shmtw}. We continue fronts in the parameter $c$ for a range of small $\eps$ values, measuring the front interface location $\mu_f(c;\eps)$ and  wave number selection curves $k_f(c;\eps)$.  Figure \ref{f:sh} depicts the results of these computations and their comparison with the leading-order absolute spectrum predictions $\mu_c$ and  $k_c$ (dashed line).  

The front interface location $\mu_f$ (upper left of Fig. \ref{f:sh}) was measured by computing a front envelope, $r = |u_\mathrm{clx}|$, the amplitude of the complex analytic extension $u_\mathrm{clx} = u_f + \ri H(u_f)$, where $H$ is the Hilbert transform; see also \cite{trans_mod_dyn}. We note the front location has difference $\mu_f - \mu_c \sim -\eps^{2/3}$ for $c = \mathcal{O}_\eps(1)$. Further, for $c$ small (in particular $c \lesssim \eps^{1/3}$), we observe a diffusive advance of the front interface, $\mu_f<0$, consistent with behavior observed in the Allen-Cahn equation.

 We also observe that the curves $k_f(c;\eps)$ appear to converge pointwise to the predicted value $k_c$ as $\eps\rightarrow0^+$ for each $c>0$ fixed. Further, in the slow ramp limit, we observe the wave number selection curves no longer exhibit a sharp local maximum for $c\gtrsim0$. This maximum is characteristic of stick-slip phase pinning observed in the opposite limit $\eps\rightarrow+\infty$ of a sharp quench \cite[\S 2.3]{goh2023growing}. Thus we do not expect stick-slip phase-pinning for slow ramps in the small $c$ regime. 
 
 Due to large computational requirements, fronts were not continued into similarly small ranges of $\eps$ as for CGL above. 
 %We remark that we cannot obtain as good of an agreement as in CGl above due large computational requirements.  
 Since the parameter ramp $\mu(\xi)$ converges weakly exponentially, $\mu(\xi) - 1\sim \mathcal{O}(\re^{-\eps L})$, a large computational domain $\xi\in [-L/2, L/2]$ with $L\gg1$ is required, subsequently requiring a large number of spatial Fourier modes. For example, continuation with $\eps = 10^{-3}$ required $L = 1024\pi$ and $2^{15}$ Fourier modes in the $\xi$ direction and $2^5$ modes in the $\tau$ direction. Even with FFT-based spectral discretization and parallelized computing in Matlab using an NVIDIA GV100 graphics card, computation of one solution family in $c$ took several days. Computations became prohibitively time-consuming for yet smaller $\eps$ and we leave more refined computations in this range to a subsequent work and algorithmic development. 

%{\color{blue} Add in smaller $\eps$ values?}

Rigorous establishment of such fronts is also the subject of future work. Here one would aim to pair infinite-dimensional center-manifold approaches \cite{goh2018pattern} with an adiabatic invariant or slowly-varying Hamiltonian approach to construct fronts for small $\rho_0.$ We expect the multi-scale analysis of \cite{kuske1997pattern} to show how the relative size of the ramp amplitude $\rho_0$ compared to the ramp slope $\eps$ helps determine the leading-order modulation equation.

\section{Discussion}\label{s:disc}
This work shows how the dynamic passage from convective to absolute instability determines the far-field asymptotic pattern wave number in the wake of a rigidly propagating parameter ramp which is slowly varying in space. Using formal multiple scales analysis from a geometric singular perturbation perspective, we derived an Airy quotient inner solution which accurately characterizes the higher-order delay in the front interface. Here, the complex dispersion coefficients in CGL cause the bifurcation delay to be determined by a $\tilde\omega$-dependent passage near zeros and poles of an Airy quotient solution to the complex projective Riccati equation, instead of a unique pole as described in \cite{krupa01}.

We confirmed our predictions with numerical continuation studies and then characterized the spectral stability of these fronts with a numerical study of the point spectrum in exponentially weighted spaces. Finally, to show the generality of our approach, we computed the $\mu$-dependent AC instability transition for a slowly-ramped Swift-Hohenberg equation and derived a prediction for the asymptotic selected wave number, both of which agree well with numerical continuation results. This is somewhat unexpected as the leading-order modulation equation for the stationary ramped Swift-Hohenberg with $c = 0$, small ramp slope $0<\eps\ll1$ and small parameter ramp height $\rho_0>0$, is given by the real-coefficient Ginzburg-Landau equation \cite{kuske1997pattern}; though we remark that the specific form of the equation is dependent on the relative size between $\eps$ and $\rho_0$. With non-zero quenching speed $c>0$ as fronts are now time-periodic, we expect oscillatory terms to arise in the center-manifold or modulation equation, possibly at higher order \cite{eckmann_wayne_1991propagating}; see also the recent work \cite{dalwadi2023universal}. Therefore, we expect this framework to apply to ramped systems where the pattern-forming instability is of Turing type, in addition to the oscillatory setting, where the complex Ginzburg-Landau equation is the natural leading-order modulation equation \cite{CGL_Lambda_Stability}.
% Thus, we expect this approach to apply to other pattern-forming systems in chemistry, biology, and fluids with slowly ramped heterogeneities as mentioned in the introduction. 

There are several avenues of current and future study which stem from this work. A rigorous study of these fronts is a natural next step, and the subject of current work. Building from the GSPT framework described in Section \ref{s:het}, one would first rigorously track the invariant manifolds $W_\eps^\mathrm{u}(r_p,z_p,1)$ and $W_\eps^\mathrm{s}(0,z_-,-1)$ into a neighborhood of the singular nilpotent point $(\tilde z, r,\tilde\mu;\tilde\omega) = (0,0,0;0)$. Then, appending $\eps_\zeta = 0, \tl\omega_\zeta = 0$, these manifolds can be tracked around the singular point using geometric desingularization, blowing the origin up into a sphere in real 6-dimensional space. The slow manifold $S_{-,\eps}$ in the $r = 0$-plane can be tracked from $\tilde \mu<0$ into this neighborhood using standard Fenichel theory and exchange lemmas. We expect the singular sphere dynamics in the rescaling chart to be given by the scaled Riccati equation \eqref{e:zhatmu} for the inner expansion in Section \ref{ss:inn}. The inner solution \eqref{e:tzin} can be used to transport $W_\eps^\mathrm{s}(0,z_-,-1)$ across the sphere to an appropriate exit chart where it can be matched with $W_\eps^\mathrm{u}(r_p,z_p,1)$ by varying $\tl\omega$. A suitable Melnikov analysis in $\tomega$ would then give the higher order wave number correction and front interface delay.  

We note this approach would only apply for $c = \mathcal{O}_\eps(1)$. We remark that such a high-dimensional slow passage problem does not, to our knowledge, appear to have been previously analyzed. It would also be of interest to investigate fronts for $c\lesssim \eps^{1/3}$ where we expect a different inner solution, namely a complex extension of the Hastings-McLeod connecting solution to Painlev\'e's second equation \cite{goh2023fronts,goh2024pitchfork}, to govern the front interface. Recall also that for small $c$, the predicted wave number $k_c$ is absolutely unstable.  

On the stability side, we expect eigenvalues to be rigorously located using a slow matrix Riccati Evans function to track slow eigenvalues near the absolute spectrum. These eigenvalues accumulate onto the origin from the left due to the absolutely unstable regime where $A\sim0$ while $\mu\in(\mu_c,\mu_f)$; see for example \cite{carter2021pulse,goh2020spectral}. To our knowledge linear and nonlinear stability have not been studied. Furthermore, it would be of interest to consider the dynamics near the absolutely unstable fronts found above for $c<c_\mathrm{eck}.$ Here we expect perturbations of the front to grow pointwise and select a different coherent structure in the wake.

It would also be of interest to consider the effect of a slowly-ramped parameter quench in other nonlinearities, such as a subcritical cubic-quintic nonlinearity $\mu(\xi) A + (\rho+\ri\gamma) A|A|^2 - (1+\ri\beta)A|A|^4)$, where the heterogeneity interacts with pushed pattern forming fronts \cite{goh2016pattern}, or a subcritical nonlinearity which supports localized structures or bi-stable fronts. Here we expect the Maxwell point selection heuristic and WKB analysis of \cite{krause2024pattern} to be relevant.

Finally, it would also be of interest to rigorously investigate the front interface and pattern-selection properties of slowly-varying parameter ramps in higher dimensional spatial settings; see for example \cite{hoyle1995steady,paul2002rayleigh,malomed1993two,malomed1998coexistence} for modulation equation studies. 

\paragraph{Acknowledgements:}   RG and BK gratefully acknowledge NSF support under Award DMS-2307650. NP and KR were also supported in a summer REU through this award.

\bibliography{patterns_param_ramp_i}

\begin{thebibliography}{10}

\bibitem{CGL_Lambda_Stability}
I.~S. Aranson and L.~Kramer.
\newblock The world of the complex {G}inzburg-{L}andau equation.
\newblock {\em Rev. Mod. Phys.}, 74:99--143, Feb 2002.

\bibitem{beck2014nonlinear}
M.~Beck, T.~T. Nguyen, B.~Sandstede, and K.~Zumbrun.
\newblock Nonlinear stability of source defects in the complex
  {G}inzburg--{L}andau equation.
\newblock {\em Nonlinearity}, 27(4):739, 2014.

\bibitem{cannell83}
D.~S. Cannell, M.~A. Dominguez-Lerma, and G.~Ahlers.
\newblock Experiments on wave number selection in rotating couette-taylor flow.
\newblock {\em Phys. Rev. Lett.}, 50:1365--1368, May 1983.

\bibitem{carter2021pulse}
P.~Carter, J.~D. Rademacher, and B.~Sandstede.
\newblock Pulse replication and accumulation of eigenvalues.
\newblock {\em SIAM Journal on Mathematical Analysis}, 53(3):3520--3576, 2021.

\bibitem{couairon99}
A.~Couairon and J.-M. Chomaz.
\newblock Fully nonlinear global modes in slowly varying flows.
\newblock {\em Physics of Fluids}, 11(12):3688--3703, 12 1999.

\bibitem{dalwadi2023universal}
M.~P. Dalwadi and P.~Pearce.
\newblock Universal dynamics of biological pattern formation in spatio-temporal
  morphogen variations.
\newblock {\em Proceedings of the Royal Society A}, 479(2271):20220829, 2023.

\bibitem{NIST:DLMF}
{\it NIST Digital Library of Mathematical Functions}.
\newblock \url{https://dlmf.nist.gov/}, Release 1.2.6 of 2026-03-15.
\newblock F.~W.~J. Olver, A.~B. {Olde Daalhuis}, D.~W. Lozier, B.~I. Schneider,
  R.~F. Boisvert, C.~W. Clark, B.~R. Miller, B.~V. Saunders, H.~S. Cohl, and
  M.~A. McClain, eds.

\bibitem{dodson2024efficient}
S.~Dodson, R.~Goh, and B.~Sandstede.
\newblock Efficient numerical computation of spiral spectra with exponentially
  weighted preconditioners.
\newblock {\em IMA Journal of Numerical Analysis}, page draf139, 2026.

\bibitem{doedel2007auto}
E.~J. Doedel, A.~R. Champneys, F.~Dercole, T.~F. Fairgrieve, Y.~A. Kuznetsov,
  B.~Oldeman, R.~Paffenroth, B.~Sandstede, X.~Wang, and C.~Zhang.
\newblock Auto-07p: Continuation and bifurcation software for ordinary
  differential equations.
\newblock 2007.

\bibitem{trans_mod_dyn}
S.~Dunn, R.~Goh, and B.~Krewson.
\newblock Transverse modulational dynamics of quenched patterns.
\newblock {\em Chaos: An Interdisciplinary Journal of Nonlinear Science},
  34(6):063104, 06 2024.

\bibitem{eckmann_wayne_1991propagating}
J.-P. Eckmann and C.~E. Wayne.
\newblock Propagating fronts and the center manifold theorem.
\newblock {\em Communications in mathematical physics}, 136(2):285--307, 1991.

\bibitem{feng21}
J.~Feng, W.-H. Hsu, D.~Patterson, C.-S. Tseng, H.-W. Hsing, Z.-H. Zhuang, Y.-T.
  Huang, A.~Faedo, J.~L. Rubenstein, J.~Touboul, and S.-J. Chou.
\newblock {COUP-TFI} specifies the medial entorhinal cortex identity and
  induces differential cell adhesion to determine the integrity of its boundary
  with neocortex.
\newblock {\em Science Advances}, 7(27):eabf6808, 2021.

\bibitem{fenichel1979geometric}
N.~Fenichel.
\newblock Geometric singular perturbation theory for ordinary differential
  equations.
\newblock {\em Journal of differential equations}, 31(1):53--98, 1979.

\bibitem{goh2020spectral}
R.~Goh and B.~de~Rijk.
\newblock Spectral stability of pattern-forming fronts in the complex
  {G}inzburg-{L}andau equation with a quenching mechanism.
\newblock {\em Nonlinearity}, 35(1):170--244, 2022.

\bibitem{goh2024pitchfork}
R.~Goh, T.~J. Kaper, and A.~Scheel.
\newblock Pitchfork bifurcation along a slow parameter ramp: Coherent
  structures in the critical scaling.
\newblock {\em Studies in Applied Mathematics}, 153(2):e12702, 2024.

\bibitem{goh2023fronts}
R.~Goh, T.~J. Kaper, A.~Scheel, and T.~Vo.
\newblock Fronts in the wake of a parameter ramp: slow passage through
  pitchfork and fold bifurcations.
\newblock {\em SIAM Journal on Applied Dynamical Systems}, 22(3):2312--2356,
  2023.

\bibitem{goh2016pattern}
R.~Goh and A.~Scheel.
\newblock Pattern formation in the wake of triggered pushed fronts.
\newblock {\em Nonlinearity}, 29(8):2196--2237, 2016.

\bibitem{goh2018pattern}
R.~Goh and A.~Scheel.
\newblock Pattern-forming fronts in a {S}wift--{H}ohenberg equation with
  directional quenching—parallel and oblique stripes.
\newblock {\em Journal of the London Mathematical Society}, 98(1):104--128,
  2018.

\bibitem{goh2023growing}
R.~Goh and A.~Scheel.
\newblock Growing patterns.
\newblock {\em Nonlinearity}, 36(10):R1, 2023.

\bibitem{hasan23}
C.~R. Hasan, R.~M. C\'{a}rthaigh, and S.~Wieczorek.
\newblock Rate-induced tipping in heterogeneous reaction-diffusion systems: An
  invariant manifold framework and geographically shifting ecosystems.
\newblock {\em SIAM Journal on Applied Dynamical Systems}, 22(4):2991--3024,
  2023.

\bibitem{HISCOCK2015408}
T.~W. Hiscock and S.~G. Megason.
\newblock Orientation of {T}uring-like patterns by morphogen gradients and
  tissue anisotropies.
\newblock {\em Cell Systems}, 1(6):408 -- 416, 2015.

\bibitem{holzer2014criteria}
M.~Holzer and A.~Scheel.
\newblock Criteria for pointwise growth and their role in invasion processes.
\newblock {\em Journal of Nonlinear Science}, 24(4):661--709, 2014.

\bibitem{hoyle1995steady}
R.~Hoyle.
\newblock Steady squares and hexagons on a subcritical ramp.
\newblock {\em Physical Review E}, 51(1):310, 1995.

\bibitem{hunt91}
R.~E. Hunt and D.~G. Crighton.
\newblock Instability of flows in spatially developing media.
\newblock {\em Proceedings: Mathematical and Physical Sciences},
  435(1893):109--128, 1991.

\bibitem{kapitula2013spectral}
T.~Kapitula, K.~Promislow, et~al.
\newblock {\em Spectral and dynamical stability of nonlinear waves}, volume
  457.
\newblock Springer, 2013.

\bibitem{konow2019turing}
C.~Konow, N.~H. Somberg, J.~Chavez, I.~R. Epstein, and M.~Dolnik.
\newblock Turing patterns on radially growing domains: experiments and
  simulations.
\newblock {\em Physical Chemistry Chemical Physics}, 21(12):6718--6724, 2019.

\bibitem{kramer1982wavelength}
L.~Kramer, E.~Ben-Jacob, H.~Brand, and M.~Cross.
\newblock Wavelength selection in systems far from equilibrium.
\newblock {\em Physical Review Letters}, 49(26):1891, 1982.

\bibitem{krause2024pattern}
A.~L. Krause, V.~Klika, E.~Villar-Sep\'{u}lveda, A.~R. Champneys, and E.~A.
  Gaffney.
\newblock Pattern localization in the {S}wift–{H}ohenberg equation via slowly
  varying spatial heterogeneity.
\newblock {\em SIAM Journal on Applied Dynamical Systems}, 24(4):2804--2847,
  2025.

\bibitem{krupa01}
M.~Krupa and P.~Szmolyan.
\newblock Extending geometric singular perturbation theory to nonhyperbolic
  points---fold and canard points in two dimensions.
\newblock {\em SIAM Journal on Mathematical Analysis}, 33(2):286--314, 2001.

\bibitem{kuske1997pattern}
R.~Kuske and W.~Eckhaus.
\newblock Pattern formation in systems with slowly varying geometry.
\newblock {\em SIAM Journal on Applied Mathematics}, 57(1):112--152, 1997.

\bibitem{malomed1993two}
B.~Malomed and A.~Nepomnyashchy.
\newblock Two-dimensional stability of convection rolls in the presence of a
  ramp.
\newblock {\em Europhysics Letters}, 21(2):195, 1993.

\bibitem{malomed1993ramp}
B.~A. Malomed.
\newblock Ramp-induced wave-number selection for traveling waves.
\newblock {\em Physical Review E}, 47(4):R2257, 1993.

\bibitem{malomed1998coexistence}
B.~A. Malomed and A.~A. Nepomnyashchy.
\newblock Coexistence of patterns on a ramp of overcriticality.
\newblock {\em Physics Letters A}, 244(1-3):92--96, 1998.

\bibitem{melrose1987boundary}
R.~B. Melrose and M.~Taylor.
\newblock Boundary problems for wave equations with grazing and gliding rays.
\newblock {\em preprint}, 1987.

\bibitem{mielke2002ginzburg}
A.~Mielke.
\newblock The {G}inzburg-{L}andau equation in its role as a modulation
  equation.
\newblock In {\em Handbook of dynamical systems}, volume~2, pages 759--834.
  Elsevier, 2002.

\bibitem{miguez2006effect}
D.~G. M{\'\i}guez, M.~Dolnik, A.~P. Mu{\~n}uzuri, and L.~Kramer.
\newblock Effect of axial growth on {T}uring pattern formation.
\newblock {\em Physical Review Letters}, 96(4):048304, 2006.

\bibitem{MishchenkoRozov1980}
E.~F. Mishchenko and N.~K. Rozov.
\newblock {\em Differential Equations with Small Parameters and Relaxation
  Oscillations}, volume~13 of {\em Mathematical Concepts and Methods in Science
  and Engineering}.
\newblock Plenum Press, New York, NY, 1980.

\bibitem{morrissey2015characterizing}
D.~Morrissey and A.~Scheel.
\newblock Characterizing the effect of boundary conditions on striped phases.
\newblock {\em SIAM Journal on Applied Dynamical Systems}, 14(3):1387--1417,
  2015.

\bibitem{paul2002rayleigh}
M.~Paul, M.~Cross, and P.~Fischer.
\newblock Rayleigh-{B}{\'e}nard convection with a radial ramp in plate
  separation.
\newblock {\em Physical Review E}, 66(4):046210, 2002.

\bibitem{pomeau1981wavelength}
Y.~Pomeau and S.~Zaleski.
\newblock Wavelength selection in one-dimensional cellular structures.
\newblock {\em Journal de Physique}, 42(4):515--528, 1981.

\bibitem{pomeau1983pattern}
Y.~Pomeau and S.~Zaleski.
\newblock Pattern selection in a slowly varying environment.
\newblock {\em Journal de Physique Lettres}, 44(4):135--141, 1983.

\bibitem{rademacher2007computing}
J.~D. Rademacher, B.~Sandstede, and A.~Scheel.
\newblock Computing absolute and essential spectra using continuation.
\newblock {\em Physica D: Nonlinear Phenomena}, 229(2):166--183, 2007.

\bibitem{rehberg1987rayleigh}
I.~Rehberg and H.~Riecke.
\newblock Rayleigh number ramps cause moving convection patterns.
\newblock In {\em The Physics of Structure Formation: Theory and Simulation},
  pages 142--152. Springer, 1987.

\bibitem{riecke1987perfect}
H.~Riecke and H.-G. Paap.
\newblock Perfect wave-number selection and drifting patterns in ramped
  {T}aylor vortex flow.
\newblock {\em Physical review letters}, 59(22):2570, 1987.

\bibitem{sandstede2000absolute}
B.~Sandstede and A.~Scheel.
\newblock Absolute and convective instabilities of waves on unbounded and large
  bounded domains.
\newblock {\em Physica D: Nonlinear Phenomena}, 145(3-4):233--277, 2000.

\bibitem{scheel2018wavenumber}
A.~Scheel and J.~Weinburd.
\newblock Wavenumber selection via spatial parameter jump.
\newblock {\em Philosophical Transactions of the Royal Society A: Mathematical,
  Physical and Engineering Sciences}, 376(2117):20170191, 2018.

\bibitem{digit}
R.~Sheth, L.~Marcon, M.~F. Bastida, M.~Junco, L.~Quintana, R.~Dahn, M.~Kmita,
  J.~Sharpe, and M.~A. Ros.
\newblock Hox genes regulate digit patterning by controlling the wavelength of
  a {T}uring-type mechanism.
\newblock {\em Science}, 338(6113):1476--1480, 2012.

\bibitem{steinberg1985pattern}
V.~Steinberg, G.~Ahlers, and D.~S. Cannell.
\newblock Pattern formation and wave-number selection by
  {R}ayleigh--{B}{\'e}nard convection in a cylindrical container.
\newblock {\em Physica Scripta}, 1985(T9):97, 1985.

\bibitem{vansarloos_unstable-states}
W.~van Saarloos.
\newblock Front propagation into unstable states.
\newblock {\em Physics Reports}, 386(2):29 -- 222, 2003.

\bibitem{van1992fronts}
W.~van Saarloos and P.~C. Hohenberg.
\newblock Fronts, pulses, sources and sinks in generalized complex
  {G}inzburg-{L}andau equations.
\newblock {\em Physica D: Nonlinear Phenomena}, 56(4):303--367, 1992.

\end{thebibliography}
\bibliographystyle{plain}

\end{document}